\documentclass[preprint,floatfix]{revtex4}

\pdfoutput=1

\usepackage{pdfpages}
\usepackage{graphicx} 
\usepackage{dcolumn} 
\usepackage{amsmath}
\usepackage{amssymb}
\usepackage{bm}
\usepackage{times}
\usepackage{ulem}
\usepackage[justification=raggedright]{caption}

  \usepackage{boondox-cal}

\usepackage{subfigure}
\usepackage{tabularx}
\usepackage{appendix}

\usepackage{amstext}
\usepackage{array}

\usepackage{bbm}
\usepackage{blkarray}

\definecolor{wine-stain}{rgb}{0.5,0,0} 
\definecolor{bblue}{rgb}{0,0.0,0.5} 

\usepackage[colorlinks=true
,urlcolor=blue
,anchorcolor=blue
,citecolor=blue 
,filecolor=blue
,linkcolor=wine-stain
,menucolor=blue
,pagecolor=blue
,linktocpage=true
,pdfproducer=medialab
,linktoc=all 
]{hyperref}

\usepackage{booktabs}  

\allowdisplaybreaks

\newcommand{\ncmd}{\newcommand}

\ncmd{\lt}{\left}
\ncmd{\rt}{\right}
\ncmd{\tr}[1]{~\mbox{tr}\lt\{ {#1}\rt\}}
\ncmd{\half}{\frac{1}{2}}
\ncmd{\eps}{\epsilon}
\ncmd{\veps}{\varepsilon}
\ncmd{\dgr}{\dagger}
\ncmd{\sig}{\sigma}
\ncmd{\gam}{\gamma}
\ncmd{\rtarw}{\rightarrow}
\ncmd{\Rt}{\Rightarrow}
\ncmd{\abs}[1]{\lt\cb{#1}\rt\cb}
\ncmd{\avg}[1]{\lt\lb{#1}\rt\rb}
\ncmd{\sgn}[1]{\mbox{sgn}\lt(#1\rt)}
\ncmd{\kap}{\kappa}
\ncmd{\wtil}[1]{\widetilde{#1}}
\ncmd{\thrfr}{\therefore}
\ncmd{\eq}[1]{Eq. \eqref{#1}}
\ncmd{\fig}[1]{Fig. \ref{#1}}
\ncmd{\ordr}[1]{\mathcal{O}\lt(#1\rt)}
\ncmd{\dsty}{\displaystyle}
\ncmd{\alert}[1]{\color{red}{#1}}
\ncmd{\mc}{\mathcal}
\ncmd{\mbf}[1]{\mathbf{#1}}
\ncmd{\Deriv}[2]{\frac{d{#1}}{d{#2}}}
\ncmd{\ParDeriv}[2]{\frac{\partial{#1}}{\partial{#2}}}
\ncmd{\step}[1]{\Theta\lt(#1\rt)}
\ncmd{\td}{\tilde} 
\ncmd{\what}{\widehat}

\ncmd{\hphi}{\hat \phi} 
\ncmd{\hpi}{\hat \pi} 
\ncmd{\hK}{\hat K} 
\ncmd{\hL}{\hat L}

\ncmd{\bqa}{\begin{eqnarray}} 
\ncmd{\eqa}{\end{eqnarray}}

\ncmd{\nn}{\nonumber \\}
\ncmd{\comment}[1]{{\color{red}{#1}}}
\definecolor{new_color}{RGB}{50,155,0}

\ncmd{\vrho}{\varrho}

\ncmd{\qstar}{\mathcal{q}}
\ncmd{\sstar}{\mathcal{ s}}
\ncmd{\pcstar}{ \mathcal{p}_{c}}
\ncmd{\tcstar}{ \mathcal{t}_{c}}

\ncmd{\ha}{\frac{1}{2}}

\ncmd{\lb}{\big<}
\ncmd{\rb}{\big>}
\ncmd{\cb}{\big|}
\ncmd{\cH}{{\cal H}}
\ncmd{\lc}{l_c}
\ncmd{\Lc}{l_c}
\ncmd{\ta}{\tilde \alpha}

\ncmd{\bt}{\bar t}
\ncmd{\bp}{\bar p}

\ncmd{\btc}{\bar t_c}
\ncmd{\bpc}{\bar p_c}
\ncmd{\bs}{\bar s}
\ncmd{\bq}{\bar q}
\ncmd{\bU}{\bar U}
\ncmd{\bQ}{\bar Q}
\ncmd{\bW}{\bar W}
\ncmd{\by}{\bar y}

\ncmd{\rot}{\br_A, \br_B}
\ncmd{\rto}{\br_B, \br_A}

\ncmd{\PhiAi}{\Phi^A_{~~i}}
\ncmd{\PhiAj}{\Phi^A_{~~j}}
\ncmd{\PhiAk}{\Phi^A_{~~k}}
\ncmd{\PhiBi}{\Phi^B_{~~i}}
\ncmd{\PhiBj}{\Phi^B_{~~j}}
\ncmd{\PhiBk}{\Phi^B_{~~k}}

\ncmd{\PiiA}{\Pi^i_{~~A}}
\ncmd{\PijA}{\Pi^j_{~~A}}
\ncmd{\PikA}{\Pi^k_{~~A}}

\ncmd{\PiiB}{\Pi^i_{~~B}}
\ncmd{\PijB}{\Pi^j_{~~B}}
\ncmd{\PikB}{\Pi^k_{~~B}}

\ncmd{\hPhi}{\hat \Phi}
\ncmd{\hPi}{\hat \Pi}
\ncmd{\hG}{{\hat G}}
\ncmd{\hH}{\hat H}
\ncmd{\hJ}{\hat J}

\ncmd{\GLL}{$GL(L, \mathbb{R})$ }
\ncmd{\SLL}{$SL(L, \mathbb{R})$ }

\ncmd{\bk}{{\bf k}}
\ncmd{\br}{{\bf r}}

\ncmd{\qw}{q^{(w)}} 
\ncmd{\qo}{q^{(1)}} 
\ncmd{\qt}{q^{(2)}}

\ncmd{\tds}{S}
\ncmd{\tdt}{T }
\ncmd{\tdp}{P}
\ncmd{\tdX}{X}
\ncmd{\tdY}{Y}

\ncmd{\hd}{\frac{r'}{2}}

\ncmd{\hPhiAi}{\hat \Phi^A_{~~i}}
\ncmd{\hPhiAj}{\hat \Phi^A_{~~j}}
\ncmd{\hPhiAk}{\hat \Phi^A_{~~k}}
\ncmd{\hPhiBi}{\hat \Phi^B_{~~i}}
\ncmd{\hPhiBj}{\hat \Phi^B_{~~j}}
\ncmd{\hPhiBk}{\hat \Phi^B_{~~k}}

\ncmd{\hPiiA}{\hat \Pi^i_{~~A}}
\ncmd{\hPijA}{\hat \Pi^j_{~~A}}
\ncmd{\hPikA}{\hat \Pi^k_{~~A}}
\ncmd{\hPiiB}{\hat \Pi^i_{~~B}}
\ncmd{\hPijB}{\hat \Pi^j_{~~B}}
\ncmd{\hPikB}{\hat \Pi^k_{~~B}}

\ncmd{\cA}{{\cal A}}
\ncmd{\cB}{{\cal B}}
\ncmd{\cC}{{\cal C}}
\ncmd{\cD}{{\cal D}}

\ncmd{\cG}{{\cal G}}
\ncmd{\cGI}{{\cal G}^{-1}}

\ncmd{\cGij}{{\cal G}^i_{~~j}}
\ncmd{\cGji}{{\cal G}^j_{~~iå}}

\ncmd{\cGIij}{({\cal G}^{-1})^i_{~~j}}
\ncmd{\cGIji}{({\cal G}^{-1})^j_{~~iå}}

\ncmd{\cstr}{\cb s, t_1, t_2 \rb}

\ncmd{\czr}{ \cb 0 \rb } 
\ncmd{\lzc}{ \lb 0 \cb }

\ncmd{\cepr}{ \cb q, \phi, \varphi \rb^{'} }

\ncmd{\tx}{\td x}
\ncmd{\ty}{\td y}

\ncmd{\cphir}{\cb \phi \rb}
\ncmd{\lphic}{\lb \phi \cb}

\ncmd{\ctr}{\cb T \rb}
\ncmd{\ltc}{\lb T \cb}
\ncmd{\cpr}{\cb p \rb}
\ncmd{\lpc}{\lb p \cb}

\ncmd{\cTr}{\cb {\cal T} \rb}

\ncmd{\ccr}{\cb \chi \rb}
\ncmd{\lcc}{\lb \chi \cb}

\ncmd{\cS}{{\cal S}}
\ncmd{\tcS}{\tilde { \cal S}}

\ncmd{\sqg}{\sqrt{|g(x)|}}
\ncmd{\gmn}{E_{\mu i}}

\ncmd{\emi}{E_{\mu i}}
\ncmd{\emj}{E_{\mu j}}
\ncmd{\eni}{E_{\nu i}}
\ncmd{\enj}{E_{\nu j}}
\ncmd{\de}{|E(x)|}

\ncmd{\gemn}{g_{E,\mu \nu}}
\ncmd{\gemnp}{g_{E',\mu \nu}}

\ncmd{\bgmn}{\bar E_{\mu i}}
\ncmd{\Fs}{F(\sigma)}
\ncmd{\cJ}{ {\cal J}\left(E',\sigma';E,\sigma \right)  }
\ncmd{\Si}{ \Sigma \left(E',\sigma';E,\sigma \right)  }

\ncmd{\grs}{g_{\alpha \beta}}
\ncmd{\pmn}{\pi^{\mu i}}
\ncmd{\prs}{\pi^{\alpha \beta}}
\ncmd{\fs}{ \frac{3}{\sqrt{2}} \sigma }
\ncmd{\fst}{ 3 \sqrt{2} \sigma }

\ncmd{\ee}{entanglement entropy }

\ncmd{\cT}{{\cal T} }
\ncmd{\cP}{{\cal P} }
\ncmd{\cV}{{\mathbcal V} }
\ncmd{\cW}{{\cal W} }

\ncmd{\ccw}{\cos( 2 \omega \delta) }
\ncmd{\ssw}{\sin (2 \omega \delta) }

\makeatletter
\newcommand*{\rom}[1]{\expandafter\@slowromancap\romannumeral #1@}
\makeatother

\begin{document}

\title{
A model of quantum gravity with emergent spacetime 
}

\author{Sung-Sik Lee\\
\vspace{0.3cm}
{\normalsize{Department of Physics $\&$ Astronomy, McMaster University, Hamilton ON, Canada}}
\vspace{0.2cm}\\
{\normalsize{Perimeter Institute for Theoretical Physics, Waterloo ON, Canada}}
}

\date{\today}

\begin{abstract}

We construct a model of quantum gravity
in which dimension, topology and geometry 
of spacetime are  dynamical.
The microscopic degree of freedom is a real rectangular matrix
whose rows label internal flavours, 
and columns label spatial sites.
In the limit that the size of the matrix is large,
the sites can collectively form a spatial manifold.
The manifold is
determined from the pattern of entanglement 
present across local Hilbert spaces 
associated with column vectors of the matrix.
With no structure of manifold fixed in the background, 
the spacetime gauge symmetry is generalized to a  
group that includes diffeomorphism in arbitrary dimensions.
The momentum and Hamiltonian 
that generate the generalized diffeomorphism
obey a first-class constraint algebra at the quantum level.
In the classical limit, the constraint algebra of the general relativity 
is reproduced as a special case.
The first-class nature of the algebra allows one to express 
the projection of a quantum state of the matrix to a gauge invariant state 
as a path integration of dynamical variables
that describe collective fluctuations of the matrix.
The collective variables describe dynamics of emergent spacetime,
where multi-fingered times arise as Lagrangian multipliers that enforce the gauge constraints.
If the quantum state has a local structure of entanglement,
a smooth spacetime with well-defined dimension, topology, signature and geometry 
emerges at the saddle-point,
and the spin two mode that determines the geometry
can be identified.
We find a saddle-point solution that describes 
a series of $(3+1)$-dimensional de Sitter-like spacetimes with the Lorentzian signature 
bridged by Euclidean spaces  in between.  
The phase transitions between spacetimes with different signatures are 
caused by Lifshitz transitions in which the pattern of entanglement
is rearranged across the system.  
Fluctuations of the collective variables are described by 
bi-local fields that propagate in the spacetime set up by 
the saddle-point solution.

\end{abstract}

\maketitle


\newpage

{
\hypersetup{linkcolor=bblue}
\hypersetup{
    colorlinks,
    citecolor=black,
    filecolor=black,
    linkcolor=black,
    urlcolor=black
}
\tableofcontents
}


\newpage

\section{Introduction}

According to Einstein's theory of general relativity,  
gravity originates from dynamical geometry\cite{Einstein:1916vd}.
While the theory has been extremely successful 
in explaining a myriad of phenomena in the classical regime, 
understanding the quantum nature of gravity 
remains an outstanding problem\cite{
PhysRev.160.1113,
Regge1961,
Weinberg:1980gg,
PhysRevLett.57.2244,
PhysRevD.79.084008,
FRANCESCO19951,
Maldacena:1997re,
Witten:1998qj,
Gubser:1998bc,
PhysRevD.55.5112,
polchinski_1998,
Oriti:2011jm,
Rovelli:2014ssa,
Carlip_2001,
book_Hamber,
2019arXiv190508669L}.
%
One notoriously difficult problem in quantum gravity is 
to tame quantum fluctuations at short distance scales 
while preserving the essential aspects of the general relativity at long distances\cite{1974AIHPA..20...69T,GOROFF1986709}.
The crucial feature that a successful quantum theory of gravity should reproduce 
in the continuum limit is the diffeomorphism invariance,
which largely fixes the theory at long distance scales.

Quantizing fluctuations of geometry in a fixed dimension,
either in the form of metric or a new degree of freedom,
has provided important insights into quantum gravity.
However, this may not give the complete picture.
If metric is dynamical, 
it is natural to posit that 
dimension and topology of spacetime  
are also dynamical.
In the presence of strong quantum fluctuations of geometry,
there is in priori no reason why topology and dimension of spacetime 
remain well-defined.
Ideally, a full theory of quantum gravity should be able to describe 
phenomena in which not only geometry but also dimension and topology
evolve dynamically. 
A background independent theory in the strongest sense
should include all of dimension, topology and geometry as 
dynamical degrees of freedom :
the entirety of spacetime should emerge\cite{2007qsst.conf..163S}.

Background independent theories 
can not be local theories\cite{PhysRev.132.2788}
because a fixed notion of locality can not be defined
without specifying dimension, topology and geometry
in the background\cite{2015PhRvL.114c1104M}.
Nonetheless, the success of local quantum field theories 
as a low-energy description of our universe implies
that local effective theories should arise as approximate descriptions 
within states that describe classical spacetimes\cite{Donoghue:1995cz,Burgess:2004aa}.
The degree of locality in the effective theory should be determined 
with respect to the metric of the classical geometry.
Because the classical geometry is state dependent,
so is the locality of the effective field theory.
Background independent theories from which local effective theories emerge,
while being non-local in the strict sense,
should have a weaker notion of locality called relative locality\cite{Lee2018}.
This may be an essential ingredient
that needs to be taken into account in understanding 
quantum gravity.

The AdS/CFT correspondence\cite{
Maldacena:1997re,Witten:1998qj,Gubser:1998bc} 
provides a  route to quantum gravity
by mapping a quantum field theory without gravity 
to a string theory that includes dynamical gravity 
in one higher dimensional space.
Although a background independent
non-perturbative formulation of the string theory
is not known yet,
the AdS/CFT correspondence already
provides an important clue on
the microscopic origin of gravity
: geometry is nothing but a coarse-grained 
variable that controls entanglement (among other things) 
of ordinary quantum matter\cite{
PhysRevLett.96.181602,
1126-6708-2007-07-062,
VanRaamsdonk:2010pw,
Casini2011,
Lewkowycz2013,
doi:10.1002/prop.201300020,
PhysRevLett.75.1260}.
This suggests an approach to quantum gravity
that we adopt in this paper.
Here, the fundamental degrees of freedom are ordinary quantum matter defined on a set.
The set is what is to become a spatial manifold, but
it does not have a fixed structure of manifold.
Dimension, topology and geometry of the set are all collective degrees of freedom of the matter defined on the set.
In particular, distances between points in the set are determined from 
the amount of entanglement between points.
Two points that are strongly (weakly) entangled are deemed to be close (far)\cite{PhysRevD.95.024031,Lee2018}.
The pattern of entanglement determines the connectivity among points in the set. 
The connectivity, in turn, determines a manifold, 
if its pattern exhibits a local structure.

Within this framework, a Hamiltonian that governs the dynamics of the underlying quantum matter naturally
induces dynamics of collective variables
that describe dimension, topology and geometry 
of the manifold.
As much as the underlying quantum matter is dynamical,
the emergent manifold is fully dynamical.
One main goal in this program is to construct a Hamiltonian for the matter 
such that the induced Hamiltonian for the collective variables
reduces to that of the general relativity in a classical limit. 
Such a Hamiltonian must be relatively local
if it is to induce background independent dynamics for the collective variables
and admit a local effective theory description
for perturbative fluctuations around semi-classical states\cite{Lee2018}.
Since geometric distance is nothing but a collective property of the underlying matter,
the effective strength of interactions between points should be state dependent.
In a state that describes a geometry 
that gives rise to a small (large) proper distance between two points,
the interaction between them should be strong (weak) in the relatively local Hamiltonian.

Recently, a simple relatively local model 
has been constructed.
In the background independent model, 
the collective variables that describe
dynamics of dimension, topology and geometry
are classical in the large $N$ limit,
where $N$ is the number of flavours of underlying quantum matter\cite{Lee2019}.  
In different states, the same Hamiltonian acts as local Hamiltonians 
defined on manifolds with different dimensions, topologies and geometries.
However, the model is not a theory of gravity yet
because it lacks the diffeomorphism invariance.  
The goal of this paper is to construct a relatively local model 
of quantum gravity.
For other related approaches to quantum gravity,
see Refs. \cite{
PhysRevD.18.3833,
PhysRevD.77.104029,
2006hep.th...11197K,
doi:10.1142/S0218271818460057,
PhysRevD.95.024031}.

\subsection{Conceptual overview}

In this section, we provide a conceptual overview 
without technical details.

\vspace{1cm}
\begin{figure}[ht]
\begin{center}
\centering
\includegraphics[scale=0.5]{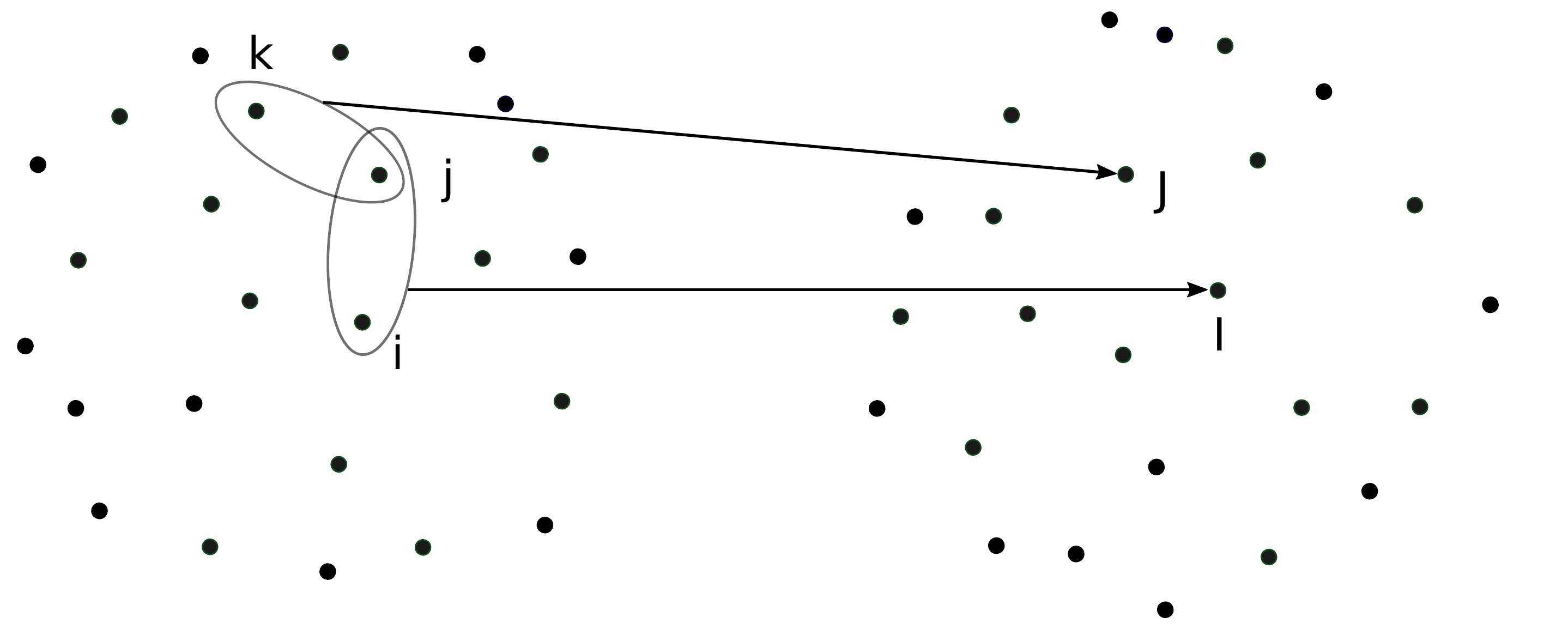}
\end{center}
\caption{
Each column of the $M \times L$ matrix represents a point in the set of $L$ sites
on the left panel.
At each site, one can define a local Hilbert space spanned by a column of $M$ scalars.
Under \SLL that acts on the matrix from the right,
column vectors are linearly superposed to form a new set of $L$ column vectors.
The rotated column vectors define a new set of sites and associated local Hilbert spaces
shown in the right panel.
A choice of local Hilbert spaces into which the total Hilbert space is decomposed
is called a frame.  
}
\label{fig:matrix}
\end{figure}

The microscopic degree of freedom of the theory 
is a real rectangular matrix 
with $M$ rows and $L$ columns with $M \gg L \gg 1$.
While the row index labels  internal flavours,
the column index plays the role of sites.
Each column defines a local Hilbert space 
spanned by states of $M$ real scalars.
A choice of such local Hilbert spaces is called a {\it frame}.
There are multiple ways to choose a frame 
as $L$ column vectors can be linearly superposed
to form a different set of column vectors.
A set of rotated vectors form a new frame.
Rotations of frame 
that preserve the norm of the Hilbert space 
are generated by the special linear group, 
$SL(L, \mathbb{R})$
that multiplies the matrix from the right.
Once a frame is chosen, 
the total Hilbert space can be written 
as a direct product of the local Hilbert spaces 
defined in that frame\footnote{
A frame naturally defines a set of local observers
who can access the local Hilbert space at each site.
}.
This is illustrated in \fig{fig:matrix}.

\vspace{1cm}
\begin{figure}[ht]
\begin{center}
\centering
\includegraphics[scale=0.5]{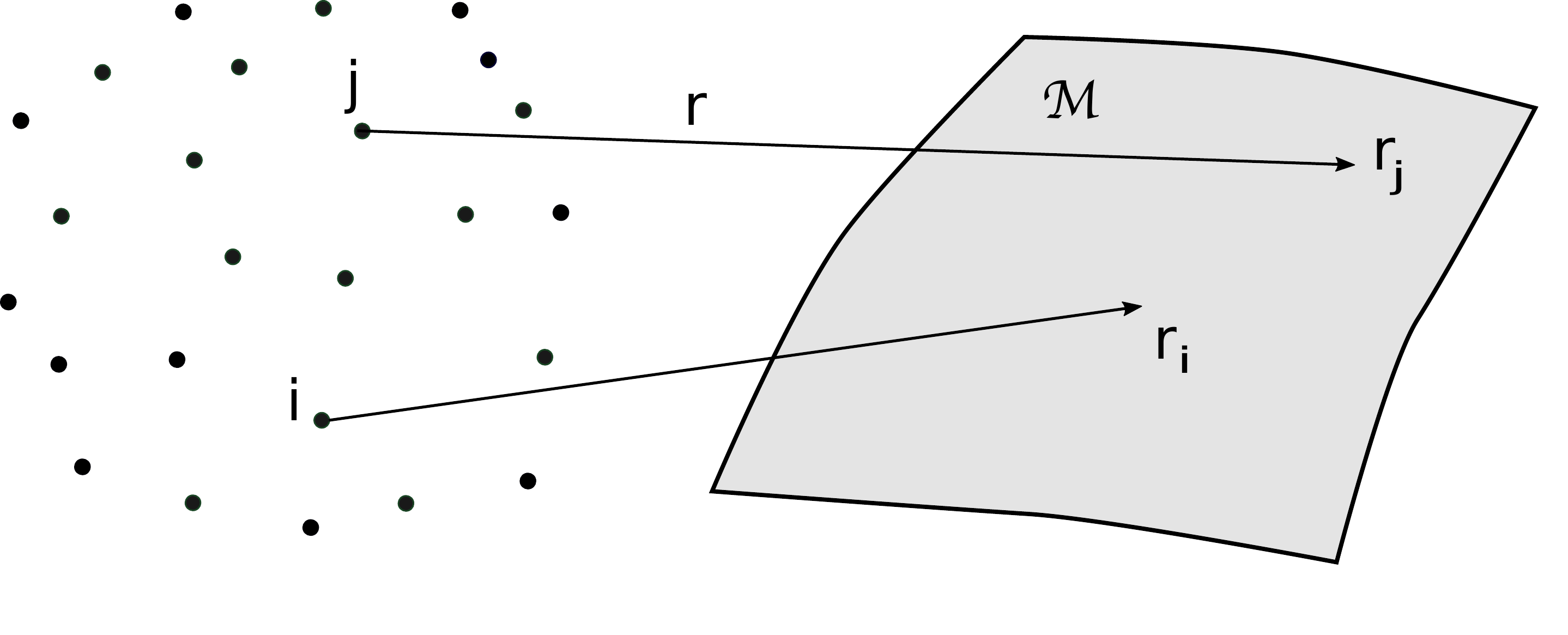}
\end{center}
\caption{
A state has a local structure in a frame
if there exists a mapping $r$ from the set of sites into a Riemannian manifold ${\cal M}$  such that
the mutual information between sites $i$ and $j$
is proportional to $e^{-\frac{d(r_i,r_j)}{\xi}}$
to the leading order in $d(r_i,r_j)$ 
for any $i$ and $j$,
where  
 $d(r_i,r_j)$ is the proper distance between 
$r_i$ and $r_j$ in the Riemannian manifold
and $\xi$ is a constant.
}
\label{fig:local_structure}
\end{figure}

In a frame, the collection of sites can form a spatial manifold 
for $L \gg 1$.
However, no structure of manifold, 
not even its existence, is fixed in the theory.
Instead, the existence of spatial manifold and its structure, if exists, 
depend on quantum state of the matrix in the following way. 
Suppose we compute the mutual informations between all pairs of sites in a frame.
The mutual information between sites $i$ and $j$ is given by
$I_{ij} = S_i + S_j - S_{i \cup j}$,
where $S_A$ is the von Neumann entanglement entropy of subset $A$.
Then we ask if there exists a Riemannian manifold 
into which the sites can be embedded
such that $- \ln I_{ij}$ is proportional to the proper distance
between the images of $i$ and $j$ in the Riemannian manifold
to the leading order in the proper distance\footnote{
The mutual information has been used to extract the bulk
geometry in a holographic mapping in Ref. \cite{2013arXiv1309.6282Q}.
}
If there exists such a Riemannian manifold, 
we say that the state has a local structure (see \fig{fig:local_structure}).
Roughly speaking, a state with a local structure is a short-range entangled state 
when viewed as a state defined on a Riemannian manifold in which sites are embedded. 
A classical notion of manifold 
exists only for states with local structures.
Any spatial manifold can emerge in this manner.
Some examples are given in \fig{fig:graph_12D}.
Here, dimension, topology and proper volume of space 
are order parameters 
that encode different patterns of entanglement.

In order to identify physical Hilbert space,
we need to construct constraints that generate gauge symmetry.
We start with the gauge group that generalizes the spatial diffeomorphism.
The immediate issue that we face is that the theory does not assume 
a manifold with fixed dimension and topology.
Because a spatial manifold with any dimension can dynamically arise,
the gauge group must be general enough to include spatial diffeomorphism in arbitrary dimensions.
With this in mind, we consider $O(M) \times SL(L, \mathbb{R})$ symmetry,
where the orthogonal group acts on the rows of the matrix,
and the special linear group on the columns.
The $O(M)$ acts on the internal flavour,
and is regarded as a global symmetry of the theory.
On the other hand, \SLL acts on the site index, and is a space symmetry. 
As \SLL can rotate a frame into any other frame,
it can generate diffeomorphism in any manifold.
\SLL generators that give rise to smooth diffeomorphisms
in the continuum limit can be identified in any dimension
as is discussed in Sec. \ref{sec:Mc}.
Therefore, we take \SLL as the generalized spatial diffeomorphism group.

An unavoidable consequence of the enlarged gauge symmetry 
is that \SLL also includes transformations that are non-local 
in a given coordinate system chosen by the local structure of a state.
Under an \SLL transformation that mixes two columns
through linear superpositions,
the set of local sites is not preserved.
A local Hilbert space in one frame is made of
linear superpositions of states defined 
in multiple local Hilbert spaces in another frame. 
Since the total Hilbert space can be written 
as direct products of local Hilbert spaces 
in any frame, 
there is no preferred frame.
Consequently, whether a state has a local structure or not is not gauge invariant.
Even if a state has a local structure of entanglement in one frame,
it does not have a local structure in a rotated frame in general.
In order to define a gauge invariant notion of local structure,
we first need to choose a frame in a gauge invariant manner.
Although there is no preferred frame fixed in the background, 
one can define a set of local Hilbert spaces
in terms of physical degrees of freedom within the theory.
This can be done in states 
in which the $O(M)$ flavour symmetry 
is spontaneously broken to a smaller group
by a non-zero expectation value  
in an $L \times L$ block of the $M \times L$ rectangular matrix.
If the square matrix formed by the $L \times L$ block is non-singular,
a frame can be defined in terms of the row vectors in the block.
Since it provides a physical reference
with respect to which the notion of local Hilbert spaces are defined,
the local structure defined in this frame is gauge invariant.

To fully specify the physical Hilbert space,
one also needs to define a Hamiltonian constraint.
In the general relativity, the Hamiltonian density 
forms the representation of a scalar density 
under the spatial diffeomorphism group.
In the present theory, the Hamiltonian should form a representation 
under the generalized spatial diffeomorphism group, \SLL.
The minimal representation that an 
$O(M)$ invariant Hamiltonian forms under \SLL 
is a rank $2$ symmetric tensorial representation.
Accordingly, the lapse function is generalized to a rank $2$ symmetric tensor,
which is called lapse tensor.

In the general relativity, the lapse function controls 
the lapse of proper time at each location in space,
and there are one scalar function worth of ways 
to generate multi-fingered time evolutions.
In the present theory, there are more ways of generating time evolutions 
because of the off-diagonal elements of the lapse tensor.
A general lapse tensor can be rotated into a diagonal form using an \SLL transformation.
The off-diagonal elements of the lapse tensor encode the information
on the frame in which lapse tensor is put into a diagonal form.
Therefore,
the lapse tensor determines not only the speed of local time evolutions
but also the frame in which  the time evolution is generated.
In priori,  one can describe time evolution in any frame,
and there is no preferred lapse tensor.
A gauge invariant notion of time evolution can only be defined
in terms of physical clocks made of dynamical degrees of freedom
within the theory.
If a set of unentangled clocks are prepared 
out of local degrees of freedom in a frame,
the Hamiltonian with a lapse tensor diagonal in that frame should evolve those clocks independently.
When the whole system that includes the clocks and other degrees of freedom
is evolved with the Hamiltonian,
the evolution of the other degrees of freedom relative to the local clocks
describes the physical time evolution. 
The correlation between the evolution of the local clocks
and the remaining degrees of freedom 
is the gauge invariant content of the theory.
Different choices of local clocks give rise to different 
relative motions.

In order to make sure that the Hamiltonian is background independent,
and  evolves unentangled set of clocks independently, 
the Hamiltonian must be relatively local
in the frames in which the lapse tensor is diagonal.
For the background independence,
any two sites can couple with each other 
as there is no fixed notion of locality.
However, the  effective strength of coupling between two sites
should be determined from the entanglement present between the sites
because the proper distance between the points is determined from the entanglement.
If two points are weakly (strongly) entangled,
the coupling between them are weak (strong).
Although this kind of state dependent evolution generally requires a non-linear action on states,
a linear operator can realize the state dependent locality approximately
in the limit that the size of the matrix is large. 
This guarantees that two sites that are unentangled remain 
decoupled to the leading order in the size of the matrix.
The momentum and Hamiltonian  
obey a first-class  constraint operator algebra.
The first-class nature of the constraint algebra allows one to define the set of gauge invariant states
without introducing additional constraints.
However,  gauge invariant states are non-normalizable
with respect to the norm defined for the microscopic degree of freedom.
All gauge invariant states have infinite norm
because the gauge group, which is non-compact,
necessarily generates gauge orbits that are unbounded in the phase space.
Any state that has a finite norm 
must break the gauge symmetry spontaneously.
In this case, a natural object to consider is a projection between a state with a finite norm
with a gauge invariant state.
This can be viewed as a wavefunction of the gauge invariant state
written in basis states with finite norms.
Alternatively, the projection of the state without gauge invariance toward the gauge invariant state
can be implemented by applying momentum and Hamiltonian constraints.
The fact that the state with a finite norm is not gauge invariant
gives rise to a non-trivial evolution 
as the state is projected toward the gauge invariant state,
where the state with a finite norm plays the role of an initial state defined on 
a Cauchy surface.
This process can be understood as a time evolution,
where multi-fingered times arise as Lagrangian multipliers 
that enforce the gauge constraints.

In order to describe the process of gauge projection,
it is useful to consider a sub-Hilbert space 
that a given `initial state' explores through the evolution
generated by gauge transformations.
A sub-Hilbert space 
that is closed under the operations of the  gauge transformations
is specified by global symmetry.
Here we consider the sub-Hilbert space in which the $O(M)$ flavour symmetry
is broken to $S_L \times O(N/2) \times O(N/2)$, 
where $S_L$ is the permutation group of the first $L$ flavours
and $N=M-N$.
This specific pattern of broken symmetry is not crucial. 
Here, this group is chosen as an example that gives rise to
a minimum number of propagating degrees of freedom 
after gauge degrees of freedom is removed.
One may choose a smaller unbroken symmetry, 
in which case there exist more physical degrees of freedom.
The sub-Hilbert space with  $S_L \times O(N/2) \times O(N/2)$ 
is the kinematic Hilbert space.
It is spanned by basis states 
which are labeled by a set of collective variables.
The collective variables, which are singlets of the unbroken symmetry,
control the pattern of entanglement within the sub-Hilbert space. 
Accordingly, those collective variables encode
the information on dimension, topology and geometry
of the emergent manifold 
if the pattern of entanglement has a local structure.
Since general  states in the sub-Hilbert space can be faithfully represented 
as linear superpositions of the basis states labeled by the collective variables,
the dynamics within the sub-Hilbert space is completely captured by the collective variables.
In particular, the momentum and Hamiltonian induce 
constraints that act on wavefunctions defined in the space of the collective variables.
The induced Hamiltonian for the collective variables is background independent
as the underlying Hamiltonian for the microscopic degree of freedom is relatively local.

The projection of a state with a finite norm 
in the kinematic Hilbert space
to a gauge invariant state can be expressed as a path integration 
over the collective variables that describe 
fluctuations of spacetime.
In the path integration for the collective variables,
spacetimes with different dimensions, topologies and geometries
are summed over non-perturbatively.
In the limit that $M \gg L \gg 1$,
the path integration can be replaced with a saddle-point.
If the state with a finite norm has a $D$-dimensional local structure, 
a $(D+1)$-dimensional spacetime manifold emerges
at the saddle-point. 
The collective variables, which are bi-local in space, can be viewed as an infinite tower of local fields
that includes a spin $2$ mode.
In the classical limit, 
the constraint algebra for the collective variables reduces 
to the hypersurface deformation algebra of the general relativity
to the leading order in the gradient expansion 
if all other modes except for the spin $2$ mode is turned off.
From the constraint algebra, 
one can  identify the emergent metric degree of freedom 
as a composite of the dynamical collective variables.
The metric identified from the algebra  confirms 
the idea that the proper distance between sites 
is determined from the entanglement such that 
strongly entangled sites are physically close to each other.
The saddle-point configuration provides a classical spacetime 
on which fluctuations of collective variables propagate.
The propagating modes are bi-local fields 
that are described by an effective theory 
whose locality is determined with respect to the
classical geometry set by the saddle-point configuration.

%
The collective variables
are kinematically non-local.
They can be viewed as an infinite tower of local fields
that include the gravitational degree of freedom and many other degrees of freedom. 
The gravitational degree of freedom is a collective mode of the matrix
that encodes the inter-site entanglement in the spin $2$ channel.
Other degrees of freedom describe collective modes with different spins.
The emergent geometry captures only partial information
of the full entanglement pattern.
The higher-spin collective modes describe entanglement of the underlying matrix 
which is not captured by the geometry.


\subsection{Outline}

\begin{table}
\centering
\begin{tabular}[t]{| c | c |}
\hline
Fundamental  & A real rectangular matrix  : $\{ \Phi^A_{~~i}~ |~ 1 \leq A \leq M, 1 \leq i \leq L\}$ \\
degree of freedom (d.o.f.)   &  $A$ : flavour index, ~ $i$ : site index ~~ $( M \gg L  \gg 1 )$ \\
\hline
Frame & A decomposition of the full Hilbert space \\
 &     into local Hilbert spaces \\
\hline
 Frame rotation & \SLL  ~ [right multiplication on $\Phi$] \\ 
\hline
Local structure &  A pattern of entanglement that exhibits locality \\
  &    across local Hilbert spaces  [Sec. \ref{sec:localstructure}] \\
\hline
Flavour symmetry & $O(M)$  ~ [left multiplication on $\Phi$]  \\
\hline
~~ Kinematic Hilbert space (~$\cV$~)  ~ & ~ Space of states with unbroken  $S_L \times O\left( \frac{M-L}{2} \right) \times  O\left( \frac{M-L}{2} \right) \subset O(M)$ ~  \\
\hline
Basis states of  $\cV$ & $ 
\cstr$  
~~[ Eqs.
(\ref{eq:qphivarphi}), (\ref{eq:st}), (\ref{symm}) 
]
 \\
 & $s$, $t_1$, $t_2$ : collective variables ($ L\times L$  matrices)  \\
\hline
 Generators of spacetime &\hspace{-1cm}  Generalized momentum  :   \SLL transformation ~~   [  Eqs. (\ref{eq:GG}), (\ref{Gy}) ]   \\
gauge symmetry &~  Generalized Hamiltonian  :  frame dependent local time translation    [  \eq{Hx} ]~ \\
\hline
Constraint algebra (C.A.) & First-class operator algebra ~[ Eqs. (\ref{GGc}), (\ref{GHc}), (\ref{HHc}) ] \\
\hline
~~C. A.  in the classical limit~~ & First-class Poisson algebra ~ [ Eqs. (\ref{eq:GGH1}), (\ref{eq:GGH2}), (\ref{eq:GGH3}) ] \\
\hline
Constraints &          Momentum constraint~ [ \eq{eq:Gyc} ]  with shift [ \eq{eq:xi} ] \\
in the continuum limit & Hamiltonian constraint~ [ \eq{eq:HvcH} ] with lapse [ \eq{eq:theta} ]  \\
\hline
Constraint algebra  &  Generalized hypersurface deformation algebra \\
in the continuum limit & ~ [ Eqs. (\ref{eq:DHDH}), (\ref{eq:DHH}), (\ref{Gmunu})  ]\\
\hline
Projection of a state in $\cV$ &  Path integration of the collective variables ~ [ \eq{eq:pathint} ] \\
to a gauge invariant state &   that represents fluctuating  spacetime \\
\hline
Emergent metric & A composite of the collective variables ~[ Eqs. (\ref{Gmunu2}), (\ref{eq:metric}) ] \\
\hline
Saddle-point equation & \eq{eq:EOM} \\
\hline
A classical solution &  A series of de Sitter-like spacetimes   \\
for a generic initial condition & bridged by Euclidean spaces~ [ \fig{fig:qsg} ] \\
\hline
A fine-tuned  &  Minkowski spacetimes   \\
classical solution & ~ [ \eq{eq:Minkowski}  ] \\
\hline
Effective theory  & Bi-local field theory ~[ \eq{Sfinal3} ] \\
\hline
\end{tabular}
\caption{
A roadmap of the paper.
}
\label{table1}
\end{table}

Here is an outline of the rest of the paper.
In Sec. \ref{sec:kinematics}, the kinematics of the theory is discussed.
We define the full Hilbert space from which 
the kinematic Hilbert space and the physical Hilbert space
are to be defined.
We also define the inner product, and introduce the notion of frame.
In Sec.  \ref{sec:gauge}, we first review the Hamiltonian formulation of the general relativity
as we will use the Hamiltonian formalism in this paper.
We then construct the generalized momentum and Hamiltonian constraints.
From an explicit computation of the commutators between the constraints, 
it is shown that the constraints obey a first-class operator algebra.
In Sec. \ref{sec:path}, 
we define the kinematic Hilbert space as the subset of states
with unbroken $S_L \times O(N/2) \times O(N/2)$ flavour symmetry.
It is shown that states that are gauge invariant have infinite norm,
and states with finite norms necessarily break the gauge symmetry.
The projection of a state with a finite norm in the kinematic Hilbert space
to a gauge invariant state
is expressed as a path integration over collective variables
that describe dynamical spacetime.
Multi-fingered time evolutions arise as a sum over possible routes 
via a state in the kinematic Hilbert space is projected to a gauge invariant state.
In Sec. \ref{sec:spacetime}, it is shown that the constraint algebra
of the present theory reduces to that of the general relativity
in a special case.
Based on the algebra that the momentum and Hamiltonian constraints obey in the continuum limit,
the emergent metric degree of freedom is identified 
in terms of the collective variables.
In the limit that the size of the matrix is large,
the dynamical collective variables become classical.
In Sec. \ref{sec:classical}, the saddle-point equation of motion
for the collective variables is derived.
The equation of motion is solved both numerically and analytically 
in Sec. \ref{sec:translationally} for an initial state that exhibits a three-dimensional local structure.
We find a solution which describes a series of $(3+1)$-dimensional 
de Sitter-like spacetimes with the Lorentzian signature
which are bridged by $4$-dimensional Euclidean spaces.
We show that the signature-changing dynamical phase transitions
are caused by Lifshitz transitions
in which the dispersion 
of the collective variables are inverted dynamically.
The  spacetime that emerges from a generic initial condition
breaks the Lorentz symmetry.
However, the Minkowski solution can be found with fine-tuning.
We derive an effective theory that describes propagating modes, 
which are small fluctuations of the collective variables.
We show that bi-local fields propagate in the spacetime determined from the saddle-point configuration,
obeying local dynamics.
In Sec. \ref{sec:conclusion}, we conclude with discussions
on connections with the dS/CFT and AdS/CFT correspondences,
and future directions.

Table \ref{table1} is a  roadmap 
to the key concepts and results of the paper.

\section{Kinematics}
\label{sec:kinematics}

\subsection{Hilbert space}

We consider an $M \times L$ real rectangular matrix, $\Phi^A_{~~i}$
with $A=1,2,..,M$ and $i=1,2,..,L$
in the $M \gg L \gg 1$ limit.
The full Hilbert space is spanned by 
$
\cb \Phi \rb 
\equiv
\otimes_{i,A} \cb \Phi^A_{~~i} \rb
$, where
$ \cb \Phi^A_{~~i} \rb $ is the eigenstate of $\hat \Phi^A_{~~i}$ 
with eigenvalue $ \Phi^A_{~~i} $.
The inner product between basis states is given by
\bqa
\lb \Phi' \cb \Phi \rb =
 \prod_{i,A} \delta \left(
\Phi_{~~i}^{'A} - \Phi_{~~i}^A
\right).
\label{eq:norm}
\eqa
We note that the Fock space spanned by $\Big\{ \cb \Phi \rb \Bigr\}$ 
is infinite-dimensional  
for any $M>0$ and $L>0$ 
because  
$\cb \Phi \rb$
and
$ \cb  \Phi' \rb$
are orthogonal unless $\Phi = \Phi'$.
The conjugate momentum denoted as $\hat \Pi^i_{~~A}$
satisfies the standard commutation relation,
$ \left[  \hat \Phi_{~~i}^A, \hat \Pi^j_{~~B} \right] = i ~\delta^j_i ~\delta^A_B$.
$\hat \Phi$  $(\hat \Pi)$ represents
the $M \times L$ $(L \times M)$ operator valued matrix.

The row index $A$ is referred to as flavour index.
In this paper, we consider a model 
that has the $O(M)$ flavour symmetry
generated by
\bqa
\hat T_{A B} = \frac{1}{2} \left( \hPhi_{Ai} \hPi^{i}_{~B} -  \hPhi_{Bi} \hPi^{i}_{~A} \right),
\label{eq:OM}
\eqa
where the flavour indices are raised or lowered 
with the Euclidean metric : $ \hPhi_{Ai} =  \hPhi^A_{~~i}$.
The flavour symmetry acts on $\Phi$ ($\Pi$) from the left (right)  as
\bqa
e^{ -i \tr{ \td o \hat T}} ~\hat \Phi  ~ e^{ i \tr{ \td o \hat T}}&=& O ~ \hat \Phi,  \nn
e^{ -i \tr{ \td o \hat T}}  ~\hat \Pi    ~ e^{ i \tr{ \td o \hat T}}&=& \hat \Pi  ~O^{-1},
\eqa
where $\td o$ is an anti-symmetric matrix 
and $O = e^{-\td o} \in O(M)$.
General $O(M)$ invariant operators can be constructed
as composites of the following bi-linears,
\bqa 
 \hPi \hPhi, 
 ~~~ \hPi \hPi^T, ~~~\hPhi^T \hPhi,
\label{eq:bilinear}
\eqa
where $\hPhi^T$ ($\hPi^T$) denotes the transpose of $\hPhi$ ($\hPi$).
Products of operator valued matrices 
are defined in the usual way, 
e.g.,  $ ( \hPi \hPhi )^i_{~j}  = \hPi^i_{~A} \hPhi^A_{~~j}$. 
Henceforth, all repeated indices are understood to be summed over 
unless mentioned otherwise.

\subsection{Frame}
\label{sec:frame}

\begin{figure}[ht]
\begin{center}
\centering
\includegraphics[scale=2.5]{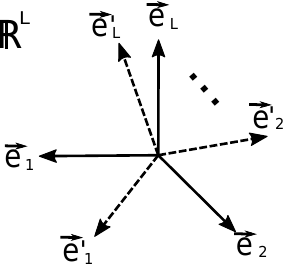}
\end{center}
\caption{
The total Hilbert space can be written as a direct product of local Hilbert spaces in a frame.
A frame is given by $L$ linearly independent basis vectors 
that form an $L$-dimensional parallelepiped 
with the unit Euclidean volume in ${\mathbb R}^L$.
Each basis vector determines a local Hilbert space. 
Under a special linear transformation, 
a frame can be rotated into another frame,
which defines a new set of local Hilbert spaces.
}
\label{fig:frame}
\end{figure}

The column index $i$ is referred to as site index
as it labels points of space
in the model of gravity to be constructed.
Once we identify $i$ as a site index,
it is natural to write the total Hilbert space as a direct product of local Hilbert spaces as 
\bqa
{\mathbb H}= \otimes_i {\mathbb H}_i,
\label{eq:frame}
\eqa
where ${\mathbb H}_i$ is the local Hilbert space
spanned by 
$\otimes_A \cb \Phi^A_i \rb$.
Such a decomposition of the total Hilbert space is called a {\it frame}.
In a given frame, each site is associated with an infinite-dimensional local Hilbert space
spanned by basis states, each of which is labeled by an $O(M)$ vector.
In the limit that both $M$ and $L$ are large,
a manifold with any spatial dimension can emerge 
as will be shown in the next section.
This is a key difference  from other approaches to quantum gravity
in which local Hilbert space is tailored for a specific space dimension.

The total Hilbert space can be decomposed in different frames.
For example, one can use a different set of basis states 
that are related to the original basis states through
\bqa
\cb \Phi \big)   \equiv  \cb \td \Phi \rb,
\label{eq:basist}
\eqa
where $\td \Phi^{A}_{~~i} =  g_i^I \Phi^{A}_{~~I}$ 
and  $g \in SL(L, \mathbb{R})$.
\eq{eq:basist} should be understood as a
change of basis in the infinite dimensional  Fock space.
The inner product is preserved because
\bqa
\big( \Phi \cb \Phi' \big) & = & \prod_A \prod_i \delta \Bigl( g_i^I ( \Phi^A_{~~I} - \Phi^{'A}_{~~I}) \Bigr) 
= \prod_{I,A} \delta \left(
\Phi_{~~I}^{'A} - \Phi_{~~I}^A
\right),
\label{eq:rinner}
\eqa
where $\det g =1 $ is used.
This allows one to represent
$\cb \Phi \big)  = \otimes_{A,I} \cb \Phi^A_{~~I} \big)$,
where 
$\otimes_A \cb \Phi^A_{~~I} \big)$ 
spans the local Hilbert space
${\mathbb H}^{'}_I$ at site $I$ in the rotated frame.
The total Hilbert space 
can be written as a direct product of the local Hilbert spaces 
in the rotated frame as
${\mathbb H}= \otimes_I {\mathbb H}^{'}_I$.

A frame, denoted as $X$, is defined by a set of $L$
linearly independent row vectors in $\mathbb{R}^L$ :
$X = \{ \vec e^{~ i}\in \mathbb{R}^L \cb  V(\vec e^{~1}, \vec e^{~2},..,\vec e^{~L})=1, i=1,2,..,L \}$,
where $V(\vec e^{~1},\vec e^{~2},..,\vec e^{~L})$ is the Euclidean volume of the parallelepiped
formed by the $L$ vectors.
This is illustrated in \fig{fig:frame}.
If $X$ is a frame, 
$X_g = \{  \vec e^{~i} g  \cb i=1,2,..,L \}$ is also a frame
for any $g \in SL(L, \mathbb{R})$.
A frame defines a set of local Hilbert spaces
of which the total Hilbert is decomposed as
a direct product.
Associated with a frame, one can define
a set of local observers :
a local observer at site $i$ in frame $X$ has access to 
${\mathbb H}^{X}_i$,
where 
${\mathbb H}^{X}_i$ is the local Hilbert space defined at site $i$ in frame $X$.
The Hilbert space accessible to 
a local observer in one frame 
is comprised of  
linear superpositions of
states accessible to multiple local observers in another frame.
There is a priori no preferred frame 
and thus no preferred set of local observers.

\subsection{Local structure}
\label{sec:localstructure}

In a given frame, one can define entanglement 
formed across local Hilbert spaces.
In the presence of a local structure of entanglement, 
a spatial manifold can be defined from the pattern 
of entanglement. 
A state is defined to have a local structure in a frame 
if there exists a mapping from the sites to a Riemannian manifold
such that the mutual information between any two sites
decays exponentially in the proper distance between the images of the sites
in the Riemannian manifold to the leading order in the proper distance
(see \fig{fig:local_structure})\footnote{
For states that have local structures, 
the von Neumann entanglement entropy of a subset of sites scales 
with the boundary of the corresponding subregion in the Riemannian manifold.
One can alternatively use this as a definition of local structure\cite{Lee2019}. 
}.
The dimension, topology and metric of the manifold are 
collective properties of a state.
A state with a local structure can be 
regarded as a short-range entangled state
with respect to the corresponding spatial manifold.
For general states, local structure does not exist.
The existence of local structure is a dynamical property 
that only a sub-set of states possess.
Dimension, topology and geometry 
are order parameters 
that differentiate different local structures.
Just as the vacuum expectation value of a field can be used
an order parameter that characterizes symmetry of states in quantum field theories,
dimension, topology and geometry in the present theory
represent a set of coarse-grained data
that characterize the macroscopic structure of quantum states.
Dimension and topology form discrete order parameters,
and two states with different dimensions or topologies
can not be smoothly deformed to each other.
On the other hand, geometry quantifies finer patterns 
of entanglement present across local Hilbert spaces.

In order to illustrate the idea, 
we consider a set of $O(M)$ invariant states  labeled by a collective variable,
\bqa
\cb t \rb  & = &  \int d \Phi ~ e^{ i t^{ij} \Phi^A_{~i} \Phi^A_{~j}  } \cb \Phi \rb,
\label{eq:t}
\eqa
where 
$t^{ij}$ is a complex $L \times L$ collective variable.
Here $\int d \Phi \equiv \prod_{i,A} \int_{-\infty}^\infty d \Phi^A_{~~i}$.
The state is normalizable as far as 
the eigenvalues of the $L \times L$ matrix, $t^{ij}$
lie within the upper half of the complex plane.
If $t^{ij}$ is diagonal,
$\cb t \rb$ represents a direct product state with no entanglement. 
In order to see how off-diagonal elements of $t$ 
is related to inter-site entanglement,
we compute the mutual information between two sites for \eq{eq:t}.
For states that are close to the direct product state,
the mutual information can be computed perturbatively in 
 $\left| \frac{t^{ij}}{Im t^{ii}} \right| \ll 1 $.
An explicit calculation shows that 
the mutual information between sites $i$ and $j$ is given by\cite{Lee2016}
\bqa
I_{ij} = 2M
 \left(
- \ln \frac{ |t^{ij}|^2 }{ 4 Im t^{ii} Im t^{jj}}
+ 1
\right)
\frac{ |t^{ij}|^2}{4 Im t^{ii} Im t^{jj} } + ...
\label{eq:EE}
\eqa
to the leading order in
 $\frac{t^{ij}}{Im t^{ii}}$.
 \eq{eq:EE}  shows that $t^{ij}$ that connects sites $i$ and $j$ 
 creates the mutual information between the sites to the lowest order.
The ellipsis in \eq{eq:EE} represents the higher order mutual information
formed through chains of 
link variables that connect $i$ and $j$ through other sites,  
$
\sum_{n+m>0} 
\sum_{k_1,..,k_n} 
\sum_{l_1,..,l_m} 
\frac{  
t^{i k_1} 
\left( \prod_{a=1}^{n-1} t^{k_a k_{a+1}}  \right)
 t^{k_n j}
t^{j l_1} 
\left( \prod_{b=1}^{m-1} t^{l_b l_{b+1}}  \right)
t^{l_m i}
}{
t^{ii}~ t^{jj} ~
\left( \prod_{a=1}^{n} t^{k_a k_{a}}  \right)~
\left( \prod_{b=1}^{m} t^{l_b l_{b}}  \right)
}
$.
Sites that are not directed connected by a non-zero collective variable
are entangled via multiple legs of the bi-local collective variables. 
The off-diagonal elements of the collective variable 
describe `bonds' that create inter-site entanglement,
where the strength of the bond between sites $i$ and $j$ 
is proportional to the magnitude of $t^{ij}$.
If the short-ranged entanglement bonds form a regular lattice
(similar to the way a lattice is formed by chemical bonds in solids),
the corresponding state has a local structure
that exhibits a manifold with a well-defined dimension and topology. 
We will later see how the emergent geometry is determined from the collective variables as well.
Intuitively, the geometry is determined such that the proper distance between
two points gets smaller if the two points are connected by stronger entanglement bonds (larger $t^{ij}$).

\vspace{1cm}
 \begin{figure}[h]
 \begin{center}
   \subfigure[]{
 \includegraphics[scale=0.4]{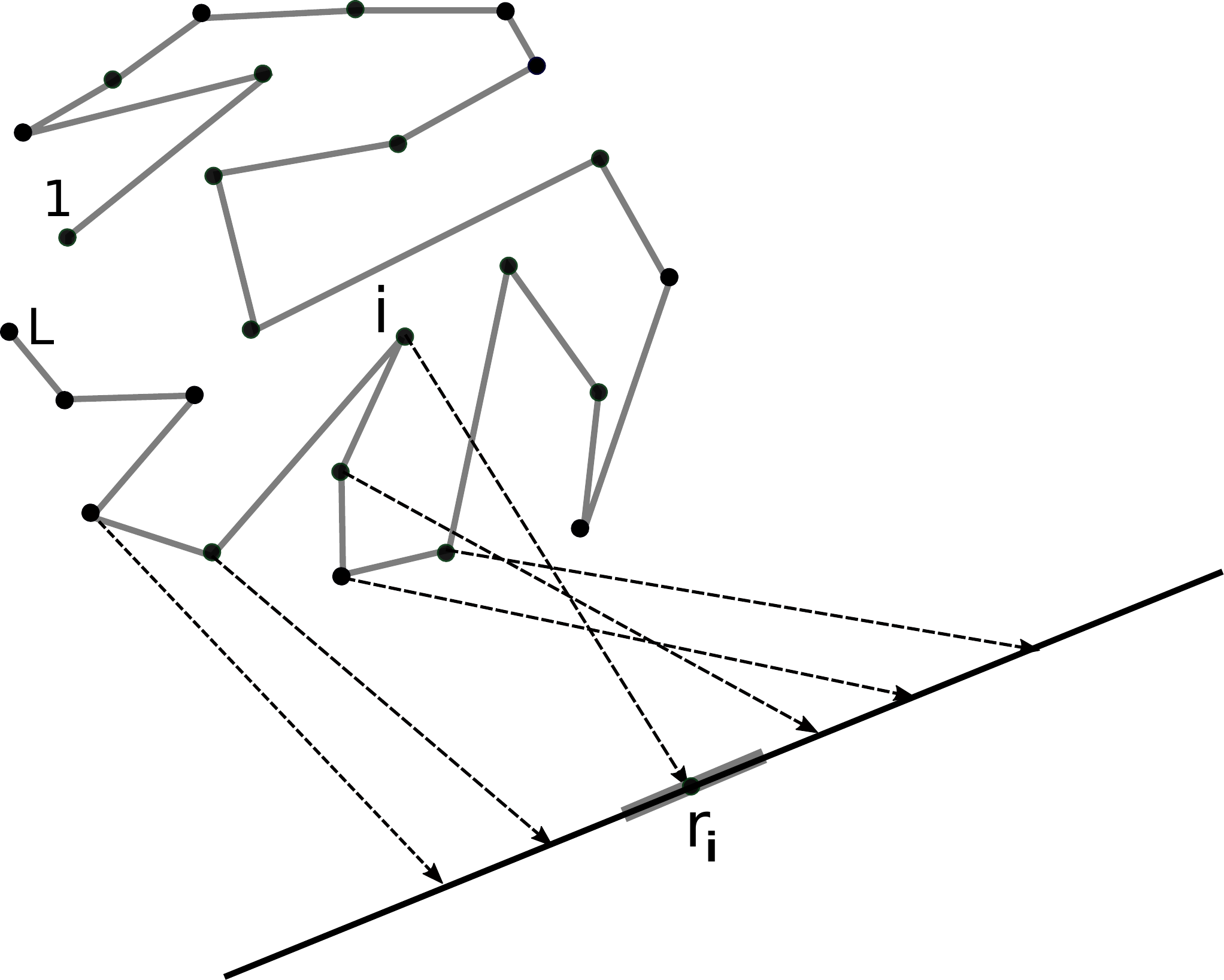} 
  \label{fig:graph_12Da}
 } 
 \subfigure[]{
 \includegraphics[scale=0.4]{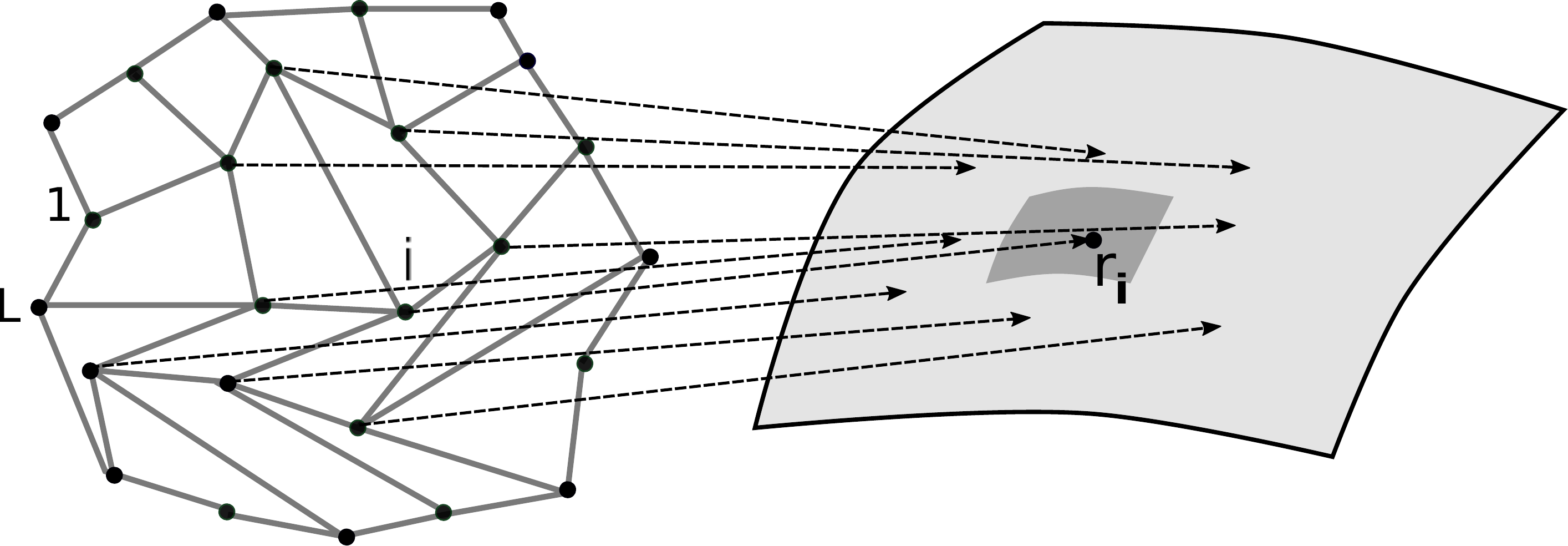} 
  \label{fig:graph_12Db}
 } 
  \end{center}
 \caption{
The panels (a) and (b) represent states in \eq{eq:t} 
with local structures given by
\eq{eq:tij1d} and \eq{eq:tij2d}, respectively.
  In the figures on the left, the dots represent sites in a frame,
 and the links represent entanglement bonds
formed by bi-local collective variable, $t^{ij}$.
The dashed lines represent mappings from  
sites to manifolds.
In the figures on the right, 
the grey area around $r_i$ denotes the region of coordinate volume $V_i$ 
assigned to site $i$ in the manifold.
The dimension and topology of the manifold 
in which the local structure is manifest are determined
from the pattern of entanglement bonds.
In the large $L$ limit, 
the proper distances between sites $1$ and $L$ in states (a) and (b) 
scale as $O(L)$ and $O(1)$ respectively 
due to different local entanglement structures.
 }
 \label{fig:graph_12D}
 \end{figure}

The same set of sites can exhibit manifolds 
with different dimensions, topologies and geometries, 
depending on the pattern of entanglement.
Let us consider a few examples
of states with local structures.
As a first example, we consider the state in \eq{eq:t} 
with  
\bqa
t^{ij} = i \left( \delta_{ij} + \epsilon  \delta_{ |i - j|, 1} \right)
\label{eq:tij1d}
\eqa
for $\epsilon \ll 1$.
In this state, nearest neighbour entanglement bonds form an open chain.
For this state, the local structure is manifest in the one-dimensional coordinate system, 
\bqa
r_j = j.
\label{eq:1dc}
\eqa
The emergent one-dimensional space has the topology of a line segment, $[0,1]$.
In particular, the physical distance between the first site and the last site
is very far due to the weak entanglement.
If one creates a direct entanglement bond between sites $1$ and $L$
by adding $ t^{1L} = i  \epsilon$ to \eq{eq:tij1d}, 
the two sites become neighbours.
As a consequence, the topology of the space changes to $S^1$.

As a second example, let us consider the collective variable given by
\bqa
t^{ij}= i \left(
\delta_{ij} + \epsilon 
\delta_{
\sqrt{
(r_i^1 - r_j^1)^2
+ ( r_i^2 - r_j^2)^2
}, 1}
\right),
\label{eq:tij2d}
\eqa
where 
\bqa
(r_j^1, r_j^2 ) = 
\left(
 j ~\mbox{mod} ~\sqrt{L} ,
\left\| \frac{j - 1}{\sqrt{L}} \right\| + 1
\right)
\label{eq:2dc}
\eqa
with $\left\| x \right\|$ 
being the largest integer equal to or smaller than $x$.
In this state, the entanglement bonds form a square lattice.
It exhibits a two-dimensional local structure
with the topology of a disk.
\eq{eq:2dc} is a natural coordinate system
in which the local structure is manifest.
These are illustrated in \fig{fig:graph_12D}.
These examples illustrate the fact that dimension and topology of space are 
nothing but collective variables of the underlying matrix.
In Sec. \ref{sec:translationally}, we will show how the geometry is 
determined from the collective variables.

The existence of  local structure depends on the choice of frame.
Under a change of frame, the collective variable is transformed 
as $t^{' IJ} = g^I_i g^J_j t^{ij}$, where $g \in SL(L, \mathbb{R})$.
Even if a state exhibits a local entanglement structure in one frame,
it does not have a local structure in another frame
if the latter is related to the former through a transformation 
that is non-local with respect to the locality defined in one frame.
Generic states with bonds that form a global network do not exhibit a local structure.

\section{Gauge symmetry}
\label{sec:gauge}

In this section, 
we construct the generators of the gauge symmetry
that generalizes the spacetime diffeomorphism
of the general relativity.
Since we are going to use the Hamiltonian formalism,
we first review the Hamiltonian formulation of the general relativity.

\subsection{Review of the Hamiltonian formalism of the general relativity}

In $(3+1)$ dimensions,
the action of the general relativity can be written as\cite{PhysRev.116.1322}
\bqa
S = \int d \tau d^3r \Bigl[
\pi^{\mu \nu} \partial_\tau g_{\mu \nu} 
- \xi^\mu(r) \cP_\mu(r) - \theta (r) \cH(r)
\Bigr].
\label{eq:Einstein}
\eqa
Here a four-dimensional spacetime is decomposed 
into a stack of three-dimensional spatial manifolds
that are labeled by coordinate time $\tau$.
A point within each time slice is labeled by $r$. 
 $g_{\mu \nu}$  and  $\pi^{\mu \nu}$
with $\mu, \nu =1,2,3$ are the spatial metric
and its conjugate variable, respectively.
The symplectic form in 
\eq{eq:Einstein}
defines the Poisson bracket,
$\{g_{\mu \nu}(r), \pi^{\rho \sigma}(r') \}_{PB} = \delta^{\rho \sigma}_{\mu \nu} \delta(r-r')$,
where $
 \delta^{\rho \sigma}_{\mu \nu}
 =\frac{1}{2} \left(
 \delta^\rho_{\mu} \delta^\sigma_{\nu}
 + 
  \delta^\rho_{\nu} \delta^\sigma_{\mu} 
\right)$.
$\cP_\mu(r)$ is the momentum density
that generates spatial diffeomorphism within a spatial manifold.
$\xi^\mu(r)$ is the shift that specifies an infinitesimal spatial diffeomorphism.
$\cH(r)$ is the Hamiltonian density
that generates local time translation. 
$\theta(r)$ is the lapse
that determines the position dependent time translation.
The shift and the lapse can be chosen arbitrarily.
Consequently, the momentum and Hamiltonian, 
\bqa
P \left[ \vec \xi \right] & = & \int d^3r ~\xi^\mu(r) \cP_\mu(r), \nn
H \left[\theta \right] & = & \int d^3r ~\theta(r) \cH(r)
\label{eq:PxHt}
\eqa
become constraints.
The key property of the general relativity is that the entire dynamics 
is generated by the constraints that satisfy the hypersurface deformation algebra\cite{TEITELBOIM1973542},
 \bqa
 \left\{ P\left[\vec \xi_1 \right], P\left[ \vec \xi_2 \right] \right\}_{PB} & = & P \left[ {\cal L}_{\vec \xi_1} \vec \xi_2 \right], \label{eq:GR1} \\
 \left\{ P \left[ \vec \xi \right], H \left[ \theta  \right] \right\}_{PB} & = & H \left[ {\cal L}_{\vec \xi} ~\theta   \right], \label{eq:GR2} \\ 
  \Bigl\{ H \left[ \theta _1 \right], H \left[ \theta _2 \right] \Bigr\}_{PB} & = & P\left[ \vec \xi_{\theta _1, \theta _2} \right].  \label{eq:GR3}
 \eqa 
Here
${\cal L}_{\vec \xi}$ represents the Lie derivative with respect to the vector field $\vec \xi$
and
\bqa
\xi^\mu_{\theta _1,\theta _2} =  -{\cal S} g^{\mu \nu} \left( \theta _1 \nabla_\nu \theta _2  - \theta _2 \nabla_\nu \theta _1 \right).
\label{xitt}
\eqa
The signature of spacetime is chosen to be $({\cal S}, +, +, +)$.
For a Lorentzian (Euclidean) spacetime, ${\cal S}=-1 (+1)$.
\eq{eq:GR1}  and \eq{eq:GR2} denote the fact that 
$\cP_\mu(r)$ and $\cH(r)$
transform as a vector density and a scalar density respectively
under a spatial diffeomorphism.
These two are purely kinematic.
On the other hand, \eq{eq:GR3} implies that two successive infinitesimal local time translations
performed in different orders
are related to each other through a spatial diffeomorphism.
Under an infinitesimal time translation generated by lapse $\theta_1$,
a general phase space function $f(g, \pi)$ evolves into
$f + \epsilon \{ f, H[ \theta_1 ] \}_{PB}
+ \frac{\epsilon^2}{2}  \{  \{ f, H[ \theta_1 ] \}_{PB}, H[ \theta_1 ] \}_{PB}
$
to the second order in $\epsilon$.
A consecutive time evolution generated by lapse $\theta_2$ gives
$f + \epsilon \{ f, H[ \theta_1 ] \}_{PB}
   + \epsilon \{ f, H[ \theta_2 ] \}_{PB}
   + \frac{\epsilon^2}{2}  \{  \{ f, H[ \theta_1 ] \}_{PB}, H[ \theta_1 ] \}_{PB}
+ \frac{\epsilon^2}{2}  \{  \{ f, H[ \theta_2 ] \}_{PB}, H[ \theta_2 ] \}_{PB}
   + \epsilon^2 \{ \{ f, H[ \theta_1 ] \}_{PB}, H [\theta_2]  \}_{PB}$.
The time translations applied in the opposite order results in a different outcome,
$f + \epsilon \{ f, H[ \theta_2 ] \}_{PB}
   + \epsilon \{ f, H[ \theta_1 ] \}_{PB}
+ \frac{\epsilon^2}{2}  \{  \{ f, H[ \theta_2 ] \}_{PB}, H[ \theta_2 ] \}_{PB}
      + \frac{\epsilon^2}{2}  \{  \{ f, H[ \theta_1 ] \}_{PB}, H[ \theta_1 ] \}_{PB}
   + \epsilon^2 \{ \{ f, H[ \theta_2 ] \}_{PB}, H [\theta_1]  \}_{PB}$.
The discrepancy between the two is generated by the shift given in \eq{xitt}
to the order of $\epsilon^2$
due to the Jacobi identity.
This relation has some dynamical information because \eq{xitt} depends on the spatial metric
and the signature of spacetime.
In other words, the discrepancy is compensated by different shifts in different states.
This will be important in identifying the emergent metric degree of freedom in our theory 
later in this paper.

The Hamiltonian and momentum form the first-class constraint algebra classically,
which is crucial to guarantee that the constraints are preserved under the 
evolution generated by the constraints themselves.
The constraint algebra largely fixes the form of the Einstein-Hilbert action.
Up to two derivative order, the Einstein-Hilbert action is the only theory 
that satisfies Eqs. (\ref{eq:GR1}), (\ref{eq:GR2}) and (\ref{eq:GR3})
\cite{TEITELBOIM1973542,PhysRevD.81.064002}.
In quantum gravity, 
the constraints are to be promoted to
operators that satisfy 
a first-class operator algebra.
The challenge is to regularize the constraints
in a way that they satisfy a first-class algebra 
at the quantum level
which is reduced to Eq. (\ref{eq:GR1})-(\ref{eq:GR3})
in the classical limit.

In the following two subsections, 
we construct momentum and Hamiltonian constraints 
that generate generalized spacetime diffeomorphism
in the absence of manifold with fixed dimension and topology.
We impose the $O(M)$ flavour symmetry,
and the constraints are built out of
the bi-linears in \eq{eq:bilinear}.

\subsection{Momentum constraint}
\label{sec:Mc}

Because dimension and topology of spatial manifold are not fixed,
spatial diffeomorphism needs to be generalized to a group
that includes diffeomorphism in any dimension 
in the limit that $L$ is large.
For any pair of sites $i$ and $j$, 
there should exist a generator
that maps  $i$ to  $j$
because there are states with local structures 
in which the two sites are close to each other.
The desired gauge group is the special linear group (\SLL)
introduced in Sec. \ref{sec:frame}
that generates rotations of frame.

The first operator in \eq{eq:bilinear} generates the general linear transformation.
The Hermitian generator of  $GL(L, \mathbb{R})$ is given by 
\bqa
\hat {\bf G}^i_{~j} = \frac{1}{2} \left( \hPiiA \hPhiAj + \hPhiAj \hPiiA \right),
\label{gij}
\eqa
where $i,j = 1,2,...,L$.
The \GLL generators
can be decomposed into
$(L^2-1)$ generators for \SLL,
\bqa
\hat G^i_{~j} &= \hat {\bf G}^i_{~j}  - \frac{1}{L} \hat {\bf G}^k_{~k} \delta^i_{~j},
\label{eq:GG}
\eqa 
and one for the global dilatation,
\bqa
\hat G_0 &= 
\frac{1}{L} \hat {\bf G}^k_{~k}.
\eqa 
Because the inner product is preserved under \SLL as is shown in \eq{eq:rinner},
we pick \SLL as the gauge group that 
generalizes the spatial diffeomorphism.
\SLL generators can be written as 
\bqa
\hat G_y = 
\tr{ \hat  G y },
\label{Gy}
\eqa
where $y$ is a traceless  $L \times L$ real matrix.
The \SLL transformations act on $\Phi$ ($\Pi$) from the right (left) as
\bqa
e^{- i \hG_y } ~\hat \Phi ~ e^{i \hG_y } &=&  \hat \Phi ~ g_y,  \nn
e^{- i \hG_y }~\hat \Pi ~e^{ i \hG_y } &=& g_y^{-1}~ \hat \Pi,
\label{SLLphipi}
\eqa
where $g_y = e^{-y} \in \mbox{ \SLL}$.
Under \SLL, $\Phi$ ($\Pi$) transforms  covariantly (contravariantly).

Now we examine how \SLL acts on the matrix 
in states which have local structures.
In the presence of a local structure,
one can define a manifold 
into which  sites are embedded. 
Let $r_i$ represent the point in the spatial manifold
associated with site $i$.
The matrix $\Phi^A_{~~i}$ 
is then viewed as a field  $\Phi^A (r_i)$
defined at position $r_i$.
For an infinitesimal \SLL transformation 
with $g_y = e^{-\epsilon y}$ in \eq{SLLphipi}, 
the field transforms as
\bqa
\Phi^{'A}(r_i) & = & 
\Phi^A(r_i) - \epsilon  \sum_j  \Phi^A(r_j) y^j_{~i}.
\label{eq:phia'}
\eqa
Let us consider field configurations that change slowly on the manifold
in the continuum limit with $L \gg 1$.
In this case, a gradient expansion can be used to write 
$ \Phi^A(r_j)  =  \Phi^A(r_i) + \partial_\mu  \Phi^A(r_i) ( r^\mu_j - r^\mu_i ) + ..$
\footnote{
One can view this as the defining expression of the gradient
in the limit that $\Phi^A(r)$ changes slowly in space.
},
and  \eq{eq:phia'} becomes
\bqa
\Phi^{'A}(r_i) -  \Phi^A(r_i)  & = & 
 - \epsilon \zeta_y(r_i)  \Phi^A(r_i) 
- \epsilon \xi_y^\mu( r_i ) \partial_\mu  \Phi^A(r_i) + ...
\label{eq:phia'2}
\eqa
Here 
\bqa
\zeta_y(r_i) & = & \sum_j y^j_{~i}, \label{eq:zeta} \\
\xi_y^\mu(r_i) & = & \sum_j y^j_{~i}( r^\mu_j - r^\mu_i ) \label{eq:xi}
\eqa
represents a scalar and a vector fields, respectively,
associated with $y$.
In \eq{eq:phia'2},
$...$ represent higher derivative terms.
The scalar field determines the position dependent rescaling of the field (Weyl transformation)\cite{Weyl:1918ib}.
The vector field describes spatial diffeomorphism.
To the first derivative order,
\eq{eq:phia'2} is precisely
how a scalar field transforms
under the Weyl transformation
and the spatial diffeomorphism.
While the former is an internal gauge symmetry,
the latter is defined 
with reference to the manifold associated with the state.
We note that the Weyl symmetry and the spatial diffeomorphism 
is a part of the larger gauge symmetry
generated by \SLL .
The generalized momentum constraint
includes other gauge transformations
associated with the higher derivative terms in \eq{eq:phia'2}. 
Some of the extra gauge symmetry act non-locally
with respect to the manifold selected by a state.
They generate smooth diffeomorphism in different
manifolds associated with states 
with different local structures.
Namely, smooth diffeomorphisms acting locally in one manifold
act non-locally in another manifold with different dimension and topology.
Because there is no pre-determined manifold,
the gauge symmetry should include all of them. 
The presence of extra gauge symmetry also plays an important role
in determining the physical degrees of freedom of the theory,
as will be discussed 
in Sec. \ref{GITA}.
We call $\hat G$ and $y$  
the momentum constraint and the shift tensor, respectively.

A few remarks are in order.
First, the diffeomorphism induced by \SLL is an active transformation.
\eq{eq:phia'2} shows how the field is actively `dragged' under \SLL
in a fixed coordinate system. 
Second, \SLL includes smooth diffeomorphism of any dimension in the large $L$ limit.
Once a $D$-dimensional coordinate system is chosen by the local structure of a state,
there exists a set of shift tensors that generate general diffeomorphism in $D$-dimensional. 
For example, the state in \eq{eq:t} 
with \eq{eq:tij1d} has the one-dimensional local structure.
The shift tensor given by
\bqa
y^{j}_{~i} = \frac{\xi_i}{2} \left( \delta_{j, i+1} - \delta_{j,i-1} \right)
\label{eq:y1d}
\eqa 
gives rise to a vector field $\xi(r_i) = \xi_i$ 
in the coordinate system given by \eq{eq:1dc}.
This generates an one-dimensional diffeomorphism in the continuum limit. 
The state with \eq{eq:tij2d} has a two-dimensional local structure
which is manifest in the coordinate system given by \eq{eq:2dc}.
The shift tensor,
\bqa
y^{j}_{~i}  &=& 
\frac{\xi^1_i}{2} \left( 
\delta_{ r_j^1, r_i^1+1  } 
- \delta_{  r_j^1, r_i^1-1} 
\right)
\delta_{ r_j^2, r_i^2} 
+
\frac{\xi^2_i}{2} \left( 
\delta_{ r_j^2, r_i^2+1  } 
- \delta_{  r_j^2, r_i^2-1} 
\right)
\delta_{ r_j^1, r_i^1} 
\label{eq:y2d}
\eqa
gives rise to a two-dimensional diffeomorphism 
generated by the vector field $\xi^\mu(r_i) = ( \xi^1_i, \xi^2_i )$
on the two-dimensional manifold
in the continuum limit.
These examples show that \SLL include general diffeomorphism
in arbitrary dimensions.
Third, $D$-dimensional diffeomorphisms act locally 
only in states with a $D$-dimensional local structure.  
In general, a \SLL transformation that generates a $D$-dimensional diffeomorphism 
acts as a non-local transformation in states 
with local structures with different dimensions.
For example, the shift tensor in \eq{eq:y2d} 
 generates a non-local transformation 
in the one-dimensional manifold given by \eq{eq:1dc},
while it generates a local diffeomorphism 
in the two-dimensional manifold in \eq{eq:2dc}.
A transformation that maps site $1$ to site $L$ 
is quasi-local in  \fig{fig:graph_12Db} but 
non-local in \fig{fig:graph_12Da}
in the continuum limit.

\subsection{Hamiltonian constraint}

Having identified the generator for the spatial diffeomorphism,
we now construct the Hamiltonian density.
In the general relativity, the Hamiltonian density forms
a representation of scalar density under spatial diffeomorphism.
Each element in the representation generates one of many-fingered time translations.
In the present theory, Hamiltonian should form a representation of \SLL.
Given that the column indices of the matrix play the role of sites in the present theory,
one may expect that Hamiltonian density forms a vectorial representation of \SLL.
However, there is no $O(M)$ invariant operator that forms a vectorial representation.
This is because all $O(M)$ invariant operators should be constructed from the bi-linears
in \eq{eq:bilinear}.
To be concrete, we first consider the usual $O(M)$ invariant kinetic operator localized at a site,
$
h^{ii} = 
\sum_A \hPi_{~A}^{i}  \hPi_{~A}^{ i} $
as a candidate for the Hamiltonian density 
(here, $i$ is not summed over).
The problem is that the set of $\{ h^{ii}  \cb i=1,2,..,L \}$ does not form a representation of \SLL. 
Under \SLL, the ultra-local kinetic term is transformed to  
$e^{- i \hG_y }~ h^{ii}  ~e^{ i \hG_y } 
= \sum_{k,l} \sum_A  \left( g_y^{-1} \right)^i_k  \left( g_y^{-1} \right)^i_l   h^{kl}  
$, which is not ultra-local any more.
This underlines the fact that the notion of locality is frame dependent.
Even if the kinetic term is local in one frame, it is generally not in other frames.
In order to construct a kinetic term that forms a representation of  \SLL,
we need to include the full $L \times L$ matrix, $\hPi  \hPi^T$
which forms a rank $2$ contravariant symmetric representation of \SLL.
The most general kinetic operator is labeled by 
a rank $2$ symmetric tensor $v$ as
\bqa
\hat h_{1,v} =\tr{ \hPi  \hPi^T v }.
\label{eq:pipi1}
\eqa
It is reminded that the trace in \eq{eq:pipi1} sums over both the flavour
and the site indices.
In the component form, \eq{eq:pipi1} reads
$
\sum_{A} \sum_{i,j}  \hPi^i_{~~A}  \hPi^j_{~~A} v_{ji}
$.
Under \SLL, $v$ transforms as $v \rightarrow  g_y^{-1 T}  v  g_y^{-1 } $.
The set of $\hat h_{1,v}$ with symmetric $v$
forms a representation under \SLL.
We refer to $v$ as the lapse tensor
as it plays the role of the lapse function in the general relativity.
Before we discuss the meaning of the lapse tensor further, 
let us complete the construction of the Hamiltonian.
We need to add a hopping operator 
that becomes gradient term ($|\nabla \Phi|^2$) in the continuum limit.
As a part of the Hamiltonian that is added to \eq{eq:pipi1}, 
it should also form 
a rank $2$ symmetric contravariant tensorial representation of \SLL. 
The bilinear,  $\Phi^A_{~i} \Phi^A_{~j}$ is a candidate of the hopping term,
but it is a covariant tensor not a contravariant tensor.
In order to convert it into a contravariant tensor,
the site indices should be raised with $\hPi \hPi^T$.
The minimal hopping term 
that includes  $\Phi^A_{~i} \Phi^A_{~j}$ 
and transforms in the desired representation is
 \bqa
\hat h_{2,v} =  \frac{ 1}{M^2}  \tr{  \hPi \hPi^T  \hPhi^T  \hPhi \hPi \hPi^T v }.
   \label{eq:pipi2}
\eqa
The factor of $M^{-2}$ is introduced in \eq{eq:pipi2} to make sure that
both $\hat h_{1,v}$ and $\hat h_{2,v}$ scale as $O(M)$ in the large $M$ limit.
We combine $\hat h_{1,v}$ and $\hat h_{2,v}$ to write the Hamiltonian 
as $\hat H_v = \tilde \alpha_1 \hat h_{1,v} + \tilde \alpha_2 \hat h_{2,v}$,
where $\tilde \alpha_1$ and $\tilde \alpha_2$ are dimensionless parameters.
We choose $\td \alpha_2 > 0$ such that the Hamiltonian is bounded from below for large $\Pi$.
Furthermore, $\tilde \alpha_1 < 0$ is chosen so that
the space of configurations that satisfy the constraint $\hat H_v = 0$
is non-trivial in the classical limit\footnote{
Both $\hPi \hPi^T$ 
and  $\hPhi^T \hPhi$
are non-negative matrices.
If both $\tilde \alpha_1$ and $\tilde \alpha_2$ are positive,
the only configuration
that satisfies $\hat H_v = 0$ 
is $\hPi \hPi^T  = \hPhi^T \hPhi = 0$
classically
}.
Without loss of generality, one can set $\tilde \alpha_1= -1$.
The full Hamiltonian with lapse tensor $v$ is written as 
\bqa
\hat H_v =   
\tr{ 
\left(  
- \hPi \hPi^T 
+ \frac{ \ta }{M^2}  \hPi \hPi^T  \hPhi^T  \hPhi \hPi \hPi^T 
\right) v }.
\label{Hx}
\eqa

\vspace{0.5cm}
\begin{figure}[ht]
\begin{center}
\centering
\includegraphics[scale=0.5]{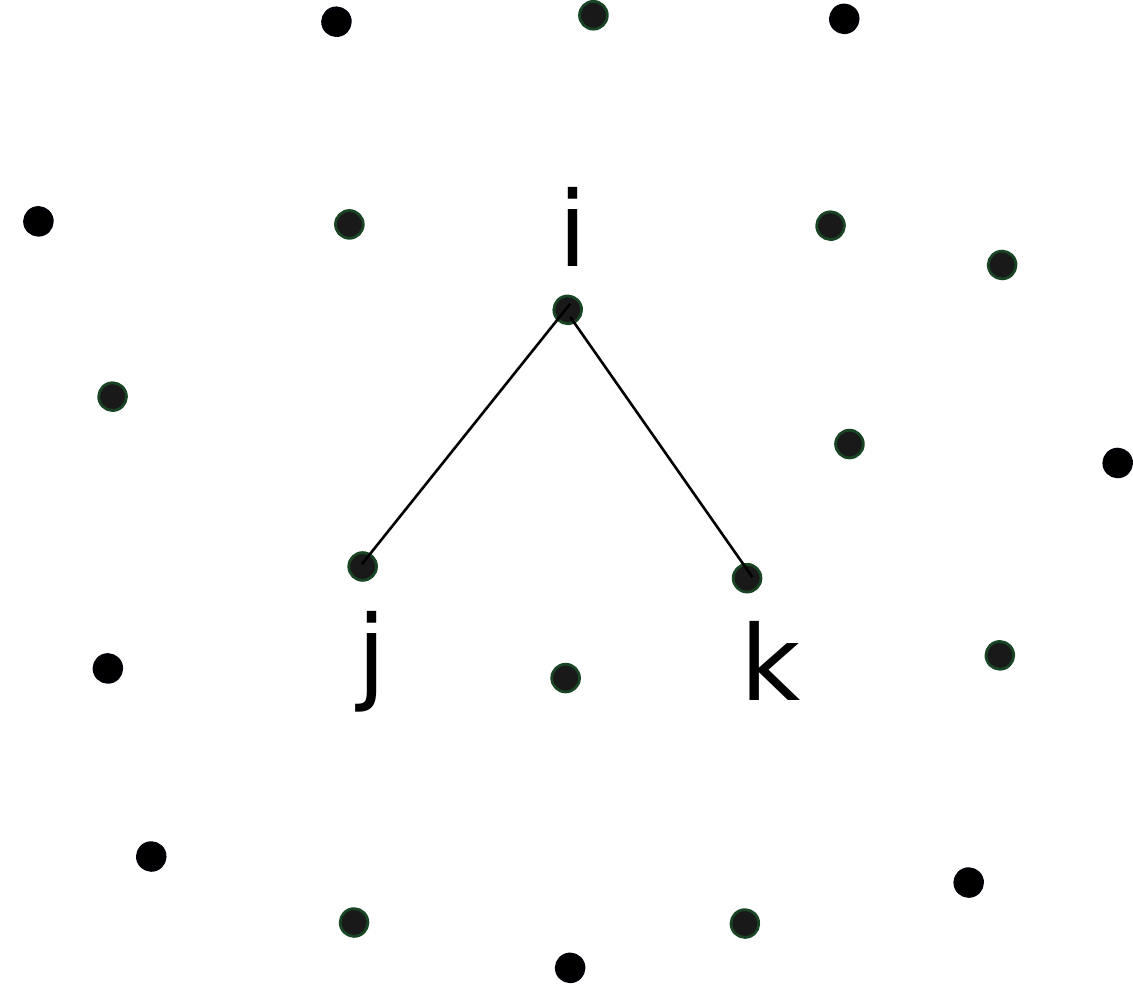}
\end{center}
\caption{
The local Hamiltonian density defined at site $i$ in \eq{eq:Hx2} 
includes hoppings between sites $j$ and $k$
which are connected through $i$ via two entanglement bonds.
The hopping amplitude is proportional 
to the product of 
$ \left(  \hPi \hPi^T \right)^{ij}$ and $\left(  \hPi \hPi^T \right)^{ki} $.
}
\label{fig:hopping2}
\end{figure}

In order to understand the meaning of the Hamiltonian,
it is convenient to go to the frame in which the lapse tensor is diagonal.
A non-singular lapse tensor can be written as  
\bqa
v =  n_v g_v^T S_v g_v,
\label{eq:SVD}
\eqa
where $n_v$ is a positive number,
$g_v \in  SL(L, \mathbb{R})$
and $S_v$ is a diagonal matrix whose elements
are either $1$ or $-1$.
See Appendix \ref{app:SVD} for the proof.
Then, \eq{Hx} can be written as
\bqa
\hat H_v =  n_v  \sum_i S_i \left[
- \hPi^{'i}_A \hPi^{'i}_A
+
\frac{ \ta }{M^2} \sum_{j,k}   
 \hPi^{'i}_{~A} \hPi^{'j}_{~A}
   \hPhi^{'B}_{~~j}  \hPhi^{'B}_{~~k}
 \hPi^{'k}_{~C} \hPi^{'i}_{~C}
 \right],
\label{eq:Hx2}
 \eqa
where 
$\hPhi' = \hPhi g_v^{-1}$ and
$\hPi' = g_v \hPi$.
Henceforth, let us omit the prime signs.
Here $S_i$ determines the direction of local time evolution at each site $i$.
The first term in the bracket of \eq{eq:Hx2} is an ultra-local kinetic term.
The second term describes a hopping process, 
where a particle jumps from sites $j$ to $k$ (and vice versa) 
with a hopping amplitude proportional to 
$(\hPi^j_{~A}  \hPi^i_{~A})   ( \hPi^i_{~C}  \hPi^k_{~C})$.
The hopping amplitude between two sites is given by 
the amplitudes of bi-local operators 
that connect the two sites through a third site $i$
as is shown in \fig{fig:hopping2}.
In the large $M$ limit,
the bi-linear operator $(\hPi^j_{~A}  \hPi^i_{~A})$ 
has a well-defined expectation value 
for $O(M)$ invariant states.
For the state in \eq{eq:t}, 
the expectation value 
of the bi-local operator 
is given by the collective variable 
\bqa
\frac{1}{M} \frac{ \lb t \cb (\hPi^j_{~A}  \hPi^i_{~A})  \cb t \rb}{\lb t \cb t \rb}
= - 2 i   (t^{-1} - t^{* -1})^{-1}_{ij}.
\label{eq:hatTT}
\eqa
The collective variable $t^{ij}$ in turn controls the mutual information 
between sites $i$ and $j$ through \eq{eq:EE}.
Therefore, \eq{eq:Hx2} describes a relatively local kinetic term
whose hopping amplitude is determined from the 
entanglement present between sites\cite{Lee2018,Lee2019}.
Because entanglement is state dependent, 
so are the hopping amplitudes and 
the graph that is formed by the network of hopping amplitudes.
Since the underlying matrix is dynamical,
the emergent manifold that is formed by the entanglement bonds
is fully dynamical.
In a globally entangled state, the Hamiltonian acts as a non-local Hamiltonian with global hoppings.
In a state with a local structure, the Hamiltonian acts as a local Hamiltonian 
in the corresponding dimension set by the local structure\footnote{
We note that this is different from the notion of
relative locality introduced in Ref. \cite{PhysRevD.84.084010}.
}.
For this reason, the dimension, topology and geometry are all dynamical in this theory\footnote{
The relatively local hopping term in \eq{eq:pipi2}
is different from the one considered in \cite{Lee2018}
in its detail form.
The difference is caused by the fact that we are imposing the condition
that the Hamiltonian should form a representation under the generalized
spatial diffeomorphism group.
}.

 \begin{figure}[ht]
 \begin{center}
   \subfigure[]{
 \includegraphics[scale=0.6]{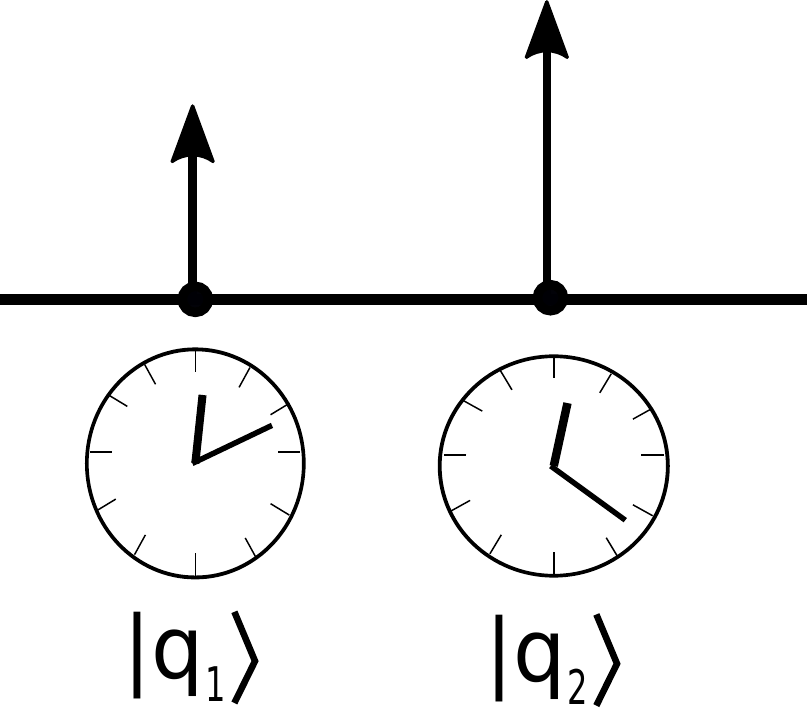} 
  \label{fig:psec1}
 } 
 \hfill
 \subfigure[]{
 \includegraphics[scale=0.6]{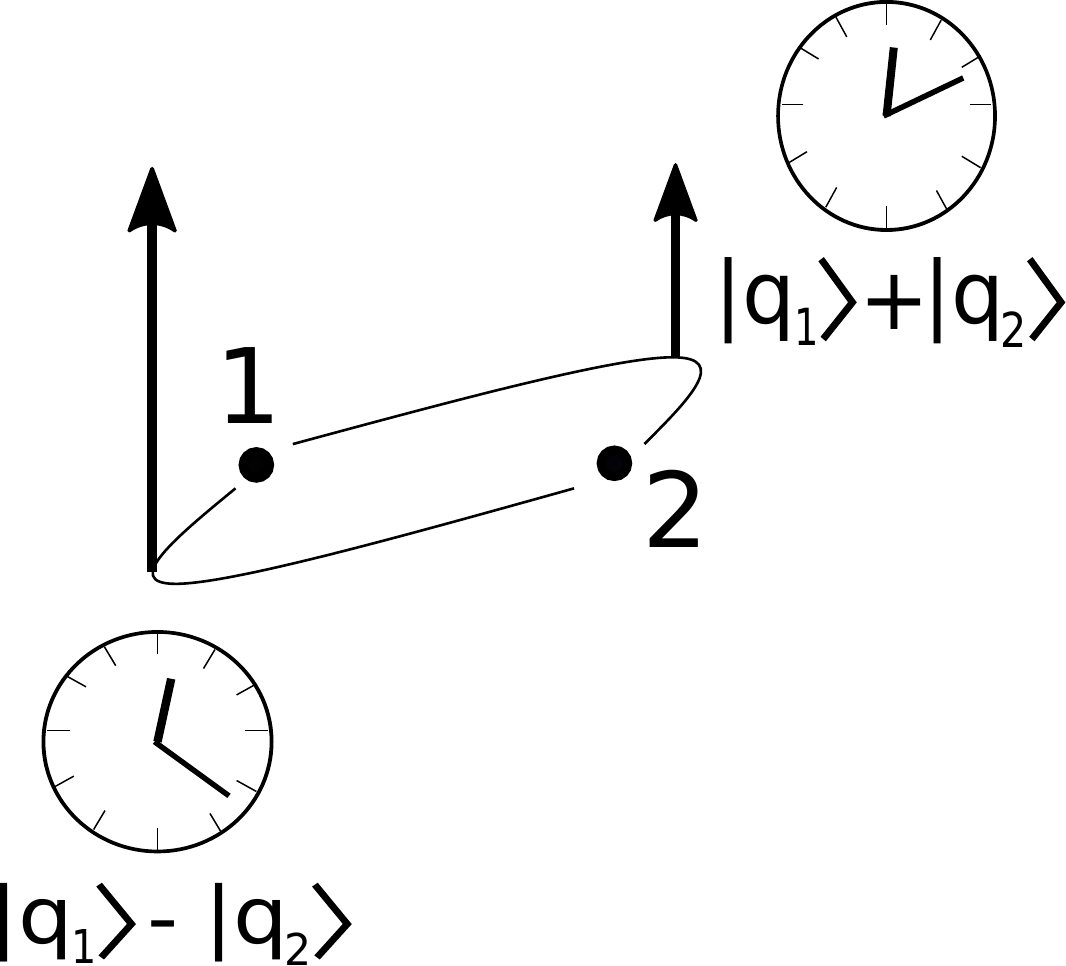} 
  \label{fig:psec2}
 } 
 \end{center}
 \caption{
 (a) In the general relativity,
 one can evolve a state defined on a Cauchy surface
 such that clocks at different locations in space run at different rates.
 This freedom of choosing space dependent lapse of time
 gives rise to the multi-fingered time evolutions.
 In the figure, $\cb q_1 \rb$ and $\cb q_2 \rb$ represent 
 the clock degrees of freedom
 localized at sites $1$ and $2$ respectively.
  (b) In the present theory, there is a larger gauge freedom.
 One not only has a freedom to choose different rates for a given set of clocks,
 but also has a freedom to construct a whole different set of clocks out of degrees of freedom
 defined at multiple sites through linear superpositions.
This extra freedom is encoded in the off-diagonal elements of the lapse tensor that rotates frame in \eq{eq:SVD}.
In priori, there is no preferred frame in which the clocks are defined. 
In order to define a time evolution in a meaningful  way,
one has to define a set of clocks out of physical degrees of freedom within the theory
and describe evolution of other degrees of freedom relative 
to the local evolution of the physical clocks.
 }
 \label{fig:clock}
 \end{figure}

For each choice of the lapse tensor,
the Hamiltonian is relatively local in the frame 
in which the lapse tensor is diagonal.
The gauge freedom in the choice of lapse tensor 
includes not only the freedom to choose site dependent speed of time evolution in a given frame\footnote{
This freedom is actually a part of redefining the frame itself
in which the the relative magnitudes of the $L$ vectors that form a frame 
is rescaled without rotating the directions of the vectors. 
}
but also the freedom to  rotate the frame in which the lapse tensor is diagonal.  
The space of lapse tensors in the present theory is much larger than 
that of lapse functions in the general relativity. 
In the general relativity, the rate of local time flow can be chosen independently at each site.
In the presence of $L$ sites, this would give rise to $L$ independent directions 
of multi-fingered time evolutions.
In the present theory, there are $\frac{L(L+1)}{2}$ independent parameters 
in the lapse tensor. 
The extra $\frac{L(L-1)}{2}$ off-diagonal elements
come from the freedom to rotate frames. 
As a result, the notion of the many-fingered time evolutions 
is generalized to a greater extent.
This is illustrated in \fig{fig:clock}.

In the general relativity, one can choose any lapse function 
to evolve a state defined on a time slice to a next time slice.
In the presence of physical degrees of freedom 
which can be used to construct a set of local clocks,
one can define a lapse function with respect to the clocks.
For example, one can use radioactive atoms distributed over space as local clocks,
and choose lapse such that the density of decayed atoms follow a specific profile in each time slices.
In the present theory, one can follow time evolution in any frame
by choosing general lapse tensor with off-diagonal elements.
This gives rise to a larger gauge freedom.
In the presence of physical degrees of freedom 
out of which a set of clocks can be constructed,
one can choose a frame in which those clocks are local.
If those clocks are initially unentangled,
the  lapse tensor that is diagonal in that frame
evolves the clocks independently to the leading order in $M$.
The time evolution of the remaining degrees of freedom generated by this Hamiltonian
describes their evolution relative to the set of local clocks that tick independently.
The relative motion of the other degrees of freedom 
measured against a chosen set of local clocks
is the gauge invariant prediction of the theory.
An example will be discussed in Sec. \ref{sec:local}.

\subsection{First-class constraint algebra}

The momentum and Hamiltonian constraints in 
Eqs. (\ref{Gy}) and (\ref{Hx})
should be compared with 
$P \left[ \vec \xi \right]$ 
and
$H \left[\theta \right]$ 
respectively in \eq{eq:PxHt}.
$y$ and $v$ in the present theory play the role of $ \xi^\mu(r)$ and $\theta(r)$ of the general relativity.
Just as $\xi^\mu(r)$ and $\theta(r)$ are the Lagrangian multipliers
that enforce 
$\cP_\mu(r) \cb \Psi \rb =0$ and
$\cH(r) \cb \Psi \rb =0$ 
for a gauge invariant state at every point $r$,
$y^j_{~i}$ and $v_{ij}$ enforce 
$\hat G^i_{~j} \cb \Psi \rb = 0$
and
$
\left(  
- \hPi \hPi^T 
+ \frac{ \ta }{M^2}  \hPi \hPi^T  \hPhi^T  \hPhi \hPi \hPi^T 
\right)^{ij} \cb \Psi \rb = 0$
for every $i$ and $j$,
where $\cb \Psi \rb$ is a gauge invariant state
of the matrix $\Phi$.

In order for the momentum and Hamiltonian constraints to generate 
consistent gauge transformations of quantum gravity,
they should satisfy a first-class constraint algebra.
In this section, we check the algebra 
that momentum and Hamiltonian satisfy. 
The commutator between an operator and \SLL generators 
is fixed by the representation that the operator forms under \SLL. 
$\Phi$ and $\Pi$ form the covariant and contravariant vectorial representations respectively
as is shown in \eq{SLLphipi}.
$\hat G$ and $\hat H$ 
are rank $2$ mixed tensor and contravariant tensors respectively.
This fixes their commutators with $\hat G$ to be
\bqa
\left[ \hG_{x}, \hG_{y} \right]  & = &     i \hG_{ ( yx-xy  ) }, \label{GGc} \\
\left[ \hG_x, \hH_v \right]  & = &    i \hH_{  v x + x^T v  }. \label{GHc}
\eqa
The commutator between Hamiltonians is more complicated.
But, the form of the Hamiltonian in \eq{Hx} suggests 
that the commutator is proportional to the generators of $GL(L, \mathbb{R})$
because the only non-trivial commutator arises from 
\bqa
\left[ (\hPhi^T  \hPhi)_{ij}, (\hPi \hPi^T)^{lm}  \right]   =
4 i  \hat {\bf G}^{[l}_{[i} \delta^{m]}_{j]},
\label{eq:PhiPi}
\eqa
where
$ \hat {\bf G}^{[l}_{[i} \delta^{m]}_{j]}
=\frac{1}{4}
\left(
\hat {\bf G}^l_{~i}
\delta^m_j
+
\hat {\bf G}^l_{~j}
\delta^m_i
+
\hat {\bf G}^m_{~i}
\delta^l_j
+
\hat {\bf G}^m_{~j}
\delta^l_i
\right)$.
An explicit calculation shows that the commutator actually depends
only on the \SLL generator and the Hamiltonian itself (see Appendix \ref{app:HH}),
\bqa
&& \left[ \hH_u, \hH_v \right]  
 =   
-i \frac{ 4  \ta}{M^2} 
\tr{ 
\left[
( \hPi \hPi^T) u  ( \hPi \hPi^T) v
-
( \hPi \hPi^T) v  ( \hPi \hPi^T) u
\right]
\hat G } \nn
&&  \hspace{0.7cm} 
+ i \frac{ 4 \ta^2}{M^4} u_{nk} v_{n' k'}
\Biggl[
- ( \hPi \hPi^T)^{kl} ( \hPhi^T \hPhi)_{li}  ( \hPi \hPi^T)^{k' i'}  ( \hPi \hPi^T)^{j' n'} \delta^n_j   
\nn && \hspace{3.5cm} 
+ ( \hPi \hPi^T)^{ki'} ( \hPhi^T \hPhi)_{jl}  ( \hPi \hPi^T)^{l n'}  ( \hPi \hPi^T)^{j' n}  \delta^{k'}_i 
 \nn &&
\hspace{3.5cm} + ( \hPi \hPi^T)^{ki'}  ( \hPi \hPi^T)^{k' l} ( \hPhi^T \hPhi)_{l i}   ( \hPi \hPi^T)^{j' n}   \delta^{n'}_j   \nn&&
\hspace{3.5cm} - ( \hPi \hPi^T)^{k' i'}  ( \hPi \hPi^T)^{j' n'} ( \hPhi^T \hPhi)_{jl}   ( \hPi \hPi^T)^{l n}  \delta^k_i   \nn &&
\hspace{3.5cm}  +  M ( \hPi \hPi^T)^{ki'}  ( \hPi \hPi^T)^{j' n'} \delta^{k'}_{i} \delta^{n}_{j}  
  + ( M+2) ( \hPi \hPi^T)^{k i'}  ( \hPi \hPi^T)^{k' n} \delta^{j'}_{i} \delta^{n'}_{j}  \nn &&
\hspace{3.5cm}  +  2 ( \hPi \hPi^T)^{k i'}  ( \hPi \hPi^T)^{j' n} \delta^{k'}_{i} \delta^{n'}_{j}  
  -2  ( \hPi \hPi^T)^{k n'}  ( \hPi \hPi^T)^{k' i'} \delta^{j'}_{i} \delta^{n}_{j}  \nn &&
\hspace{3.5cm}  -2 ( \hPi \hPi^T)^{k n'}  ( \hPi \hPi^T)^{j' i'} \delta^{k'}_{i} \delta^{n}_{j}  
  -2 ( \hPi \hPi^T)^{nk}  ( \hPi \hPi^T)^{n' i'} \delta^{j'}_{i} \delta^{k'}_{j}  
\Biggr] 
\hat G^{[i}_{[i'} \delta^{j]}_{j']}     \nn
&&  \hspace{0.7cm} 
+ \frac{4 \ta}{M^2} 
  \left[  \tr{  \hPi \hPi^T v} \tr{ \hat H u} -   \tr{  \hPi \hPi^T u} \tr{ \hat H v} \right],
\label{HHc} 
\eqa
where
$\hat G^{[l}_{[i} \delta^{m]}_{j]}
=\frac{1}{4}
\left(
\hG^l_{~i}
\delta^m_j
+
\hG^l_{~j}
\delta^m_i
+
\hG^m_{~i}
\delta^l_j
+
\hG^m_{~j}
\delta^l_i
\right)$.
It is noted that $ (\hPi \hPi^T), ( \hPhi^T \hPhi), \hat G, \hat H \sim O(M)$ in the large $M$ limit.
The first two terms in \eq{HHc} are $O(M)$.
The last term that depends on $\hat H$ is $O(1)$, 
and is sub-leading in the large $M$ limit.
The last term is generated as operators are ordered such that 
$\hat G$ appears at the far right in the first two terms.
This ordering makes it manifest 
that states annihilated by $\hat G$ and $\hat H$ 
are automatically annihilated by their commutators.
Therefore, no additional constraints are needed
to define the space of gauge invariant states.
In short, the momentum and Hamiltonian constraints form a first class algebra\cite{dirac_1950}.
We note that the first class algebra summarized
in Eqs. (\ref{GGc}), (\ref{GHc}) and (\ref{HHc}) 
is the operator algebra 
that holds independent of states.
For example, physical states that are annihilated by the constraints
may or may not have the $O(M)$ flavour symmetry.

\section{Path integral representation of state projection}
\label{sec:path}

\subsection{Projection}

The momentum and Hamiltonian are generators of gauge transformations.
The physical Hilbert space is given by 
the set of gauge invariant states.
Let $\czr$ be a gauge invariant state which satisfies
\bqa
\hG_y \czr  = \hH_v  \czr = 0 
\label{constraints}
\eqa
for any choice of $y, v$.
An example of gauge invariant states is 
$\int  D \Phi ~  \cb \Phi \rb$.
However, this state has infinite norm
with respect to the inner product defined in \eq{eq:norm}.
This is a property that all gauge invariant states share.
Gauge invariant states are non-normalizable because 
the Hilbert space and the gauge group are both non-compact.
Wavefunctions of gauge invariant states 
are necessarily extended over unbounded regions in the phase space
as is proven in Appendix \ref{app:nonnormalizability}.

On the other hand, 
quantum states to which probabilities can be assigned 
should be normalizable. 
Wavefunctions of normalizable states are localized 
within compact regions in the gauge orbit,
breaking gauge symmetry\cite{Lee2018}.
The fact that all normalizable states  
break the gauge symmetry has a few consequences.
First, there exists no normalizable state that is frozen in time in this theory.
All physical states, in the sense that they are normalizable,
must evolve non-trivially in time.
This provides one possible explanation 
for why we physically perceive the continuous passage of time 
although time evolution is merely a gauge transformation. 
Second, the emergent time should be non-compact.
Perfect periodic orbits are impossible
because the existence of a periodic gauge orbit
of normalizable states implies a normalizable gauge invariant state. 

The incompatibility between gauge invariance and normalizability 
gives rise to a non-trivial evolution as normalizable states
are projected toward a gauge invariant state\cite{1992gr.qc....10011I,1992grra.conf..211K}.
Let us denote a normalizable state as $\ccr$.
The projection of $\ccr$ to a gauge invariant state, $\czr$
is given by
\bqa
\lzc \chi \rb.
\label{eq:overlap}
\eqa
This can be viewed as the wavefunction 
of  $\czr$ written in the basis of $\ccr$\cite{PhysRevD.28.2960,Maldacena_2003}.
In the following, we show that \eq{eq:overlap} can be written as
a path integration of collective variables that describe fluctuations of 
emergent spactime.

\subsection{Gauge invariant local structure}

\label{sec:local}

Under the generalized spatial diffeomorphism,
the definition of local Hilbert spaces and the local structure associated 
with the local Hilbert spaces change. 
There is no preferred frame in priori.
A gauge invariant local Hilbert spaces
can only be defined with reference to physical degrees of freedom within the theory.
This is possible within a sub-Hilbert space
in which $O(M)$ flavour symmetry is spontaneously broken down to a smaller group.
We consider states in which the $O(M)$ symmetry is broken as
$\Phi$ acquires non-zero expectation values 
in an  $L \times L$ block.
If the $L \times L$ block of $\Phi$ is non-singular,
$O(M)$ symmetry is broken down to $O(N)$, where $N = M-L$.
We consider the sub-Hilbert space in which the $O(N)$ flavour symmetry is further 
broken down to $O(N/2) \times O(N/2)$. 
In order to distinguish the first $L$, the next $N/2$ and the remaining $N/2$ flavours,
we write
\bqa 
 \Phi^A_{~i} & = & \left\{
 \begin{array}{cl} 
 \sqrt{N}  q^{A}_{~i} & \mbox {~~for ~~$A=1,2,..,L$}  \\
\phi^{A-L}_{~i} &  \mbox {~~for ~~$A= L+1, L+2, .., L+N/2$} \\
\varphi^{A-(L+N/2)}_{~i} &  \mbox {~~for ~~$A= L+N/2+1, L+N/2+2, .., L+N$} 
 \end{array}
 \right. .
 \label{eq:qphivarphi}
 \eqa
$q$ represents  an $L \times L$ matrix
that acquire non-zero expectation values.
A factor of $\sqrt{N}$ is introduced  
as we will consider the large $N$ limit with $\lb q  \rb \sim O(1)$.
Under generalized spatial diffeomorphisms, 
$q$ is transformed as $q \rightarrow q ~ g$, where $g \in  SL(L, \mathbb{R})$,
and plays the role of a Stueckelberg field. 
If $q$ contains $L$ independent row vectors,
one can define a frame 
in terms of those vectors.
The local structure defined in this frame is gauge invariant.
$\phi$ and $\varphi$ represent $N/2 \times L$ matrices that have zero expectation value.
Henceforth, we use $ \cb q, \phi, \varphi \rb$ in place of $\cb \Phi \rb$.

In the frame chosen by the Stueckelberg field,
sites are labeled by distinguishable physical flavour.
However, it is not necessary to have distinguishable sites to define a frame.
It is sufficient to have $L$ {\it unordered} independent vectors.
In order to define an unordered set of row vectors,
we consider states in which $S_{L}^f$ symmetry is unbroken,
where $S_{L}^f$ is the permutation that acts on the first $L$ flavours. 
We refer to $S_{L}^f$ as the flavour permutation group.
We denote the sub-Hilbert space with unbroken $S_{L}^f \times O(N/2) \times O(N/2)$ as  $\cV$.

The choice of the sub-Hilbert space with this particular flavour group is not crucial.
If one chooses different sub-Hilbert spaces, 
the collective modes that describe fluctuations 
within the sub-Hilbert spaces changes.
The number of collective variables increases
as the symmetry of the sub-Hilbert space is lowered.
Here, we choose  $S_{L}^f \times O(N/2) \times O(N/2)$ as an example
that gives rise to a minimal set of physical degrees
of freedom including dynamical gravity. 
The counting of the propagating degrees of freedom 
in this sub-Hilbert space is given at the end of Sec.  \ref{GITA}.

General states in $\cV$ can be parameterized in terms of collective variables.
In order to construct basis states for $\cV$, 
it is convenient to introduce
\bqa
\cepr =  \sum_{P^f \in S_L^f }  \cb P^f q, \phi, \varphi \rb
\label{symm}
\eqa
that are symmetric under permutations of the first $L$ flavours.
While the expectation value of $\phi^a_{~i}$ and $\varphi^a_{~i}$ is zero,
$O(N/2) \times O(N/2)$ invariant operators can still have non-zero expectation values.  
Any wavefunction in $\cV$ written in the basis of \eq{symm} can be expressed as a function of 
$q^\alpha_{~~i}$,
$(\phi^a_{~i} \phi^a_{~j})$ and
$(\varphi^a_{~i} \varphi^a_{~j})$,
where the flavour $a$ is summed from $1$ to $N/2$ in the last two operators.
Therefore, states in $\cV$ can be spanned by 
the following basis states labeled by three collective variables, 
\bqa
\cstr & = & \int D q D \phi D \varphi~ ~ e^{  i  \tr{ N s q + t_1 ( \phi^T \phi) + t_2 ( \varphi^T \varphi)}   }  \cepr,
\label{eq:st}
\eqa
where
$ \tr{ s q  }  = s^i_{~\alpha}  q^\alpha_{~i}$,
$  \tr{  t_1 ( \phi^T \phi) } = t_1^{ij} \phi^a_{~j} \phi^a_{~i}$,
$  \tr{  t_2 ( \varphi^T \varphi) } = t_2^{ij} \varphi^a_{~j} \varphi^a_{~i}$.
$s^i_{~\alpha}$ is the conjugate variable of $q^\alpha_{~i}$.
$t_c^{ij}$ with $c=1,2$ are bi-local variables that are conjugate to 
$\phi^a_{~i} \phi^a_{~j}$
and 
$\varphi^a_{~i} \varphi^a_{~j}$,
respectively.
Both $t_c$ and $s$ are invariant under $O(N/2) \times O(N/2)$.
Because $\cepr$ in \eq{symm} is invariant under flavour permutations, 
so is $\cb s, t_1, t_2 \rb$,
\bqa
\cb s, t_1, t_2 \rb = \cb s P^f, t_1, t_2 \rb
\label{eq:SLf}
\eqa
for any $P_f  \in S_{L}^f$.
For a later use, we also introduce permutations of sites
which act on the site index of the collective variables as
\bqa
s \rightarrow  P_g s, ~~~~~~
t_c  \rightarrow  P_g t_c P_g^T.
\label{eq:SLg}
\eqa
The site permutation group with the even parity, denoted as $S_L^g$, 
is a subgroup of the generalized spatial diffeomorphism group, \SLL.

General states in $\cV$  can be written as 
\bqa
\ccr  = \int D s Dt  ~ \cstr  \chi(s,t_1,t_2).
\label{ccr}
\eqa
Here 
$D s \equiv \prod_{i, \alpha} d s^i_{~\alpha}$, $D t \equiv \prod_{i \geq j } \left[ d t_1^{ij} dt_2^{ij} \right]$,
and the integrations of $ds^i_{~\alpha}$
and $dt_c^{ij}$ are defined along the real axis.
$ \chi(s,t_1,t_2)$ is a wavefunction of the collective variables.
The states in \eq{ccr} 
form a complete basis of $\cV$.
This sub-Hilbert space forms 
the kinematic Hilbert space of the present theory.

\subsection{Path integration of collective variables}
\label{subsec:path}

Now we consider the projection of a normalizable state in $\cV$
to a gauge invariant state in \eq{eq:overlap}
in the limit that $N \gg L \gg 1$.
Thanks to the gauge invariance of $\czr$,
the overlap in \eq{eq:overlap} is invariant under gauge rotation\cite{Lee2016},
\bqa
&& \lzc \chi \rb  = 
\lzc
e^{-i \epsilon \left(
\hH_{v^{(1)}} +  \hG_{y^{(1)}}
\right)} \cb  \chi \rb,
\label{eq:overlap2}
\eqa
where $v^{(1)}$ is a lapse tensor
and $y^{(1)}$ is a shift tensor.
If  $\hG$ and $\hH$ are applied to the right in  \eq{eq:overlap2},
they generate a non-trivial evolution of  $\cb \chi \rb$,
\bqa
&& e^{-i \epsilon \left(
\hH_{v^{(1)}} +  \hG_{y^{(1)}}
\right)} \cb  \chi \rb \nn
&&
= 
 \int D s^{(0)} Dt^{(0)} 
 \int D q D \phi D \varphi
  ~~ 
\cepr
 ~e^{
i  \tr{  N s^{(0)} q + t_1^{(0)}  ( \phi^T \phi) + t_2^{(0)}  ( \varphi^T \varphi)  } 
}  \times \nn
&& 
~~ e^{ 
-i  \epsilon N  \tr{
v^{(1)}   {\cal H}\left[ q, s^{(0)}, \frac{(\phi^T \phi)}{N}, t_1^{(0)}, \frac{(\varphi^T \varphi)}{N}, t_2^{(0)}  \right]
+ y^{(1)}  {\cal G}\left[ q, s^{(0)}, \frac{(\phi^T \phi)}{N}, t_1^{(0)}, \frac{(\varphi^T \varphi)}{N}, t_2^{(0)}  \right]
}
}
~
\chi( s^{(0)}, t_1^{(0)}, t_2^{(0)}  ),
\label{chi0chi_1}
\eqa
where
\bqa
&& \cH[q,s,p_1,t_1,p_2,t_2]  = 
-\left( s s^{T} + \sum_c \left[ 4  t_c p_c t_c - i t_c \right] \right) \nn
&& + \ta 
\left( s s^{T} + \sum_c \left[ 4  t_c p_c t_c - i t_c \right] \right)
\left( q^T q   +  p_1+p_2 \right)   
\left( s s^{T} + \sum_{c'} \left[ 4  t_{c'} p_{c'} t_{c'} - i t_{c'} \right] \right)
+ O \left( \frac{1}{N} \right),  \nn
 \label{H}
\\
&& \cG[q,s,p_1,t_1,p_2,t_2]  =  
  \left( s q  +  2 \sum_c  t_c p_c - i \frac{M}{2N}  I  \right),
 \label{G}
 \eqa
 and $s$ and $t_c$ in \eq{ccr} are relabeled as 
 $s^{(0)}$ and $t_c^{(0)}$ in \eq{chi0chi_1}.
The evolution generated by the constraints is a manifestation of the fact
that $\ccr$ is not gauge invariant.
The resulting state in \eq{chi0chi_1} is also in $\cV$,
and can be written as a linear superposition of 
\eq{eq:st}.
This is expressed as an integration over another set of collective variables
and their conjugate variables
$\left(
q^{(1)}, 
s^{(1)}, 
p_c^{(1)},
t_c^{(1)}
\right)$,
\bqa
\lzc \chi \rb & = & 
 \int D s^{(0)} Dt^{(0)} 
 D s^{(1)} Dt^{(1)} 
 D q^{(1)} Dp^{(1)} 
~
\lzc s^{(1)}, t_1^{(1)},   t_2^{(1)}  \rb
e^{
-i N \tr{
q^{(1)} ( s^{(1)} - s^{(0)} )
+  p_c^{(1)} ( t_c^{(1)} - t_c^{(0)} )
}} \times \nn
&&
e^{
-i \epsilon N  \tr{
v^{(1)} \cH[q^{(1)},s^{(0)}, p_1^{(1)}, t_1^{(0)} , p_2^{(1)}, t_2^{(0)}     ]
+y^{(1)} \cG[q^{(1)},s^{(0)}, p_1^{(1)}, t_1^{(0)} , p_2^{(1)}, t_2^{(0)}     ]   
}
} 
~
\chi( s^{(0)}, t_1^{(0)}, t_2^{(0)}  ).
\label{chi0chi_2}
\eqa
It is straightforward to check that 
\eq{chi0chi_2} is reduced to
\eq{chi0chi_1} upon integrating out 
the collective variables.
Upon integrating over  $s^{(1)}$ and $t_c^{(1)}$,
which play the role of dynamical sources,
one obtains the delta functions that enforce
the constraints for the conjugate variables,
$q^{(1)} = q$,
 $p_1^{(1)} = \frac{ \phi^T \phi}{N}$,
  $p_2^{(1)} = \frac{ \varphi^T \varphi}{N}$.
The following integration over $q^{(1)}$ and $p_c^{(1)}$,
which represent dynamical operators,
 reproduces \eq{chi0chi_1}. 
The new set of dynamical collective variables  
in \eq{chi0chi_2}
removes terms that are non-linear in 
$q^\alpha_{~~i}$,
 $\phi^a_{~i} \phi^a_{~j}$ and
 $\varphi^a_{~i} \varphi^a_{~j}$
in \eq{chi0chi_1}.
\eq{H} and \eq{G} represent the Hamiltonian and momentum constraints
induced for the collective variables.
Since \eq{chi0chi_1} is independent of the lapse and shift tensors, 
$v^{(1)}$ and $y^{(1)}$ can be integrated over in \eq{chi0chi_2}.
They can be viewed as Lagrangian multipliers
that enforce the constraints. 

\vspace{0.5cm}
\begin{figure}[ht]
\begin{center}
\centering
\includegraphics[scale=0.6]{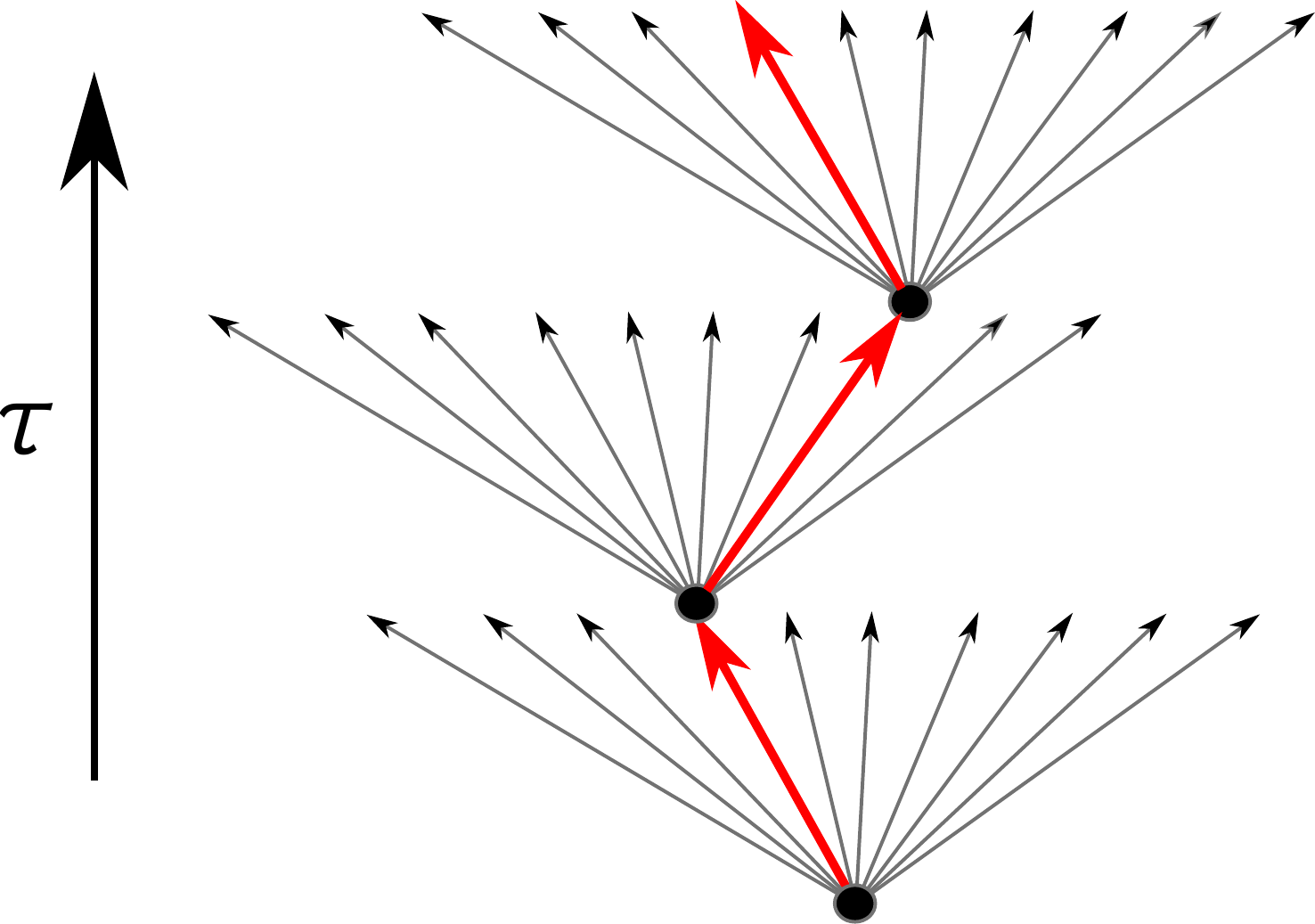}
\end{center}
\caption{
A state can be evolved by the constraints 
with different choices of lapse and shift tensors.
A specific choice represents one of multi-fingered time evolutions.
}
\label{fig:branches}
\end{figure}

We repeat the procedure by inserting 
$e^{-i \epsilon \left(
\hH_{v^{(2)}} +  \hG_{y^{(2)}}
\right)}$
between 
$ \lzc$ and 
$\cb s^{(1)}, t_1^{(1)},   t_2^{(1)}  \rb$ in \eq{chi0chi_2}.
This gives rise to an evolution of 
$\cb s^{(1)}, t_1^{(1)},   t_2^{(1)}  \rb$.
The resulting state can be again expressed as a linear superposition of
$\cb s, t_1, t_2 \rb$,
which is expressed as an integration over 
a yet another set of dynamical collective variables.
Repeated insertions of the Hamiltonian and momentum give
rise to a path integration of the collective variables\cite{Lee:2013dln},
\bqa
\lzc \chi \rb & = & 
\int
 D s^{(0)} Dt^{(0)} 
 \int {\cal D} s {\cal D} t {\cal D} q {\cal D} p
 {\cal D} v {\cal D} y
~
\lzc s^{(\infty)}, t_1^{(\infty)},   t_2^{(\infty)} \rb
e^{ i S } 
~
\chi( s^{(0)}, t_1^{(0)}, t_2^{(0)}  ),
\label{eq:pathint}
\eqa
where
\bqa
S & = & 
N \int_0^\infty d \tau ~
\mbox{tr} \Biggl\{
- q \partial_\tau  s
-  p_c \partial_\tau t_c 
- v(\tau) \cH[q(\tau),s(\tau), p_1(\tau),t_1(\tau),   p_2(\tau),t_2(\tau) ] \nn
&& 
- y(\tau) \cG[q(\tau),s(\tau), p_1(\tau),t_1(\tau),   p_2(\tau),t_2(\tau) ]
\Biggr\}. \label{Sfinal} 
\eqa
Here ${\cal D} s \equiv \prod_{l=1}^\infty ds^{(l)}$
and $s(\tau) = s^{(l)}$ with $\tau = l \epsilon$.
${\cal D} q$, ${\cal D} t$, ${\cal D} p$, ${\cal D}v$, ${\cal D} y$,
and $q(\tau)$, $t_c(\tau)$, $p_c(\tau)$, $v(\tau)$, $y(\tau)$
are similarly defined. 
Here $\tau$ is a parameter time that labels different
stages of evolution generated by the momentum
and Hamiltonian constraints.
Here $q$ and $p_c$ play the role of generalized coordinates
and $s$ and $t_c$ are their conjugate momenta\footnote{
An integration by part in \eq{Sfinal} gives the standard symplectic form, 
$\int d \tau \tr{ s \partial_\tau q + t_c \partial_\tau p_c}$ with a boundary term.
}.
The sum over the shift and lapse tensors in \eq{eq:pathint}
represents different paths in which the state in the kinematic Hilbert space
is projected to the gauge invariant state.
A particular path of the shift and lapse tensor represents
one of the multi-fingered time evolutions.
This is illustrated in \fig{fig:branches}.
The collective variables $t_c(\tau)$, $p_c(\tau)$, $s(\tau)$, $q(\tau)$ are generalized gravitational degrees of freedom
that describe fluctuations of dimension, topology and geometry of spacetime. 
In the following section,
we derive the precise relation between
the collective variables and the metric
of the emergent geometry.

\subsection{Gauge invariance of the  action}
\label{GITA}

Because $\hat H$ and $\hat G$ obey the first-class constraint algebra,
the action in \eq{Sfinal} is invariant under gauge trasformations 
generated by the constraints.
To see this, we first note that the symplectic form in \eq{Sfinal}
defines the Poisson bracket,
\bqa
\{ A, B \}_{PB} & = & 
\left(
    \frac{\partial A}{\partial q^\alpha_{~i } }  \frac{\partial B}{\partial s^i_{~\alpha} } 
-
\frac{\partial A}{\partial s^i_{~\alpha} }  \frac{\partial B}{\partial q^\alpha_{~i } }
      \right)
+ 
\delta^{kl}_{ij}
\left(
  \frac{\partial A}{\partial p_{c, ij}} \frac{\partial B}{\partial t_c^{kl} }
- \frac{\partial A}{\partial t_c^{kl} }  \frac{\partial B}{\partial p_{c,ij}}
\right),
\label{Poisson}
\eqa
where 
$\delta_{ij}^{kl} = \frac{1}{2}
\left(
   \delta_{i}^{k} \delta_{j}^{l} 
+ \delta_{i}^{l} \delta_{j}^{k}  \right)$. 
To the leading order in $1/N$, 
the Poisson brackets of \eq{H} and \eq{G} are given by
(see Appendix \ref{app:Poisson})
\bqa
 \{  \cG^i_{~j}, \cG^k_{~l} \}_{PB} & = & \cA^{ikn}_{jlm}~ \cG^m_{~n}, \label{eq:GGH1} \\
 \{  \cG^i_{~j}, \cH^{kl} \}_{PB} & = & \cB^{ikl}_{jmn}~ \cH^{mn}, \label{eq:GGH2} \\
 \{  \cH^{ij}, \cH^{kl} \}_{PB} & = & \cC^{ijkl n}_{m}~ \cG^m_{~n},
\label{eq:GGH3}
\eqa 
where
\bqa
\cA^{ikn}_{jlm} & = & \delta^k_j \delta^i_m \delta^n_l - \delta^i_l \delta^k_m \delta^n_j, \nn
\cB^{ikl}_{jmn} & = & \delta^k_j \delta^{il}_{mn} +  \delta^l_j \delta^{ki}_{mn}, \nn
\cC^{ijkln}_{m} & = &
-4 \ta
\Bigl[
U^{n[j} U^{i][l} \delta^{k]}_m - U^{n [l} U^{k][j} \delta^{i]}_m 
\Bigr]  \nn
&& + 4 \ta^2 \Bigl[
U^{n[j} U^{i]m'} Q_{m' n'} U^{n' [l} \delta^{k]}_m
+U^{n[j} U^{i][l}   U^{k] m'} Q_{ m' n'}  \delta^{n'}_m \nn
&& \hspace{1cm}
-U^{n[l} U^{k]m'} Q_{m' n'} U^{n' [j} \delta^{i]}_m
-U^{n[l} U^{k][j} U^{i] m'} Q_{ m'  n'}  \delta^{n'}_m
\Bigr]
\label{eq:ABC}
\eqa
with
\bqa
U^{ij} &=& \left( s s^T + \sum_c \left[ 4 t_c p_c t_c - i t_c \right] \right)^{ij}, \nn
Q_{ij} & = & \left( q^T q + \sum_c p_c  \right)_{ij}.
\label{UQ }
\eqa
In the last equation of \eq{eq:ABC}, 
the indices within brackets are symmetrized, 
e.g., 
$U^{n[j} U^{i][l} \delta^{k]}_m =
\frac{1}{4} \left(
 U^{nj} U^{il} \delta^{k}_m 
+U^{ni} U^{jl} \delta^{k}_m 
+ U^{nj} U^{ik} \delta^{l}_m 
+U^{ni} U^{jk} \delta^{l}_m 
\right)$.
Eqs. (\ref{eq:GGH1})-(\ref{eq:GGH3})
have the same structure as 
Eqs. (\ref{GGc}), (\ref{GHc}) and \eq{HHc}
to the leading order in $1/N$.
The Hamiltonian that appears on the right hand side of 
\eq{HHc} is sub-leading.

The action in \eq{Sfinal} is invariant under the time-dependent gauge transformations,
\bqa
\delta F & = & \eta^j_{~i} \{  F, \cG^i_{~j} \}_{PB} +  \rho_{ji} \{  F, \cH^{ij} \}_{PB}, \nn
\delta v_{mn} & = & \partial_\tau \rho_{mn} 
+  \eta^j_{~i} v_{lk}  \cB^{i k l}_{j  m n} 
- \rho_{kl} y^{j}_{~i}  \cB^{ ikl}_{  j mn}, \nn
\delta y^n_{~m} & = & \partial_\tau \eta^n_{~m} + 
 \eta^j_{~i} y^{l}_{~k}  \cA^{i k n}_{j  l m} 
 +   \rho_{ji} v_{lk}  \cC^{ijkl n}_{ m}, 
 \label{eq:spacetimegauge}
 \eqa 
where  $F = \{ s,q,t,p \}$  denote the collective variables,
and $\eta^j_{~i}(\tau)$ and $\rho_{ij}(\tau)$ are infinitesimal gauge parameters.
The action is invariant off-shell as
the equation of motion is not needed for the invariance of the action.
Besides the spacetime diffeomorphism,
the theory is also invariant under the reversal of the parameter time,
\bqa
 i & \rightarrow & -i, \nn
 \Bigl\{ p_c(\tau), q(\tau), v(\tau) \Bigr\} & \rightarrow &   \Bigl\{ p_c(-\tau), q(-\tau), v(-\tau) \Bigr\}, \nn
 \Bigl\{ t_c(\tau), s(\tau), y(\tau) \Bigr\} & \rightarrow &   - \Bigl\{ t_c(-\tau), s(-\tau), y(-\tau) \Bigr\}.
\label{eq:TR}
\eqa

Let us count the number of physical degrees of freedom.
$q$ and $s$ are $L \times L$ matrices,
and $t_1, t_2, p_1, p_2$ are $L \times L$ symmetric matrices.
This give $D_k = 2 L^2 + 2 L (L+1)$ phase space kinematic variables.
On the other hand, $\cG$ is traceless $L \times L$ matrix 
and $\cH$ is $L \times L$ symmetric matrix.
The total number of constraints is $D_c = (L^2-1) + \frac{L(L+1)}{2}$,
and the dimension of the constraint hypersurface is 
$ 
D_k - D_c
$.
The first-class constraints generate gauge orbits of dimension $D_c$
within the constraint hypersurface.
Since points on a gauge orbit are physically identical,
the total number of physical phase space variables is
\bqa
D_k - 2 D_c 
= L(L+1)+2.
\label{eq:NDOF}
\eqa

\section{Spacetime diffeomorphism and emergent geometry}
\label{sec:spacetime}

The path integral in \eq{eq:pathint} consists of two parts.
The first is the integration over $v(\tau)$ and $y(\tau)$.
Each path of the lapse and shift tensors 
selects one of the multi-fingered time evolutions.
Because of the gauge invariance in \eq{eq:spacetimegauge},
the shift and lapse tensors need to be fixed through a gauge fixing condition.
This will be discussed in the next section.
The remaining path integration is over the collective variables.
Each path describes a history of the collective variables
that represents a spacetime which emerges
dynamically.
In the large $N$ limit, 
the fluctuations of the collective variables become small,
and the saddle-point approximation can be made.
If the initial state $\ccr$ has a local structure in a frame, 
a spacetime manifold with well-defined 
dimension, topology and geometry emerges
at the saddle point.
In this section, we discuss how the geometry 
of the emergent manifold is determined from the collective variables.
For this, we first extract the constraint algebra of the general relativity in Eqs. (\ref{eq:GR1})-(\ref{eq:GR2})
from Eqs. (\ref{eq:GGH1})-(\ref{eq:GGH3}).

\subsection{ Momentum density }

We first identify the generators of smooth spacetime diffeomorphism 
for states that have local structures.
Let $r^\mu_i$ be the mapping from sites to a manifold that 
is determined from the local structure in a frame.
The tensor $\cG^i_{~j}$ can be viewed as a bi-local field that depends
on two positions on the manifold. 
If the collective variables change slowly in the manifold,
$\cG^i_{~j}$ varies slowly as a function of $r_i$ and $r_j$,
and can be expanded in coordinates.
By expanding $\cG^i_{~j}$ around $j=i$,
we write the \SLL generator with shift tensor $y$  as
\bqa
\cG_y 
& = & \cG^i_{~j} y^j_{~i}.  \nn
& = & \left[ \cG^i_{~i} + \left.  \frac{\partial \cG^i_{~j}}{\partial r^\mu_j} \right|_{j=i} ( r^\mu_j - r^\mu_i ) + .. \right] y^j_{~i} \nn
&=&   \cG^i_{~i} \zeta_y(r_i) +  \left.  \frac{\partial \cG^i_{~j}}{\partial r^\mu_j} \right|_{j=i} \xi_y^\mu(r_i) + ... ,
\label{eq:Gy}
\eqa
where $\zeta_y$ and $\xi^\mu_y$ represent the scalar and  vector fields
 defined in Eqs. (\ref{eq:zeta})-(\ref{eq:xi}) associated with shift tensor $y$.
In the continuum limit, \eq{eq:Gy} is written as
\bqa
\cG_y & = & \int dr \Bigl(
\cD(r) \zeta_y(r) +  \cP_\mu(r)  \xi_y^\mu(r) + .. 
\Bigr),
\label{eq:Gyc}
\eqa
where
\bqa
\cD(r_i) & = & V_i^{-1} \cG^i_{~i}, \label{eq:cD} \\
\cP_\mu(r_i) & = &  V_i^{-1}  \left.  \frac{\partial \cG^i_{~j}}{\partial r^\mu_j} \right|_{j=i}. \label{eq:cHmu}
\eqa
$V_i$ is the coordinate volume 
assigned to site $i$ in the manifold (see \fig{fig:graph_12D}).
$V_i^{-1}$ in \eq{eq:cD} and \eq{eq:cHmu} densitizes
the objects defined at sites.
$\cD(r_i)$ is the local dilatation density 
which generates the Weyl transformation.
$\zeta_y$ corresponds to the temporal component of the 
non-compact Weyl U(1) gauge field
in the Hamiltonian formalism. 
$\cP_\mu(r_i) $ is identified as the momentum density,
and $\xi_y$ becomes the shift. 
Now let us check that  \eq{eq:cHmu} satisfies the constraint algebra 
of the general relativity given in \eq{eq:GR1}.
The Poisson brackets of  $\cD(r_i)$  and $\cP_\mu(r_i) $ are fixed by 
those of \SLL generators.
 \eq{eq:GGH1} implies $\{ \cG_x, \cG_y \}_{PB} = \cG_{yx - xy}$,
which can be written as
\bqa
&&
 \left\{ 
\int dr \Bigl( \cD(r) \zeta_x(r) +  \cP_\mu(r)  \xi_x^\mu(r) + .. \Bigr), 
\int dr' \Bigl( \cD(r') \zeta_y(r') +  \cP_\nu(r')  \xi_y^\nu(r') + ..\Bigr)
\right\}_{PB} \nn
&=&
\int dr 
\Bigl( \cD(r) \zeta_{yx-xy}(r) +  \cP_\mu(r)  \xi_{yx-xy}^\mu(r) + ..\Bigr),
\label{eq:DHDH}
\eqa
where 
$\zeta_{yx-xy}(r)$ and $\xi^\mu_{yx-xy}(r)$ 
are the scalar and vector fields associated with the shift tensor, $yx-xy$.
They are given by 
(see  Appendices \ref{app:fields1} and  \ref{app:fields2} for derivation)
\bqa
\zeta_{yx-xy}(r) & = & 
\mathcal{L}_{\xi_x} \zeta_y(r) 
-\mathcal{L}_{\xi_y} \zeta_x(r) 
+ O( \partial^2) ,
\label{eq:GGcc1} \\
\xi^\mu_{yx-xy}(r) & = & 
\left( \mathcal{L}_{\xi_x} \xi_y (r) \right)^\mu  + O( \partial^2).
\label{eq:GGcc2}
\eqa
$ O( \partial^2) $ denotes terms that involve two or more derivatives.
Eqs. (\ref{eq:GGcc1}) and (\ref{eq:GGcc2})
imply that $\cP_\mu$ defined in \eq{eq:cHmu} 
generates spatial diffeomorphism 
under which $\cD$ and $\cP_\mu$ transform
as a scalar density and a vector density, respectively.
\eq{eq:GR1} is indeed reproduced 
to the leading order in the derivative expansion.

\subsection{ Hamiltonian density }

Now we write the Hamiltonian in the continuum limit.
In a frame chosen by the local structure of a state,
we divide the lapse tensor into the diagonal components
and the off-diagonal components as
\bqa
\cH_v & = & 
\sum_i \cH^{ii} v_{ii} +
\sum_{i \neq j} \cH^{ij} v_{ij}.
\label{eq:Hcon}
\eqa
In the continuum limit,
\eq{eq:Hcon} can be written as
\bqa
\cH_v & = & \int dr ~
  \theta_v(r) 
 \cH(r) 
+ \int_{r' \neq r} dr dr' ~ 
 \lambda_v(r,r')
\cH^{(2)}(r,r').
\label{eq:HvcH}
\eqa
Here,
\bqa
 \cH(r_i) & = & V_i^{-1}  \cH^{ii}, \label{eq:Hdensity} \\
  \theta_v(r_i) & = &   v_{ii}  \label{eq:theta}
\eqa
are identified as the Hamiltonian density and the lapse function of the general relativity.
The off-diagonal Hamiltonian
and the off-diagonal lapse function are given by
\bqa
\cH^{(2)}(r_i,r_j) & = & V_i^{-1} V_j^{-1}  H^{ij}, \label{eq:offHdensity}  \\
\lambda_v(r_i,r_j) & = &   v_{ij}.
\label{eq:lambda}
\eqa
In the second term
of \eq{eq:HvcH}, 
the integration over $r'$ excludes the region near $r' = r$ with coordinate volume $V_{r}$.
If one chooses the lapse that is diagonal in the frame, $\lambda_v(r_i,r_j)=0$.
The off-diagonal contribution
encodes the information about the mismatch between
the frame chosen by the local structure
and the frame in which the lapse tensor is diagonal.

While $\theta_v$ and $\lambda_v$ mix with each other 
under general \SLL transformations,
they don't mix to the leading order in the derivative expansion
under smooth spatial diffeomorphism.
\eq{eq:GGH2} implies 
$\{ \cG_x, \cH_v \}_{PB} = \cH_{vx + x^T v}$.
This determines how the lapse function transforms 
under the spatial diffeomorphism in the continuum,
\bqa
&&
 \left\{
\int dr'' \Bigl( \cD(r'') \zeta_x(r'') +  \cP_\mu(r'')  \xi_x^\mu(r'') + .. \Bigr), 
\int dr
  \theta_v(r)  \cH(r)
  + \int_{r \neq r'} dr dr' ~ 
\cH^{(2)}(r,r') \lambda_v(r,r')
\right\}_{PB}
\nn
&&=
\int dr ~
\theta_{v x + x^T v}(r)   \cH(r) 
+ \int_{r \neq r'} dr dr' ~ 
\cH^{(2)}(r,r') \lambda_{vx + x^T v}(r,r'),
 \label{eq:DHH}
\eqa
where the lapse function and the off-diagonal lapse function associated with $vx + x^T v$ are given by
\bqa
\theta_{vx + x^T v}(r) 
&=& 2 
\zeta_x(r) 
\theta_v(r) 
+  \mathcal{L}_{\xi_x}  \theta_v(r)
 + O(\partial^2), \nn
 \lambda_{vx + x^T v}(r,r') 
&=& 
\left[ 
\zeta_x(r) 
+ \zeta_x(r') \right]
\lambda_v(r,r') 
+  \mathcal{L}_{\xi_x}  \lambda_v(r,r')
+  \mathcal{L}^{'}_{\xi_x}  \lambda_v(r,r')
 + O(\partial^2),
\label{eq:zetavx}
\eqa
where $\mathcal{L}_{\xi_x} $
and $\mathcal{L}_{\xi_x}^{'}$ represent the Lie derivative 
acting on $r$ and $r'$, respectively. 
This is shown in Appendix \ref{app:fields3}.
The Hamiltonian density 
in \eq{eq:Hdensity}
carries charge $2$ under the
Weyl transformation generated by $\cD$,
and transforms as a scalar density of weight $1$ 
under the spatial diffeomorphism as expected.
The off-diagonal Hamiltonian density
in \eq{eq:offHdensity}
 carries the same charge 
under the Weyl transformation, 
but transforms as a bi-local scalar density with weight $2$
under the spatial diffeomorphism.
The Hamiltonian density does not mix with the off-diagonal 
Hamiltonian density under the spatial diffeomorphism 
to the leading order in the derivative expansion.
If $\lambda_v$ and $\zeta_x$ are turned off,
 \eq{eq:GR2} is reproduced from \eq{eq:zetavx}.

\subsection{ Emergent metric }

In the general relativity,
the Poisson bracket between two Hamiltonians
is proportional to the momentum constraint
with the structure factor given by the spatial metric 
and the signature of the metric as is shown in \eq{xitt}.
This implies that one can extract the spatial metric
and the signature from the Poisson bracket of Hamiltonians.
In the present theory, \eq{eq:GGH3} implies
that the Poisson bracket between two Hamiltonians with different lapse tensors 
is proportional to $\cD$ and $\cP_\mu$ to the leading order
in the derivative expansion
in the present theory.
This allows us to identify the spatial metric unambiguously 
in terms of the collective variables.
Combining Eqs.   (\ref{eq:GGH3}), (\ref{eq:HvcH}) and (\ref{eq:Gyc}),
we obtain 
\bqa
&&
 \left\{
\int dr \Bigl(
  \theta_u(r)  \cH(r) + ..  \Bigr), 
\int dr' \Bigl( 
  \theta_v(r')  \cH(r') + .. 
\Bigr) 
\right\}_{PB}
\nn
&=&
\int dr 
\Bigl(
F^\nu(r)   \cD(r)  + G^{\mu \nu}(r)   \cP_\mu(r) 
 + ..\Bigr)
 \Bigl(
 \theta_u(r) \nabla_\nu \theta_v(r)
 -
  \theta_v(r) \nabla_\nu \theta_u(r) \Bigr)
  + O( \partial^2 ),
 \label{Gmunu}
\eqa
where
\bqa
F^\nu(r_m) & = & \frac{1}{2} \sum_{i,k,n} \cC^{ i i k k n }_{m} \left( r^\nu_k - r^\nu_i \right), 
\label{Fnu} \\
G^{\mu \nu}(r_m) & = & \frac{1}{2} \sum_{i,k,n} \cC^{ i i k k n }_{m}  \left( r^\mu_n - r^\mu_m \right) \left( r^\nu_k - r^\nu_i \right).
\label{Gmunu2}
\eqa
The derivation can be found in Appendix \ref{app:fields4}.
The difference between two evolutions generated by Hamiltonians with different lapse tensors
is given by an Weyl transformation and a spatial diffeomorphism.
In order to identify the metric tensor,
we decompose $G^{\mu \nu}$ into the symmetric and anti-symmetric parts as
\bqa
-{\cal S} g^{\mu \nu} & = & \frac{G^{\mu \nu} + G^{\nu \mu}}{2}, \label{eq:metric} \\
b^{\mu \nu} & = & \frac{G^{\mu \nu} - G^{\nu \mu}}{2}. \label{eq:antisym}
\eqa
The symmetric tensor is identified as 
$-{\cal S} g^{\mu \nu}$ in \eq{xitt}.
Here $g^{\mu \nu}$ is the spatial metric 
whose overall sign is chosen such that 
the signature of the first spatial component 
is  positive.
${\cal S}$ is the signature of time relative that of the first spatial component.
$b^{\mu \nu}$ is the anti-symmetric component of $G^{\mu \nu}$.

In the generalized constraint algebra summarized in Eqs. (\ref{eq:DHDH}), (\ref{eq:DHH}) and (\ref{Gmunu}),
$\zeta_x$, $F^\nu$, $ b^{\mu \nu}$ appear as extra dynamical fields
besides the metric.
The presence of such extra modes is expected 
because the bi-local collective variables can be viewed as
an infinite tower of local fields with arbitrarily large spins
once expanded on a manifold.
The full theory in \eq{eq:pathint}  also includes 
non-perturbative modes associated with fluctuations of topology and dimension.
In states in which the extra fields are turned off,
the generalized constraint algebra in Eqs. (\ref{eq:DHDH}), (\ref{eq:DHH}) and (\ref{Gmunu})
reduces to the hypersurface deformation algebra of the general relativity 
in Eqs. (\ref{eq:GR1}), (\ref{eq:GR2}) and (\ref{eq:GR3})
up to the two derivative order in the gradient expansion.


\section{Classical equation of motion}
\label{sec:classical}

\subsection{Symmetry of semi-classical states}

We view the path integration in \eq{eq:pathint}
as the evolution of the initial state $\ccr$ in \eq{ccr}
under the change of parameter time $\tau$.
Since $N$ plays the role of the inverse of the Planck constant in \eq{Sfinal}, 
the path integration in \eq{eq:pathint} can be approximated 
by the saddle-point solution in the large $N$ limit, 
provided that the initial state is chosen
to be a semi-classical state.
The semi-classical wavefunction for the collective variable
can be written as
\bqa
&& \chi_{\qstar, \sstar, \pcstar, \tcstar}(s,t_1,t_2)  = \nn
&&  
\exp\left(
- i N 
\tr{
 \qstar s 
+ \sum_c  \pcstar t_c
}
-\frac{
\sum_{i,\alpha} \left[   \left( s \right)^i_{~\alpha}  -  \sstar^{i}_{~\alpha} \right]^2
+ \sum_c \sum_{ij} \left[ t_c^{ij}-  \tcstar^{ ij} \right]^2
}{
\Delta^2
}
\right). \nn
\label{eq:wp}
\eqa
In the $\frac{1}{N} \ll \Delta \ll 1$ limit,
the wave packet has well-defined collective variables and conjugate momenta peaked at 
$s=  \sstar$, 
$t_c=  \tcstar$,
$q=   \qstar$, 
$p_c =  \pcstar$.
While the semi-classical states are labeled by
 $\{ \sstar, \tcstar, \qstar, \pcstar \}$,
there is a redundancy in labeling states
in terms of the classical variables. 
Because the basis state in \eq{eq:SLf}
is chosen to be invariant under the permutations
that exchange the first $L$ flavours,
wavefunctions obtained by applying 
the flavour permutations to $\qstar$, $\sstar$
represent the same physical state,
\bqa
\chi_{  P_f  \qstar ,  \sstar P_f^T,  \pcstar,  \tcstar }(s, t_1, t_2) 
& \sim & 
 \chi_{\qstar, \sstar, \pcstar, \tcstar}(s,t_1, t_2),
\label{chiSLf}
\eqa 
where $P_f^T$ is the transpose of $P_f$.

The symmetry of the semi-classical state is determined from the symmetry
of the classical variables.
If $\qstar$ contains $L$ linearly independent row vectors,
the vectors can be used as a frame.
Since the frame can not be invariant 
under any infinitesimal \SLL transformation,
there is no unbroken continuous \SLL symmetry.
However, a discrete subgroup of \SLL can be still preserved.
Under the even site permutation group,
which is a discrete subgroup of \SLL,
 the wavefunction is transformed as
\bqa
&&  \chi_{\qstar, \sstar, \pcstar, \tcstar}( s,  t_1 , t_2  )  
\rightarrow  \chi_{\qstar, \sstar, \pcstar, \tcstar}(P_g s, P_g t_1 P_g^T, P_g t_2 P_g^T )
\label{eq:change}
\eqa
as is shown in  \eq{eq:SLg}.
The right hand side of \eq{eq:change} can be written as
\bqa
 \chi_{\qstar, \sstar, \pcstar, \tcstar}(P_g s, P_g t_1 P_g^T, P_g t_2 P_g^T )  
 = 
\chi_{\qstar P_g, P_g^T  \sstar, P_g^T  \pcstar P_g , P_g^T  \tcstar P_g }(s,t_1, t_2)
\label{eq:change2}
\eqa
because \eq{eq:wp} satisfies
$ 
\chi_{\qstar P^T, P \sstar, P \pcstar P^T, P \tcstar P^T} (P s, P t_1 P^T, P t_2 P^T )  
 = 
\chi_{\qstar ,   \sstar,   \pcstar  ,   \tcstar  }(s,t_1, t_2)$
for any site permutation, $P$.
In \eq{eq:change2}, $P^T = P^{-1}$ is used.
Because of the redundancy in labeling physical states 
in terms of the collective variables,
 the state is invariant 
under  \eq{eq:change} 
if there exists a flavour permutation, $P_f$ that satisfies
\bqa
 \qstar P_g  =  P_f  \qstar, &&
P_g^T  \sstar  =   \sstar P_f^T, \nn
P_g^T  \pcstar P_g   =   \pcstar, &&
P_g^T  \tcstar P_g  =   \tcstar.
\label{unbroken}
\eqa 
If there exists such $P_f$,
the site permutation 
can be canceled by a flavour permutation
through \eq{chiSLf}.
In this case, \eq{eq:change} is a symmetry of  the state,
\bqa
 \chi_{\qstar, \sstar, \pcstar, \tcstar}(P_g s, P_g t_1 P_g^T, P_g t_2 P_g^T )  
& \sim & 
 \chi_{\qstar, \sstar, \pcstar, \tcstar}(s,t_1, t_2).
\eqa
Therefore, the unbroken subgroup of \SLL  is given by the set of $P_g$
that satisfies \eq{unbroken} for some $P_f$.
This  subgroup is denoted as
${\cal I}_{\qstar, \sstar, \pcstar, \tcstar}$.

Let us examine the symmetry of the classical state
in which $\qstar$ is non-degenerate. 
In this case, it is convenient to choose the gauge in which
$\qstar$ is proportional to the identity matrix.
With a generic choice of the remaining variables, $ \sstar, \pcstar, \tcstar$,
\SLL is completely broken,
and
${\cal I}_{\qstar, \sstar, \pcstar, \tcstar} = \emptyset$.
The maximal subset of \SLL can be preserved 
if $\sstar$, $\pcstar$, $\tcstar $
are all proportional to the identity matrix.
This corresponds to a direct product state because all collective variables are diagonal.
In this case, the full site permutation group remains unbroken : 
${\cal I}_{\qstar, \sstar, \pcstar, \tcstar} = S_L^g$.
This is because 
$P_f = P_g$
satisfies  \eq{unbroken}
for any $P_g \in S_L^g$.
In order to have a non-trivial classical manifold,
one needs to turn on off-diagonal elements of the classical variables
that generate inter-site entanglement.  
%
If $(\sstar \qstar)$, $\pcstar$ and $\tcstar$ have non-zero off-diagonal elements
but $( \sstar  \qstar )^i_{~~j}, (\pcstar)_{ij}, (\tcstar)^{ij}$ depend only on $r_i - r_j$
in a manifold associated with a local structure,
the state has an unbroken discrete translational symmetry.
%
This discrete space symmetry can be enhanced to 
the continuous translation in the long distance limit.
In Sec. \ref{sec:translationally},
it will be shown that 
a discrete translational and rotational symmetry 
can be enhanced to a continuous group at long distances. 
We will also discuss an example
in which the full Lorentz symmetry 
emerges as an isometry of the spacetime.


\subsection{Saddle-point equation of motion}

\begin{figure}[ht]
\begin{center}
\centering
\includegraphics[scale=0.4]{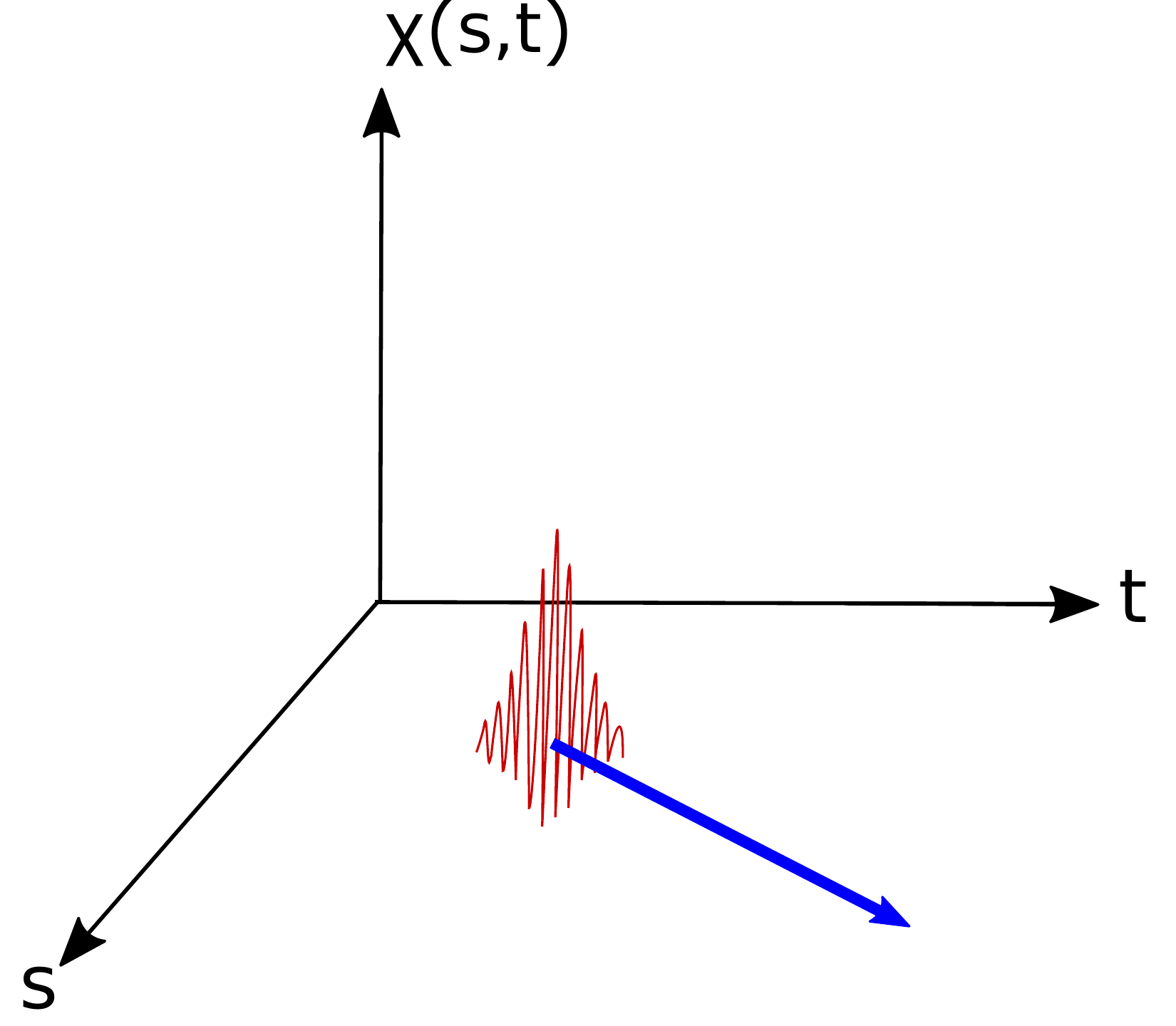}
\end{center}
\caption{
A wave packet that has a well-defined collective variables
and the conjugate momenta.
Under the time evolution, the classical variables evolve obeying the saddle-point equation.
}
\label{fig:chi_sym}
\end{figure}

In the large $N$ limit,
the path integration
is dominated by 
the semi-classical path
that satisfies the saddle-point equation of motion,
\bqa
\partial_\tau \btc & = & 4 \btc v \btc - 4 \ta  \left( \btc \bar Q\bar U v \btc + \btc v \bar U \bar Q\btc  \right) 
- \ta \bar U v \bar U - y \btc - \btc y^T, \nn
\partial_\tau \bp_c & = &  -\Bigl[
4 \bpc \btc v + 4 v \btc \bpc -  i v 
- 4 \ta \left( \bpc \btc \bar Q\bar U v + v \bU \bQ \btc \bpc \right) \nn
&& ~~~ - 4 \ta \left( \bQ \bU v \btc \bpc + \bpc \btc v \bU \bQ \right)
+  i \ta \left( \bQ \bU v + v \bU \bQ \right) \Bigr] 
+ \bpc y + y^T \bpc, \nn
\partial_\tau \bs & = & 
-2 \ta \bU v \bU \bq^T - y \bs, \nn
\partial_\tau \bq & = & 
-  2 \bs^T v + 2 \ta \left(
\bs^T \bQ \bU v + \bs^T v \bU \bQ
\right) 
+ \bq y
\label{eq:EOM}
\eqa
with the initial condition,
$\btc(0)= \tcstar$,
$\bpc(0)=\pcstar$,
$\bs(0)=\sstar$ and
$\bq(0)=\qstar$.
$y(\tau)$ and $v(\tau)$ are the time dependent shift
and lapse tensors, respectively. 
They are free parameters.
$y(\tau)$ sets the generalized spatial diffeomorphism 
within the constant time slice at $\tau$.
The lapse tensor at each $\tau$ 
determines the rate of time evolution
and the frame in which 
time evolution is generated.

In the large $N$ limit,
\eq{eq:wp} can have a non-zero overlap with a gauge invariant state in \eq{eq:overlap}
only if the collective variables satisfy the constraints classically.
This implies that one can not choose the classical values of the collective variables
and their conjugate momenta arbitrarily.
The classical collective variables should satisfy
\bqa
\tr{ 
\left( \bs \bq  +  2   \sum_c \btc \bpc -  \frac{i}{2}  I  \right) y } & = & 0, \label{eq:constraintbar1}  \\
\tr{ \left( -\bU + \ta \bU \bQ \bU \right)  v} & = & 0 \label{eq:constraintbar2}
\eqa 
for any traceless tensor $y$ and symmetric tensor $v$ at all $\tau$.
In the phase space of the collective variables,
the classical constraint hypersurface is defined by 
\bqa
\bs \bq + 2 \sum_c  \btc \bp_c - \frac{i}{2} I & = & \beta I, \label{SLLconstraint} \\
\ta \bQ \bU  & = &  I, \label{Hconstraint}
\eqa
where $\beta$ is an arbitrary constant. 
Thanks to the first-class nature of the constraints,
 Eqs. (\ref{eq:constraintbar1}) and (\ref{eq:constraintbar2})  
are conserved under the time evolution within the constraint hypersurface.
Furthermore, $\beta$ is a constant of motion on the constraint hypersurface :
the equations of motion in \eq{eq:EOM} leads to 
\bqa
&&  \frac{ \partial}{\partial \tau} \left( \bs \bq + 2 \sum_c \btc \bp_c - \frac{i}{2} I \right) = \nn
 &&
2 (- \bU + \ta \bU \bQ \bU ) v - y \left( \bs \bq + 2 \sum_c \btc \bp_c \right) 
+  \left( \bs \bq + 2  \sum_c \btc \bp_c \right) y.
\eqa
This vanishes on the constraint hypersurface for any value of $\beta$.

A few remarks are in order regarding the constraints that classical variables satisfy.
First, the fact that the initial classical data should satisfy the constraints
is a common feature of constraint systems
including the general relativity.
In the present theory, this is imposed through the condition
that the state in the kinematic Hilbert space has a nonzero projection
with a gauge invariant state in the large $N$ limit.
Second, this is closely related to the well-known 
source-operator relation in the AdS/CFT correspondence\cite{Witten:1998qj,Gubser:1998bc,Witten:2001ua}.
In the AdS/CFT correspondence,
sources fixed at a UV boundary 
(say through the Dirichlet boundary condition)
fix their conjugate variables 
to avoid singularity in the bulk.
In the present case, the collective variables (sources)
and their conjugate momenta (operators) 
should satisfy the constraints 
to make sure that 
the overlap between the late time state given by
\bqa
\cb \chi(\tau) \rb  
=
{\cal T}
e^{-i \int_0^\tau d \tau' 
\left[
\hH_{v^{(\tau')}} +  \hG_{y(\tau')}
\right]}
 \cb  \chi \rb,
\label{chiinf}
\eqa
and the gauge invariant state
is non-zero in the large $N$ limit,
where ${\cal T}$ time-orders the operators.
If the collective variables do not
satisfy the constraints classically,
the state in \eq{chiinf}  will have a 
zero overlap with 
$\lb 0 \cb$
in \eq{eq:pathint}
in the large $N$ limit.
This is because the initial state and the final state 
have different `energies' and 'momenta',
and the two states are orthogonal to each other
in the large $N$ limit.
In this case, there is no classical path
that connects the initial state to the final state.
Alternatively, the vanishing overlap between the state at large $\tau$
and the gauge invariant state can be translated into  
a divergent boundary action present at future infinity in the path integration.
A smooth boundary condition
that avoids the divergent action (a catastrophic collapse of the wavefunction) at future infinity
can be imposed only
if the initial state satisfies the constraints classically.
%
Third, the fact that the classical variables satisfy the constraints guarantees
that $\cb \chi \rb$  is invariant under the gauge transformation
to the leading order in the large $N$ limit.
The gauge symmetry is broken at the sub-leading order.
This sub-leading gauge symmetry breaking is 
what makes sure that the state is normalizable
and evolves non-trivially under time evolution\cite{Lee2018}.

\subsection{Gauge fixing}

The shift and lapse tensors are arbitrary,
and we can choose them through a gauge fixing.
It is convenient to use some physical degrees of freedom as a set of local clocks 
that fixes a frame in which 
a gauge invariant local structure is defined and time evolution is generated.
To be concrete, let us consider an initial state with  $\det \bq (0) > 0$.
Without loss of generality, we can choose our initial frame such that $\bq (0)$ is proportional to the identity matrix.
Furthermore, we can make sure that $\bq (\tau)$ remains proportional to the identity matrix at all $\tau$
by choosing the shift as 
\bqa
y &=  & \frac{1}{ \bq }  \left( \bW - < \bW > I \right),
\label{eq:gauge1}
\eqa
where
\bqa
\bW = 
2 \bs^T v -  2 \ta ( \bs^T \bQ \bU v + \bs^T v  \bU \bQ ),
\label{eq:W}
\eqa
and 
$<\bW > =  \frac{ \tr{ \bW} }{L}$.
In this gauge, 
$\bq (\tau)$ remains proportional to the identity matrix at all $\tau$,
and  $\bq (\tau)$ can be written as
\bqa
\bq (\tau) = \bq_d(\tau) I,
\label{eq:defq1}
\eqa
where $q_d$ is a single variable that represents the diagonal elements.
This amounts to choosing a frame at each moment of time 
in terms of the $L$ independent vectors of $\bq$.
This frame defines a gauge invariant set of local Hilbert spaces
and a local structure associated with them.
The  gauge freedom associated with the lapse tensor
can be fixed so that the lapse tensor is diagonal in the frame
in which $q $ is diagonal.
This describes the time evolution of the system relative to 
a set of local clocks defined in the frame chosen by $q$.
Here, we choose a uniform lapse tensor as
\bqa
v & = & I.
\label{eq:gauge2}
\eqa
We call  \eq{eq:gauge1} and \eq{eq:gauge2} the unitary gauge condition.
In the unitary gauge, the equations of motion become
\bqa
\partial_\tau \btc & = & 4 \btc^2 - 4 \ta  \left( \btc \bQ \bU  \btc + \btc  \bU \bQ \btc  \right) - \ta \bU^2  
- \frac{1}{\bq_d}  \left( \bW \btc + \btc \bW^T  - 2 < \bW > \btc \right), \label{eqt} \\
\partial_\tau \bpc & = &  -\Bigl[
4 \bpc \btc  + 4 \btc \bpc - i  
- 4 \ta \left( \bpc \btc \bQ \bU +  \bU \bQ \btc \bpc \right) 
- 4 \ta \left( \bQ \bU  \btc \bpc + \bpc \btc  \bU \bQ \right)
+  i \ta \left( \bQ \bU  +  \bU \bQ  \right) \Bigr]  \nn
&& +  \frac{1}{\bq_d} \left( \bpc \bW  +  \bW^T \bpc  - 2 < \bW > \bpc \right), \label{eqp} \\ 
\partial_\tau \bs & = & 
-2 \ta \bq_d \bU^2   -  \frac{1}{\bq_d} \left( \bW -   < \bW >  \right) \bs, \label{eqs} \\
\partial_\tau \bq_d &=& - < \bW >. \label{eqq1} 
\eqa
The equation of motion is simplified in terms of a new variable,
\bqa
\tilde t_c = \btc - \frac{i}{8} \bpc^{-1}.
\label{eq:tildet}
\eqa
On the constraint hypersurface,
$W = -2 s^T$ in \eq{eq:W},
and the equation of motion for $\td t_c$ becomes 
\bqa
\partial_\tau \tilde  t_c & = & -4 \tilde t_c^2  + \frac{1}{16} \bpc^{-2} - \ta \bU^2
+ \frac{2}{ \bq_d }   \left(  \bs^T \td t_c + \td t_c  \bs - 2 <\bs> \td t_c \right), \label{eqt2} \\
\partial_\tau \bpc & = &   4 \bpc \td t_c  + 4 \td t_c \bpc  
-  \frac{2}{\bq_d} \left( \bpc \bs^T + \bs \bpc   - 2 < \bs > \bpc \right)
, \label{eqp2} \\ 
\partial_\tau \bs & = & 
-2 \ta \bq_d \bU^2   +  \frac{2}{\bq_d} \left( \bs^T -   <\bs > \right) \bs, \label{eqs2} \\
\partial_\tau \bq_d &=& 2 < \bs >, \label{eqq12} 
\eqa
where
$<\bs > \equiv  \frac{ \tr {\bs}}{L}$,
 $\bU = \bs \bs^T + \sum_c \left[ 4 \td t_c \bpc \td t_c +  \frac{1}{16} \bpc^{-1} \right]$.

\section{Translationally invariant solution}
\label{sec:translationally}

In this section, we solve Eqs.
(\ref{eqt2})-(\ref{eqq12})
for an initial state that exhibits 
a $D$-dimensional local structure
with a translational invariance.
We choose a state with
\bqa
\td t_1, \td t_2 & = &  \td t, \nn
\bar p_1, \bar p_2 &=& \frac{\bar p}{2},
\label{eq:saddletp}
\eqa
where $\td t$ and $\bar p$
describe a $D$-dimensional manifold with topology, $T^D$.
For those states, it is natural to label sites in terms of $D$ integers
that form the $D$-dimensional hyper-cubic lattice,
\bqa
 \br = (n_1, n_2, .., n_D)
 \label{eq:coord}
 \eqa
 for $1 \leq n_i \leq L^{1/D}$
with the identification $n_i \sim n_i + L^{1/D}$.
If the state has the $D$-dimensional translational invariance,
collective variables can be represented in the momentum space,
\bqa
 \td t_{\bk} & = & \sum_{\br} e^{ -i \bk \br } ~ \td t^{\br'+\br, \br'}, \nn
 \bp_{\bk} & = & \sum_{\br} e^{ -i \bk \br } ~ \bp_{\br'+\br, \br'}, \nn
\bs_{\bk} & = & \sum_{\br} e^{ -i \bk \br } ~ \bs^{\br'+\br}_{~\br'},
\eqa
where $\bk = \frac{2 \pi}{L^{1/D}} \left( l_1, l_2, .., l_D   \right)$
with $ -\frac{L^{1/D}}{2} \leq  l_i < \frac{L^{1/D}}{2}$.
For simplicity, we consider the case with the reflection symmetry
and the discrete rotational symmetry.
This guarantees 
$\td t_\bk = \td t_{\bk'}$,
$\bp_\bk = \bp_{\bk'}$,
$\bs_\bk = \bs_{\bk'}$,
and
$\td t_\bk = \td t_{\bf R \bk}$,
$\bp_\bk =  \bp_{\bf R \bk}$,
$ \bs_\bk =  \bs_{\bf R \bk}$.
Here $\bk' = (k_1, .., k_{l-1}, -k_l, k_{l+1},..,k_D)$ for some $l$.
$\bf R$ is a $\frac{\pi}{2}$-rotation  
on any of the principal planes 
in the $D$-dimensional hypercubic lattice.

The hyper-cubic lattice 
breaks the continuous rotational symmetry
to the discrete group.
This is a spontaneously broken symmetry
that is determined from the pattern of entanglement in the state.
Nonetheless, we expect that the continuous rotational symmetry
emerges at long distance scales.
To see this, one can expand the collective variables in powers of $\bk$,
\bqa
\td t_\bk = \sum_n \td t^{\mu_1 \mu_2 .. \mu_n} k_{\mu_1} k_{\mu_2}..k_{\mu_n},
\eqa
where
$t^{\mu_1 \mu_2 .. \mu_n} $ represents a field with spin $n$.
Due to the reflection symmetry, all odd spin fields vanish.
Furthermore, the discrete rotational symmetry guarantees that
$t^{\mu \nu} \propto \delta^{\mu \nu}$,
and any spin $2$ field should respect the full rotational symmetry.
This implies that the spatial metric should be invariant under the 
full rotational symmetry as well.
Higher spin fields do break the continuous rotational symmetry,
but they become less important in the small $\bk$ limit.

The collective variables at each $\bk$ satisfy 
\bqa
\partial_\tau \tilde  t_\bk & = & -4 \tilde t_\bk^2  + \frac{1}{4} \bp_\bk^{-2} - \ta \bU_\bk^2
+ \frac{2}{ \bq_d }   \left(  \bs_{-\bk} \td t_\bk + \td t_\bk \bs_\bk - 2 < \bs > \td t_\bk \right), \label{eqt3} \\
\partial_\tau \bp_\bk & = &   8 \bp_\bk \td t_\bk   
-  \frac{2}{\bq_d} \left( \bp_\bk \bs_{-\bk} + \bs_\bk \bp_\bk   - 2 < \bs > \bp_\bk \right)
, \label{eqp3} \\ 
\partial_\tau \bs_\bk & = & 
-2 \ta \bq_d \bU_\bk^2   +  \frac{2}{\bq_d} \left( \bs_{-\bk} - < \bs > \right) \bs_\bk, \label{eqs3} \\
\partial_\tau \bq_d &=& 2 < \bs >, \label{eqq13} 
\eqa
where
\bqa
\bU_\bk = \bs_\bk \bs_{-\bk} + 4 \td t_\bk^2 \bp_\bk  +  \frac{1}{4} \bp_\bk^{-1},
\label{eq:Uk}
\eqa
$< \bs > = \frac{1}{L} \sum_\bk \bs_\bk$,
and we use $( \bp^{-1} )_\bk = ( \bp_\bk )^{-1}$.

The constraints in Eqs. (\ref{SLLconstraint}) and (\ref{Hconstraint}) imply
\bqa
&& \bq_d \bs_\bk  + 2 \td t_\bk \bp_\bk   =  \beta, \label{SLLconstraint2} \\
&& \ta \left( \bq_d^2 + \bp_\bk \right)  \left( \bs_\bk \bs_{-\bk} + 4 \td t_\bk^2 \bp_\bk  +  \frac{1}{4} \bp_\bk^{-1} \right)  
 =   1 \label{Hconstraint2}
\eqa
for all $\bk$.
We can use the constraints  to solve $s_\bk$ and $\td t_\bk$ in terms of $\bp_\bk$ and $\bq_d$ as
\bqa
s_{\bk} & = & \frac{1}{\bq_d} \left( \beta - 2 \td t_\bk \bp_\bk  \right), \label{eq:sk} \\
\td t_{\bk} & = & T_{\pm}[ \bp_\bk, \bq_d ],
 \label{eq:tk} 
\eqa
where 
\bqa
T_{\pm}[ \bp_\bk, \bq_d ] & = &
\frac{
2 \beta \bp_\bk   \pm \bq_d \sqrt{ \gamma^2 \bp_\bk -   \bq_d^2}
 }
{4  \bp_\bk \left( \bp_\bk + \bq_d^2 \right)}
\label{eq:Tpm}
\eqa
with
$\gamma \equiv \sqrt{ \frac{1}{\ta} \left( 4 - \ta - 4 \ta \beta^2 \right) }$.
Here we consider $\ta > 0$ and $\beta>0$ 
with $ \frac{1}{\ta} \left( 4 - \ta - 4 \ta \beta^2 \right) > 0$.
Here
 $T_+[ \bp_\bk, \bq_d ]$ and  $T_-[ \bp_\bk, \bq_d ]$
represent two possible branches of $\td t_\bk$ 
that satisfy the constraints for given $\bp_\bk$ and $\bq_d$.
The time evolutions of $\bp_{\bf k}$ an $\bq_d$
can be determined by  \eq{eqp} and \eq{eqq1}, 
\bqa
\partial_\tau \bq_d & = & \frac{2}{\bq_d} \left( \beta - \frac{2}{L} \sum_{\bk} \td t_{\bk} \bp_{\bk} \right), \label{eq:taUQ 1} \\
\partial_\tau \bp_{\bk} & = & 8 \td t_{\bk} \bp_{\bk} + \frac{8}{\bq_d^2} \left( \td t_{\bk} \bp_{\bk} - \frac{1}{L} \sum_{\bk'} \td t_{\bk'} \bp_{\bk'}  \right) \bp_{\bk}.  \label{eq:taupk}
\eqa

The metric in \eq{eq:metric},
which is independent of $\br$,
becomes (see Appendix \ref{app:Sgmn})
\bqa
-{\cal S} g^{\mu \nu} & = &  4 \ta \left( 
\frac{ \partial \bU_{\bk}}{\partial k_\mu}  \frac{ \partial \bU_{\bk}}{\partial k_\nu} 
+ \bU_{\bk}  \frac{ \partial^2 \bU_{\bk}}{\partial k_\mu \partial k_\nu}
\right)_{\bk=0}.
\label{eq:Sgmn}
\eqa
By using \eq{Hconstraint}, \eq{eq:Sgmn} can be written as
\bqa
-{\cal S} g^{\mu \nu} & = &
-\left. \frac{4}{\ta} \frac{1}{(\bp_\bk + \bq_d^2)^3} \frac{ \partial^2 \bp_\bk}{\partial k_\mu \partial k_\nu} \right|_{\bk=0}.
\label{eq:Sgmn3}
\eqa
Here we use \eq{eq:Uk}
and $ \left.  \frac{ \partial \bp_\bk}{\partial k_\mu } \right|_{\bk=0} = 0$ 
in the presence of the inversion symmetry.
This fixes the signature of spacetime upto the overall sign.
Due to the emergent rotational symmetry, 
one can write 
\bqa
g^{\mu \nu}(\tau) = a(\tau)^{-1} \delta^{\mu \nu},
\eqa
where $a(\tau)$ is the scale factor.
With the choice of the positive signature for the spatial metric $(a>0)$,
the signature of  time is given by
\bqa
{\cal S} = \mbox{sgn}
\left( 
\left. \frac{1}{\ta (\bp_0 + \bq_d^2)^3}  \frac{ \partial^2 }{\partial k_1^2 }  \bp_\bk  \right|_{\bk=0}
\right),
\label{eq:Strans}
\eqa 
and the inverse of the scale factor becomes
\bqa
 a^{-1} & = & 
\left| 
\left.
 \frac{4}{\ta } \frac{1}{(\bp_0 + \bq_d^2)^3}  \frac{ \partial^2 }{\partial k_1^2 }  \bp_\bk \right|_{\bk=0}
 \right|.
\label{eq:gmunutrans}
\eqa
Due to the rotational symmetry,
one can use any spatial direction to define ${\cal S}$ and $a$
in \eq{eq:Strans} and \eq{eq:gmunutrans}.

According to \eq{eq:gmunutrans},
the proper size of the emergent space becomes smaller
for $\bp_\bk$ that varies more sharply in the momentum space.
This can be understood intuitively.
$\bp_\bk$ that changes sharply in the momentum space 
leads to $\bp_{\br, \br'}$  that decays slowly in $\br-\br'$.
The slowly decaying collective variables in the real space
creates entanglement bonds that connect sites that are far from each other in coordinate.
The long-ranged entanglement bonds bring those sites close in physical distance 
because they become strongly coupled  under the relatively local Hamiltonian.
This results in the decrease of the scale factor of  space.

For $\ta > 0$ and $\bp_0 + \bq_d^2 > 0$, 
the signature of time  
is given by the sign of $ \frac{ \partial^2 }{\partial k_1^2 }  \bp_\bk $ at $\bk=0$.
The signature changes sign  
when the collective variable undergoes 
a Lifshitz transition at which   
the second derivative of 
$\bp_\bk$ changes sign.
This can happen if $\bp_\bk$ becomes flat as a function of $\bk$ 
either locally near $\bk=0$ or globally at all $\bk$.
If $\bp_\bk$ becomes flat at all $\bk$, we call the transition a global Lifshitz transition.
If only the second derivative of $\bp_\bk$ vanishes without a global Lifshitz transition,
we call it a local Lifshitz transition.
At both local and global Lifshitz critical points,
$g^{\mu \nu}$ vanishes, 
and the scale factor of the space diverges.
This can be understood intuitively.
When the second derivative of the collective variable vanishes,
the dispersion of the collective variables becomes flat near $\bk =0$ 
in the momentum space.
As the band becomes  flatter in the momentum space,
$p_{\br, \br'}$ and $t^{\br, \br'}$ decay faster
as a function of $\br-\br'$ in real space.
This, in turn, reduces inter-site entanglement,
which results in an increase in the proper distance in space.
The proper distance between two given points in the manifold increases 
as the collective variables lose dispersion in $\bk$.
If the band becomes completely flat,
the collective variables become ultra-local,
and the state becomes unentangled to the leading order in $1/N$.
This is a fragmented space\cite{Lunts2015,Lee2016}.
According to \eq{eq:gmunutrans}, the proper volume of space diverges 
as far as its second derivative vanishes at $\bk=0$
with or without a global flattening.
At local Lifshitz critical points, the scale factor of the universe diverges
although there still exists non-zero inter-site entanglement 
mediated by  higher order $\bk$ dependence of $\bp_\bk$.
In this case, the non-trivial spatial entanglement is not encoded in the metric but in higher spin fields.
This shows that the metric carries only a partial information on the pattern of entanglement.
The full structure of entanglement is encoded in the complete set of collective variables.
It is also these higher spin fields that carry the information
that the continuous rotational symmetry is spontaneously broken
to a discrete symmetry by the hypercubic lattice.
Colloquially speaking, the pattern of entanglement fixes the emergent geometry, 
but not the other way around.

\subsection{Numerical solution}

Collective variables with different $\bk$ remain coupled with each other through $< \bs >$
in Eqs. (\ref{eqt3})-(\ref{eqq13}).
This makes it hard to solve the equations of motion in a closed form.
In order to gain some insight, we first solve the equations of motion numerically
for a finite system with $L=10^6$ and $N=\infty$.
We choose the Hamiltonian with $\ta = 1$.
For an initial state, we consider a state  
which has a three-dimensional local structure 
with the discrete translation,
the $\frac{\pi}{2}$-rotation
and the reflection symmetry
of the cubic lattice.
We choose $\bp_{\br', \br}(0)$ that is non-zero 
only for nearest neighbour $\br'$ and $\br$. 
In the momentum space, this gives
\bqa
\bp_\bk(0) = \bp_d  - \epsilon \left( \cos(k_1) + \cos(k_2) + \cos(k_3) \right),
\label{eq:pbk0}
\eqa
where $\bp_d$ represents the ultra-local part of $\bp_{\br', \br}(0)$,
and $\epsilon$ determines the strength of the nearest neighbour entanglement bonds.
The initial conditions for $\td t_\bk$ and $\bs_\bk$ are fixed by the constraints
with $\td t_k(0) = T_+[\bp_\bk(0), \bq_d(0)]$ 
and  $\beta=0.1$
in \eq{eq:sk} and \eq{eq:tk}.
The numerical results that follow is obtained for 
 $\bp_d=1$, 
 $\bq_d(0)=1$ 
 and $\epsilon=0.1$.

 \begin{figure}[ht]
 \begin{center}
 \includegraphics[scale=0.2]{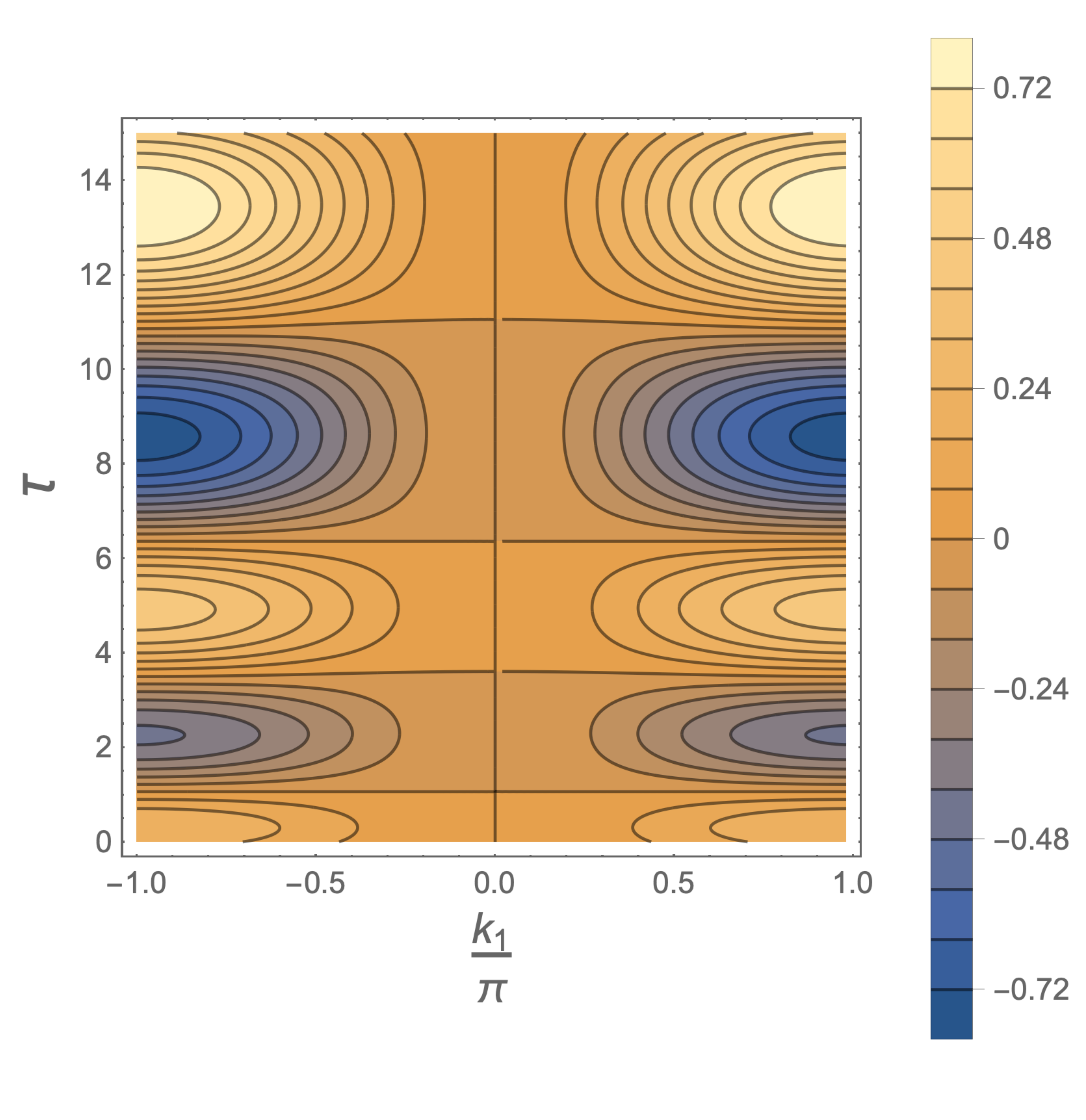} 
 \end{center}
 \caption{
 A contour plot of $\bp_{\bk}(\tau)$ in the plane of $\tau$ and  $\bk=(k_1,0,0)$.
 }
 \label{fig:Psec0}
 \end{figure}

 \begin{figure}[ht]
 \begin{center}
   \subfigure[]{
 \includegraphics[scale=0.6]{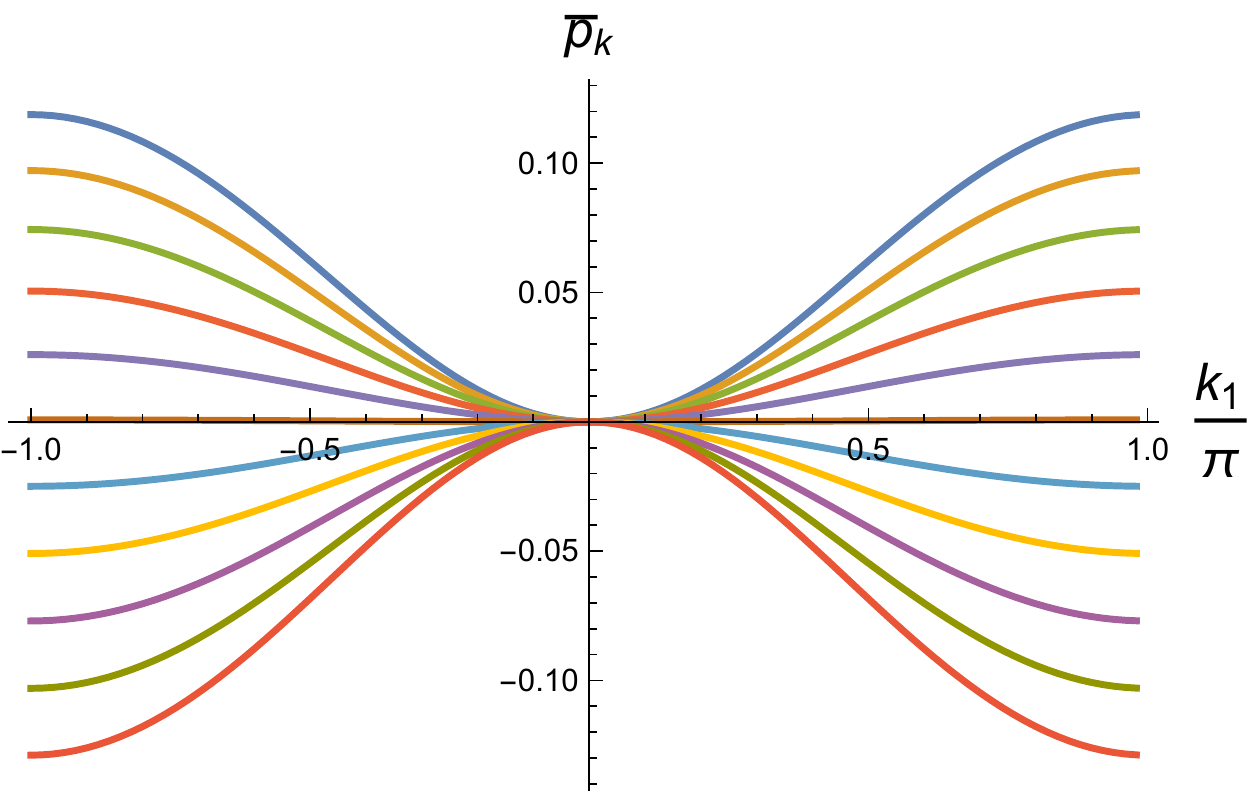} 
  \label{fig:psec1}
 } 
 \hfill
 \subfigure[]{
 \includegraphics[scale=0.6]{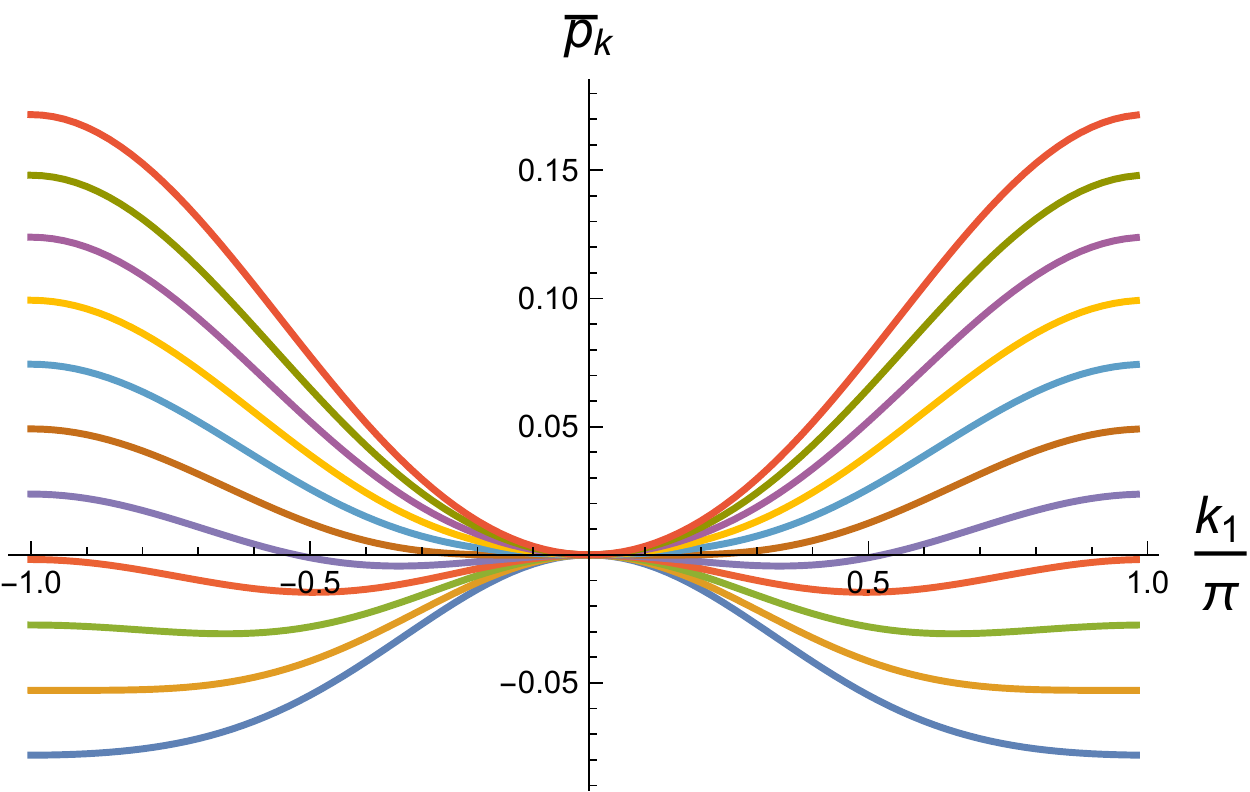} 
  \label{fig:psec2}
  } 
   \subfigure[]{
   \includegraphics[scale=0.6]{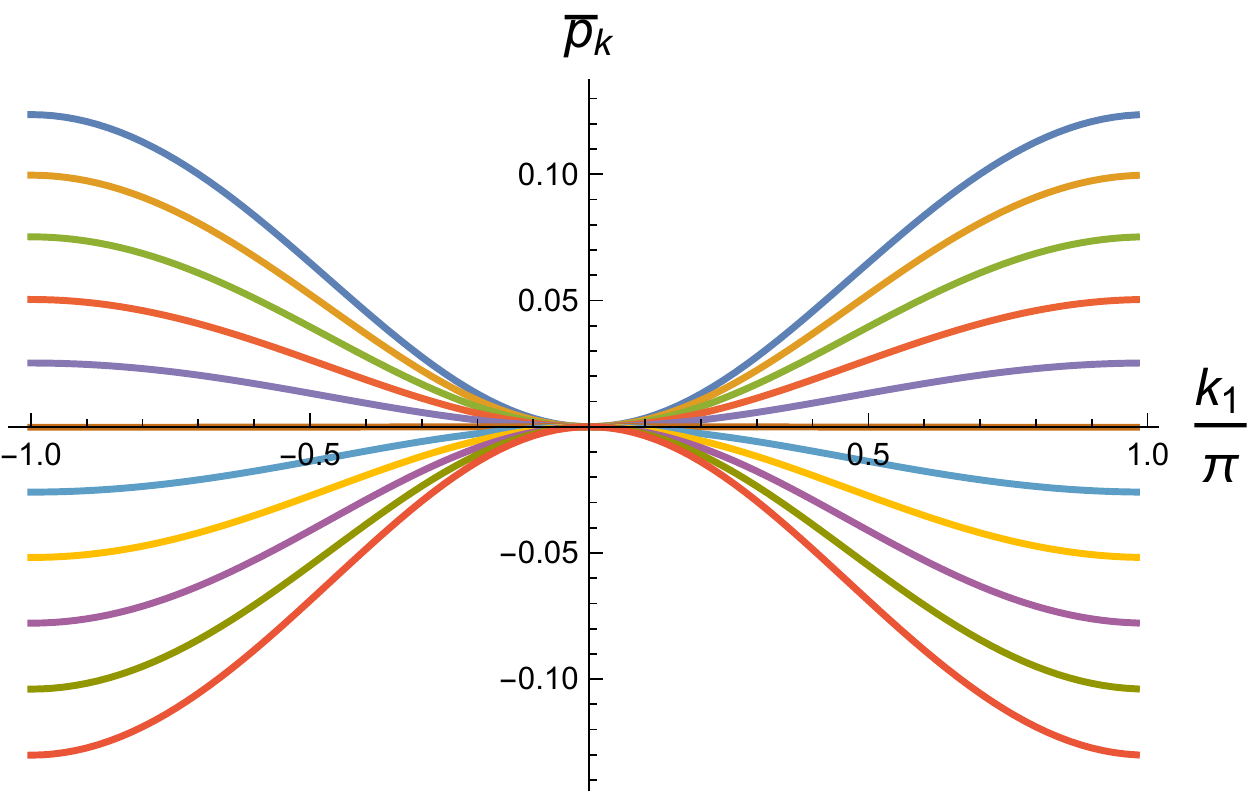} 
  \label{fig:psec3}
 } 
 \hfill
 \subfigure[]{
 \includegraphics[scale=0.6]{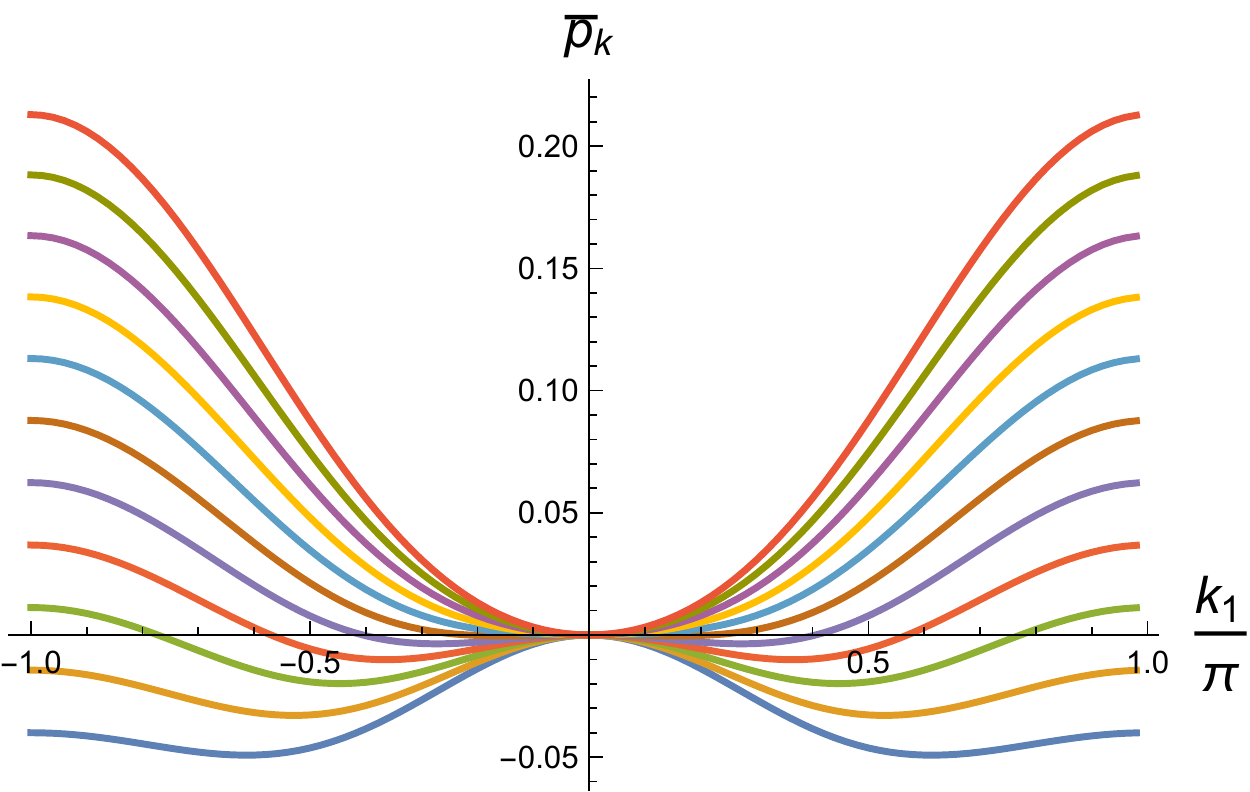} 
  \label{fig:psec4}
 } 
 \end{center}
 \caption{
 $\bp_\bk(\tau)$ plotted as a function of  $\bk=(k_1,0,0)$ at different time slices.
 (a) From the top to bottom curves, $\tau$ changes from $0.815$ to $1.315$ in the step size of $0.05$.
 At $\tau=\tau^*_1 \approx 1.065$, the collective variable that describes the pattern of entanglement undergoes a global Lifshitz transition,
 where $\bp_\bk$ becomes flat at all $\bk$.
 (b) From the bottom to top curves, $\tau$ changes from $3.35$ to $3.85$ in the step size of $0.05$.
 At $\tau= \tau^*_2 \approx 3.6$, the collective variabl  undergoes a local Lifshitz transition,
 where the second derivative of $\bp_\bk$ vanishes at $\bk=0$
 while higher derivatives remain non-zero.
 (c) From the top to bottom curves, $\tau$ changes from $6.115$ to $6.615$ in the step size of $0.05$.
 At $\tau=\tau^*_3 \approx 6.365$, the collective variable undergoes the second global Lifshitz transition.
  (d) From the bottom to top curves, $\tau$ changes from $10.785$ to $11.285$ in the step size of $0.05$.
 At $\tau=\tau^*_4 \approx 11.035$, the collective variable undergoes the second local Lifshitz transition.
 }
 \label{fig:psec}
 \end{figure}

In \fig{fig:Psec0}, we show the evolution of $\bp_\bk(\tau)$ as a function of $\tau$
along one direction of $\bk$.
At $\tau=0$, $\bp_\bk$ is convex near $\bk=0$.
As $\tau$ increases, the $\bk$ dependence becomes weaker,
and $\bp_\bk$ becomes flatter as a function of $\bk$.
At a critical time $\tau^*_1 \approx 1.065$, 
$\bp_\bk$ becomes independent of $\bk$
and completely flat at all $\bk$.
As time increases further, $\bp_\bk$ becomes concave in $\bk$.
This is a global Lifshitz transition in which
the second derivative of $\bp_\bk$ at $\bk=0$ flips the sign.
At the critical point, the band becomes globally dispersionless. 
The evolution of $\bp_\bk$ in $\tau$ near the critical time is shown in \fig{fig:psec1}.
As $\tau$ increases further,
$\bp_\bk$ undergoes a second Lifshitz transition
at $\tau^*_2 \approx 3.6$.
This time, the Lifshitz transition is local :  
 $\frac{\partial^2}{\partial k_\mu \partial k_\nu} \bp_\bk$ at $\bk=0$ changes sign from negative to positive
while the band does not become globally flat.
The profile of $\bp_\bk$ near the local Lifshitz transition is shown in  
\fig{fig:psec2}.
As time keeps increasing, 
the second set of global and local 
Lifshitz transitions occur
at $\tau^*_3 \approx 6.365$
and $\tau^*_4 \approx 11.035$,
respectively
as is shown in 
\fig{fig:psec3}
and
\fig{fig:psec4}. 
This evolution of $\bp_\bk$ near the second global Lifshitz transition is almost identical to 
 the evolution near the first transition.
This `universality' can be understood from an analytic solution
that is valid near the global Lifshitz transitions.
This will be discussed in the following section.

 \begin{figure}[ht]
 \begin{center}
   \subfigure[]{
 \includegraphics[scale=0.6]{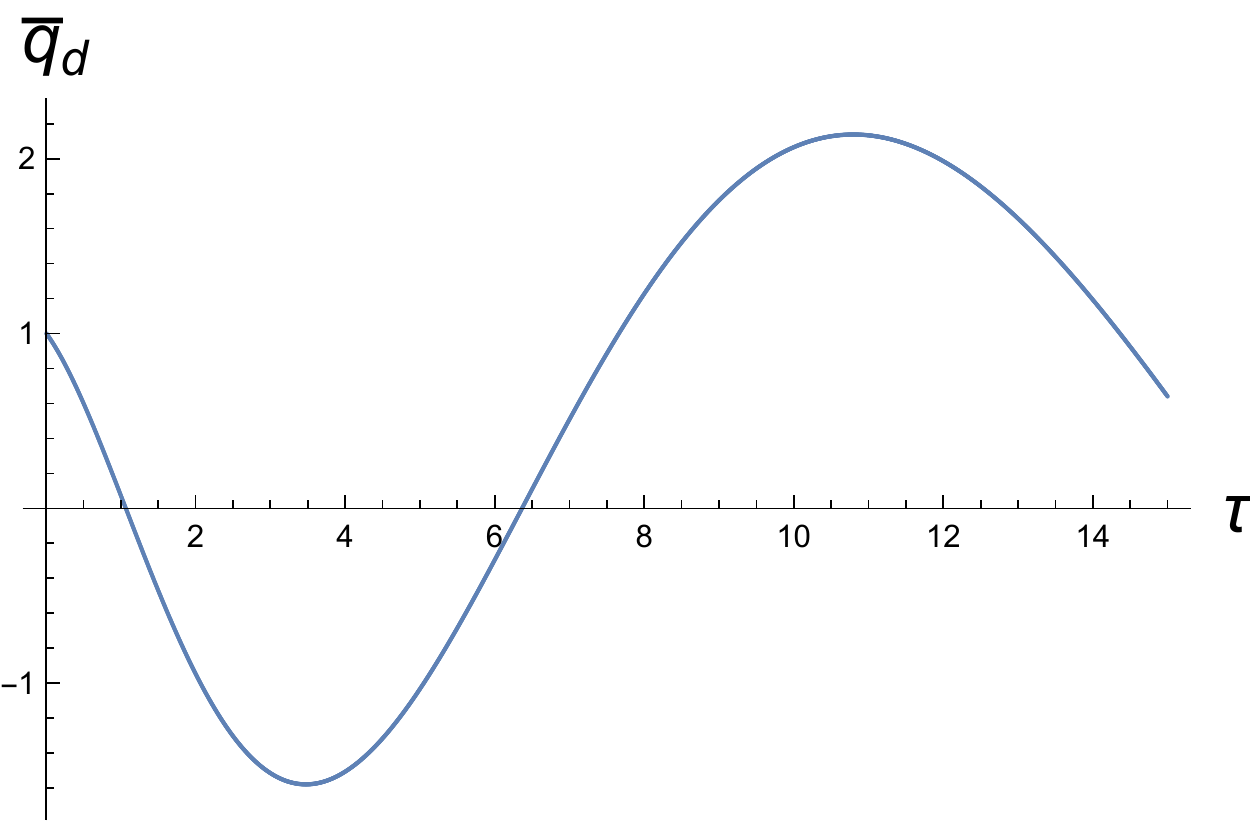} 
  \label{fig:q}
 } 
 \hfill
 \subfigure[]{
 \includegraphics[scale=0.6]{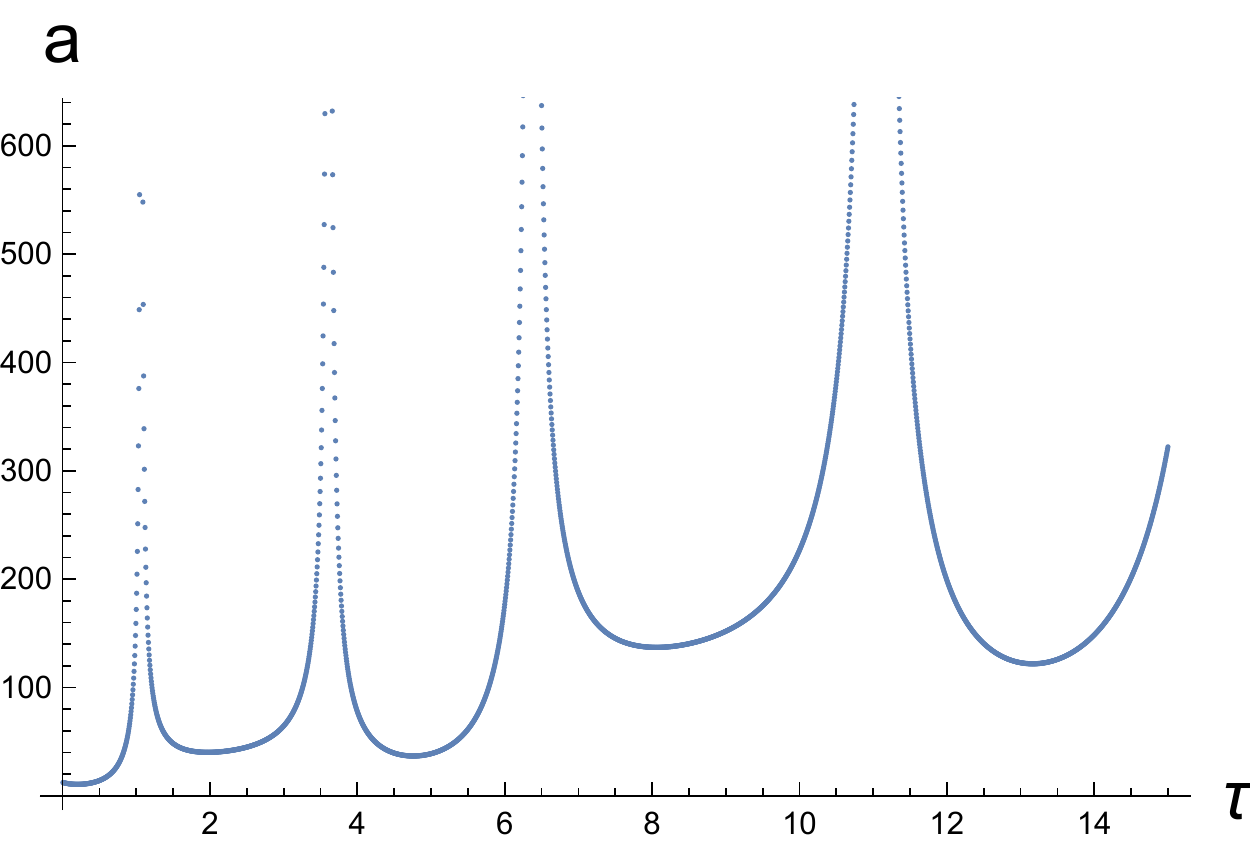} 
  \label{fig:g}
 } 
  \subfigure[]{
 \includegraphics[scale=0.6]{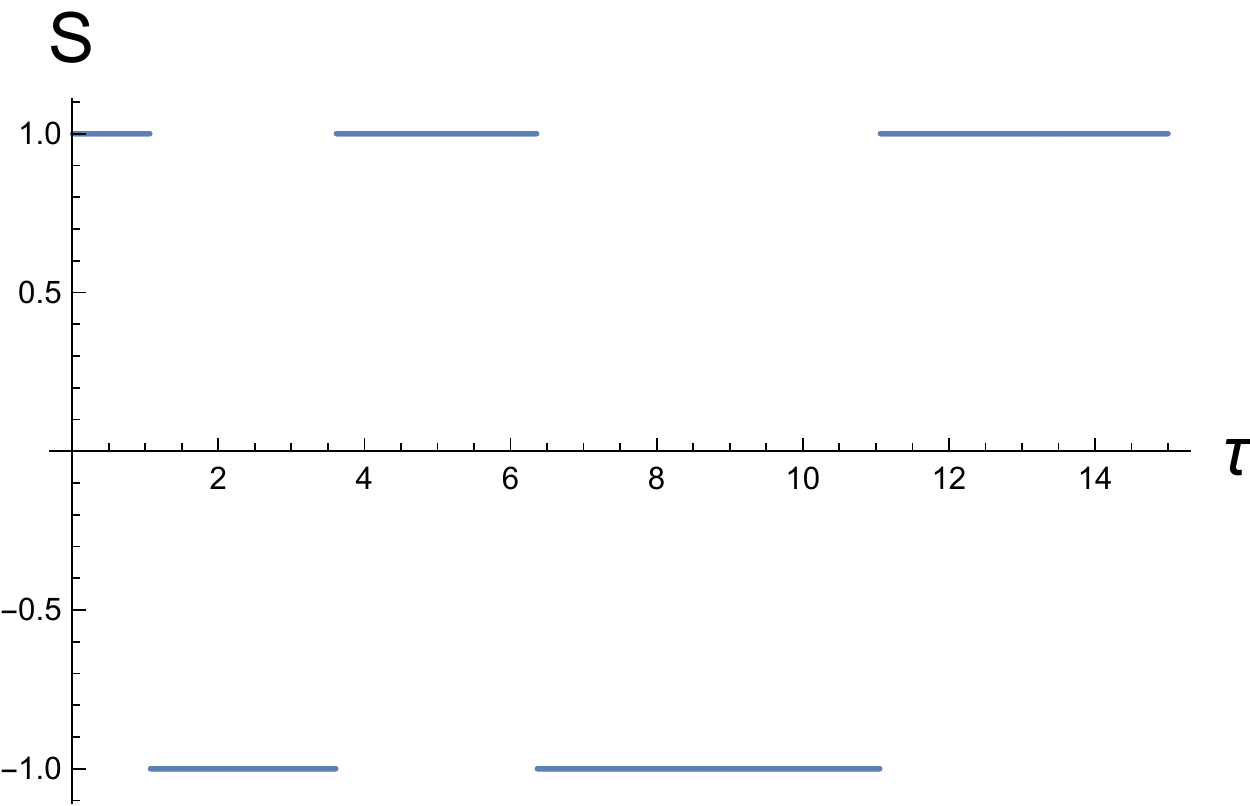} 
  \label{fig:s}
 } 
 \hfill
 \end{center}
 \caption{
$\bq_d$, $a$, ${\cal S}$ plotted as a function of 
$\tau$.
Here, $\bq_d$ is the scalar 
defined in \eq{eq:defq1}.
$a$ is the scale factor of the space defined by
$g_{\mu \nu} = a \delta_{\mu \nu}$
for the space with the translational symmetry and the reflection symmetry.
${\cal S}$ is the signature of time.
 }
 \label{fig:qsg}
 \end{figure}

Although the profile of $\bq_d(\tau)$ is not exactly periodic,
it follows an oscillatory pattern.
In \fig{fig:q}, we plot $\bq_d(\tau)$ as a function of $\tau$.
During  one oscillation of $\bq_d$,
the collective variable undergoes four Lifshitz transitions,
alternating between global and local Lifshitz transitions 
as is shown in \fig{fig:Psec0}.
The global Lifshitz transitions coincide with the points at which $\bq_d$ vanishes.
This concurrence will be explained through analytic solutions in the next section.
At the Lifshitz critical points (either global or local), 
the scale factor diverges, 
and the signature of time changes as is shown 
in  \fig{fig:g} and  \fig{fig:s}.
During the periods in which ${\cal S}<0$, 
we have de Sitter-like spacetimes 
with the Lorentzian signature\cite{2001JHEP...10..034S}.
Within one epoch of the de Sitter spacetime,
the universe initially starts with the infinite size, 
contracts to reach a minimum size,
and bounces back to expand to the infinite size again.
In one epoch,  
$\bq_d$ varies monotonically,
and can be used as a clock.
Multiple de Sitter-like spacetimes are connected by 
four dimensional Euclidean spaces.

\subsection{Analytic solution}

The global Lifshitz transitions occur when $\bq_d$ vanishes.
When $\bq_d$ is small, the equation of motion can be solved analytically
using $\bq_d$ as a small parameter.
In this section, we present the analytical solution 
valid when $\bq_d$ is small.
The analytic solution confirms the features observed in the numerical solution. 

In the small $\bq_d$ limit,  Eqs. (\ref{eq:Tpm}) and (\ref{eq:sk})  becomes
\bqa
T_\pm[ \bp_\bk, \bq_d] &=& \frac{ \beta }{ 2 \bp_{\bk} }  \pm \frac{ \gamma}{ 4 \bp_{\bk}^{3/2} } \bq_d + O(\bq_d^2), \nn
\bs_\bk &=& \mp \frac{\gamma}{2 \bp_\bk^{1/2}} + O(\bq_d).
\label{eq:TSk}
\eqa
To the leading order in $\bq_d$, \eq{eq:taUQ 1} and \eq{eq:taupk} become
\bqa
\partial_\tau \bq_d & = & \mp \gamma < \bp^{-\ha} >, \label{eq:taUQ 12} \\
\partial_\tau \bp_{\bk} & = & \pm \frac{2 \gamma}{\bq_d} \left( \bp_{\bk}^{-\ha} - < \bp^{-\ha} > \right) \bp_{\bk}\label{eq:taupk2}
\eqa
for branches $T_+[ \bp_\bk, \bq_d ]$ and  $T_-[ \bp_\bk, \bq_d ]$, respectively, 
where $< \bp^{-\ha} > \equiv  \frac{1}{L} \sum_{\bk}  \bp_{\bk}^{-\ha}$.
If $\bq_d(\tau)=0$  at $\tau = \tau^*$,
the solution to \eq{eq:taUQ 12} is given by
\bqa
\bq_d(\tau) & = &  \mp  \gamma <  \bp(\tau^*)^{-\frac{1}{2}} > ( \tau - \tau^*) + O\left(  ( \tau - \tau^*)^2 \right).
 \label{eq:solq1} 
\eqa
To keep track of momentum dependece of $\bp_{\bk}$,
it is convenient to consider the equation of motion for $\delta \bp_\bk = \bp_\bk - \bp_0$.
To the leading order in $(\tau-\tau^*)$ and $\delta \bp_\bk$, the equation of motion for $\delta \bp_\bk$ is given by
\bqa
\partial_\tau \delta \bp_\bk & = & \mp \gamma \left( 2 < \bp^{-\frac{1}{2}}>  - \bp_0^{-\frac{1}{2}}    \right) \frac{ \delta \bp_\bk }{\bq_d}
\label{eq:taudeltapk}
\eqa
To the leading order in $( \tau - \tau^*)$,
the solutions to  \eq{eq:taudeltapk} is obtained to be
\bqa
p_{\bk}(\tau) &= & \bp_0(\tau) + ( \tau - \tau^*) f_{\bk}, 
 \label{eq:sols} 
\eqa
where $f_{\bk}$ is a function of $\bk$
that is determined by matching $\bp_\bk(\tau)$ away from $\tau=\tau^*$.
In \eq{eq:sols} we use the fact that $< p(\tau^*)^{-\frac{1}{2}} > =   \bp_0(\tau^*)^{-\frac{1}{2}}$,
which holds because  $\delta \bp_\bk(\tau^*)=0$. 
\eq{eq:sols} implies that near $\tau=\tau^*$
the momentum dependence in $\delta \bp_\bk$ linearly vanishes in $\tau-\tau^*$
so that $\bp_\bk$ becomes independent of $\bk$ at $\tau^*$. 
For $\ta > 0$ and $\bp_0(\tau^*) > 0$, 
the signature and the scale factor of the universe are given by
\bqa
{\cal S} & = & 
\mbox{sgn}  \left(
( \tau - \tau^*)
\frac{\partial^2}{\partial k_1^2}  f_\bk
\right), \nn
a(\tau) &=&
\left|
\frac{1}{
  \frac{  4  }{ \ta   \bp_0(\tau^*)^{3} }
  \left.   
\frac{\partial^2}{\partial k_1^2}  
f_\bk  \right|_{\bk=0} 
   (\tau-\tau^*) 
 }
 \right|.
\eqa
This explains why the signature changes
and the scale factor diverges
at the global Lifshitz critical points.

Near the local Lifshitz transitions,
the collective variables varies in time such that only the second derivative in $\bk$ 
vanishes linearly in time as
\bqa
p_\bk(\tau) = \bp_0(\tau) + B (\tau - \tau^*) \bk^2 + O(\bk^4),
\label{eq:pkexp}
\eqa
where $B$ is a constant and $\tau^*$ is the critical point.
This also causes the change in the signature of time,
and the divergence of the scale factor as $a \propto \frac{1}{| \tau-\tau^*|}$
at the local Lifshitz critical points.
The emergent geometry is insensitive to the terms 
that are quartic and higher order in $k$,
and contains only a partial information 
on the pattern of entanglement of the microscopic degree of freedom.


\subsection{Emergent Lorentz symmetry}
\label{sec:Lorentz}

Does the spacetime 
with the Lorentzian signature
respect the Lorentz symmetry?
We answer this question for the saddle point solution 
of the form in \eq{eq:saddletp}
with the discrete translational symmetry.
The full saddle-point solution 
is not invariant under the Lorentz transformation
because the terms that are quartic or higher in ${\bf k}$
in \eq{eq:pkexp} break the rotational symmetry.
Nonetheless, the full rotational symmetry is restored
in the small $\bk$ limit
as the higher order terms are suppressed.
Since the translational and rotational symmetries emerge
in the long wavelength limit,
we only need to check the boost 
to see whether the full Lorentz symmetry emerges.

We consider a boost generated by
\bqa
K_{\vec \zeta} & = & 
\cG_\eta + \cH_{\rho}
+ \cT_{\td o_L}.
\label{eq:Kz}
\eqa
Here $\cG$ and $\cH$ are the generators
of the generalized spatial diffeomorphism
and time translation defined in 
Eqs. (\ref{G}) and
(\ref{H}), respectively.
$\eta$ and $\rho$ are the shift and lapse tensors given by
\bqa
\eta^j_{~i} & = & \frac{\varepsilon}{2} \left( \delta_{r_j-r_i, \tau \zeta} -   \delta_{r_j-r_i, -\tau \zeta}    \right),  
\label{eq:etaji} \\
\rho_{ij} & = & \varepsilon \left( \vec r_i \cdot \vec \zeta \right) \delta_{ij},
\label{eq:rhoij}
\eqa
where $\varepsilon$ is an infinitesimal parameter.
$ \cG_\eta $ generates the time-dependent  spatial translation by $\varepsilon \tau \vec \zeta$,
where $\vec \zeta$ is a spatial vector.
$ \cH_{\rho}$ generates the space-dependent
temporal translation by 
$\varepsilon \vec r \cdot \vec \zeta \equiv \varepsilon g_{\mu \nu} r^\mu \zeta^\nu$.
Combined, they generate the boost
along $\vec \zeta$
in the continuum limit.
The boost is augmented with an $O(L) \subset O(M)$ flavour transformation,
\bqa
\cT_{\td o_L} = \tr{ q s \td o_L},
\label{eq:cT}
\eqa
where 
$\td o_L$ is an $L \times L$ anti-symmetric matrix.
\eq{eq:cT} is independent of $t_c$ and $p_c$
because they are singlets under the $O(L)$ flavour symmetry.
The flavour rotation is included in \eq{eq:cT} because
some solutions are invariant under a combination
of the boost and an internal flavour rotation,
but not under the boost alone.
From now on, we refer to the combined transformation as boost.

The boost is a part of the full spacetime gauge symmetry and the global symmetry,
and the gauge constraints remain satisfied under the boost.
Since $s$ and $t_c$ are determined from 
$q$, $p_c$ through the gauge constraints in 
Eqs. (\ref{eq:sk}) and (\ref{eq:tk}),
we only need to check how 
$y$, $v$, $p_c$ and $q$ transform under the boost.
In order to understand the isometry of the emergent spacetime, 
we focus on 
the shift vector,
the lapse function
and the spatial metric, 
which together form the spacetime metric.
Since the spatial metric is determined from $p_c$ and $q$ through \eq{eq:Sgmn3},
we consider the transformations of
$p_c$,  $q$,
the shift vector ($\xi^\mu$)
and
the lapse function ($\theta$)
generated by the boost.
It is straightforward to show that the boost
generates the following transformations
(see Appendix \ref{app:Lorentz} for derivation),
\bqa
\delta \xi^\mu  =   \varepsilon \left[ 1 +  \cS  \right]  \zeta^\mu, &&~~~~~
\delta \theta  =  0, \nn
\delta p_{c}  =  4  \left( p_c \tilde t_c  \rho +  \rho \tilde t_c p_c \right),  &&~~~~~
\delta q  = 2   s^T \rho ,
\label{eq:deltaxtpq}
\eqa
where   $\td o_L = - \eta$ is chosen.
The shift vector and the lapse function are invariant under the boost
for the spacetime with the Lorentzian signature ($\cS = -1$). 
$p_c$ and $q$ remain invariant only
if $\tilde t_c= 0$ and $s=0$.
It is not surprising that the Lorentz symmetry is broken
by non-zero $\tilde t_c$ and $s$ 
because they are odd under the time-reversal symmetry
(see \eq{eq:TR}).
What is less trivial is the fact that the Lorentz symmetry
indeed emerges as an isometry of the spacetime
as far as the  time-reversal symmetry are kept
along with the discrete translational and rotational symmetry.

The saddle-point solution that follows from a generic initial condition  
is not static, and does not respect the Lorentz symmetry.
However, Lorentz-invariant static solutions can also be found.
For example, with the choice of $\td \alpha=4$,
\bqa
&& y = 0, ~~ v = I, \nn
&& s = 0, ~~ q = 0, \nn
&& \tilde t_{c} = 0,  ~~ p_c = \frac{1}{2}
\Bigl[ \bp_d  - \epsilon \left( \cos(k_1) + \cos(k_2) + \cos(k_3) \right) \Bigr],
\label{eq:Minkowski}
\eqa
with $\bar p_d > 0$ 
satisfies the equation of motion in \eq{eq:EOM}.
This is a static solution
with the time-reversal symmetry. 
For $\epsilon < 0$ ,
the emergent spacetime metric is
invariant under the Lorentz symmetry.
This gives rise to the Minkowski spacetime in the continuum.
For $\epsilon > 0$ , 
the resulting spacetime is Euclidean.
The Minkowski spacetime is a fine-tuned solution
within the present theory.


 \subsection{Effective theory}
 \label{sec:effective}

In this section, we derive the effective theory
that describes fluctuations of the collective variables 
around the general time-dependent saddle-point configuration.
The microscopic theory does not have a fixed background.
However, the saddle-point configuration of the collective variables
provides the spacetime on which their fluctuations propagate.
We expand the collective variables around the saddle point 
in \eq{eq:saddletp},
 \bqa
p^{'}_c  = p_c - \frac{\bar p}{2}, &&
 t^{~'}_c  =  t_c   - \frac{i}{8} p_c^{-1} - \tilde t,  \nn
  q^{'}  = q - \bq_d I, &&
s^{'}  = s -  \bar s ,
 \eqa
where $\bar p,  \td t, \bq_d, \bar s$
satisfy the saddle-point equations in 
Eqs. (\ref{eqt3})-(\ref{eqq13}).
Here $t'_c$ represents the fluctuation of 
$t_c  - \frac{i}{8} p_c^{-1}$ whose saddle point value
is $\tilde t$ (see  \eq{eq:tildet}).
With the shift and lapse tensors given by
Eqs. (\ref{eq:gauge1}) and (\ref{eq:gauge2}),
the quadratic action for the fluctuations of the collective variables is written as
\bqa
S_{2} 
 & = & 
N \int_0^\infty d \tau ~
\mbox{tr} \Biggl\{
 s' \partial_\tau  q'
+  t'_c \partial_\tau  p'_c 
-  \cH_2 -  \cG_2 \Biggr\}.
\label{Sfinal2} 
\eqa
Here boundary terms are not shown.
The quadratic Hamiltonian and the \SLL generator are given by
\bqa
 \cH_2 & = & 
 s' s^{'T} 
 + \ta q' \bar U^2 q^{'T} 
 + \ta \left( 2 \bar U U' Q' + \bar QU' U' \right) \nn
&& + \sum_{c=1,2} \left[
  4   \td  t \left( t'_c p_c' + p_c'  t'_c  \right) 
  + 2 \bar p t'_c  t'_c + \frac{1}{2} \frac{1}{\bar p} p'_c  \frac{1}{\bar p}  p'_c \frac{1}{\bar p} \right],   \nn
   \cG_2 & = &
   -\frac{2}{\bq_d} \left(
   \bar s - < \bar s >  
   \right) 
   \left(
      s' q'  +  2 \sum_{c=1,2}  t'_c p'_c
      \right).
\eqa
\label{eq:H2G2}
$
\bU = 
\bar s \bar s^T + 4 \td t \bar p \td t  +
 \frac{1}{4}   \frac{1}{ \bar p} $
 and
$\bQ  = 
 \left( \bq_d^2 I +  \bar p \right)$
 represent the saddle-point of $U$ and $Q$
 defined in \eq{UQ }.
$U'$ and $Q'$ represent fluctuations of $U$ and $Q$ 
which are linear superpositions of $t_c', p_c', s'$ and $q'$ :
$U'  =  s' \bar s + \bar s s^{'T} 
+ \sum_{c=1,2} \left[
2 t_c' \bar p \td t + 4 \td tp_c'  \td  t + 2  \td  t \bar p t_c' 
- \frac{1}{4} \frac{1}{\bar p} p_c' \frac{1}{\bar p} \right]$
and
$Q' = \bq_d ( q^{'T}  + q' ) + \sum_{c=1,2} p_c' $.
The shift and lapse tensors are fixed to be  
\eq{eq:gauge1} and \eq{eq:gauge2}, respectively.
In \eq{eq:H2G2}, we use 
$y =   -\frac{2}{\bq_d} \left(
   \bar s - < \bar s >  
   \right)$ on the constraint hypersurface. 
Since $U'$ and $Q'$ depend on $t_1'$ and $t_2'$ (and $p_1'$ and $p_2'$) 
only through their symmetric combinations, it is useful to introduce 
\bqa
t'_+ = \frac{ t'_1 +  t'_2}{\sqrt{2}}, &&
t'_- = \frac{  t'_1 -  t'_2}{\sqrt{2}},  \nn
p'_+ = \frac{ p'_1 + p'_2}{\sqrt{2}}, &&
p'_- = \frac{ p'_1 - p'_2}{\sqrt{2}}
\eqa
to rewrite $\cH_2$ and $\cG_2$  as
\bqa
 \cH_2 & = & 
 s' s^{'T}  + \ta q' \bar U^2 q^{'T} + \ta \left( 2 \bar U U' Q' + \bar QU' U' \right) \nn
&& +  \left[
  4   \td  t \left( t'_+ p_+' + p_+'  t'_+  \right) 
  + 2 \bar p t'_+  t'_+ + \frac{1}{2} \frac{1}{\bar p} p'_+  \frac{1}{\bar p}  p'_+ \frac{1}{\bar p} \right] \nn 
 && +  \left[
  4   \td  t \left( t'_- p_-' + p_-'  t'_-  \right) 
  + 2 \bar p t'_-  t'_- + \frac{1}{2} \frac{1}{\bar p} p'_-  \frac{1}{\bar p}  p'_- \frac{1}{\bar p} \right], \nn
   \cG_2 & = &
   -\frac{2}{\bq_d} \left(
   \bar s - < \bar s >  
   \right) 
   \left(
      s' q'  +  2  t'_+ p'_+  +  2  t'_- p'_-
      \right),
      \eqa
where 
\bqa
U' & = & s' \bar s + \bar s s^{'T} 
+ \sqrt{2} \left[ 2 t_+' \bar p  \td  t + 4   \td   t p_+'  \td  t + 2  \td  t \bar p t_+' 
- \frac{1}{4} \frac{1}{\bar p} p_+' \frac{1}{\bar p} \right], \nn
Q' &=& \bq_d ( q^{'T}  + q' ) + \sqrt{2} p_+'. 
\eqa
$t'_\pm, p'_\pm, s', q'$ are all $L \times L$ matrices
which become bi-local fields in the continuum.
Due to the local structure,
$\td t^{\br_1, \br_2}, \bp_{\br_1, \br_2}, \bs^{\br_1}_{~~\br_2}$
decay exponentially in $\br_1 - \br_2$,
and $\td t_\bk, \bp_\bk, \bs_\bk$ 
are analytic functions of momentum.
This guarantees that the effective theory 
that describes propagation of the fluctuating modes is local.

 \begin{figure}[ht]
 \begin{center}
 \includegraphics[scale=0.3]{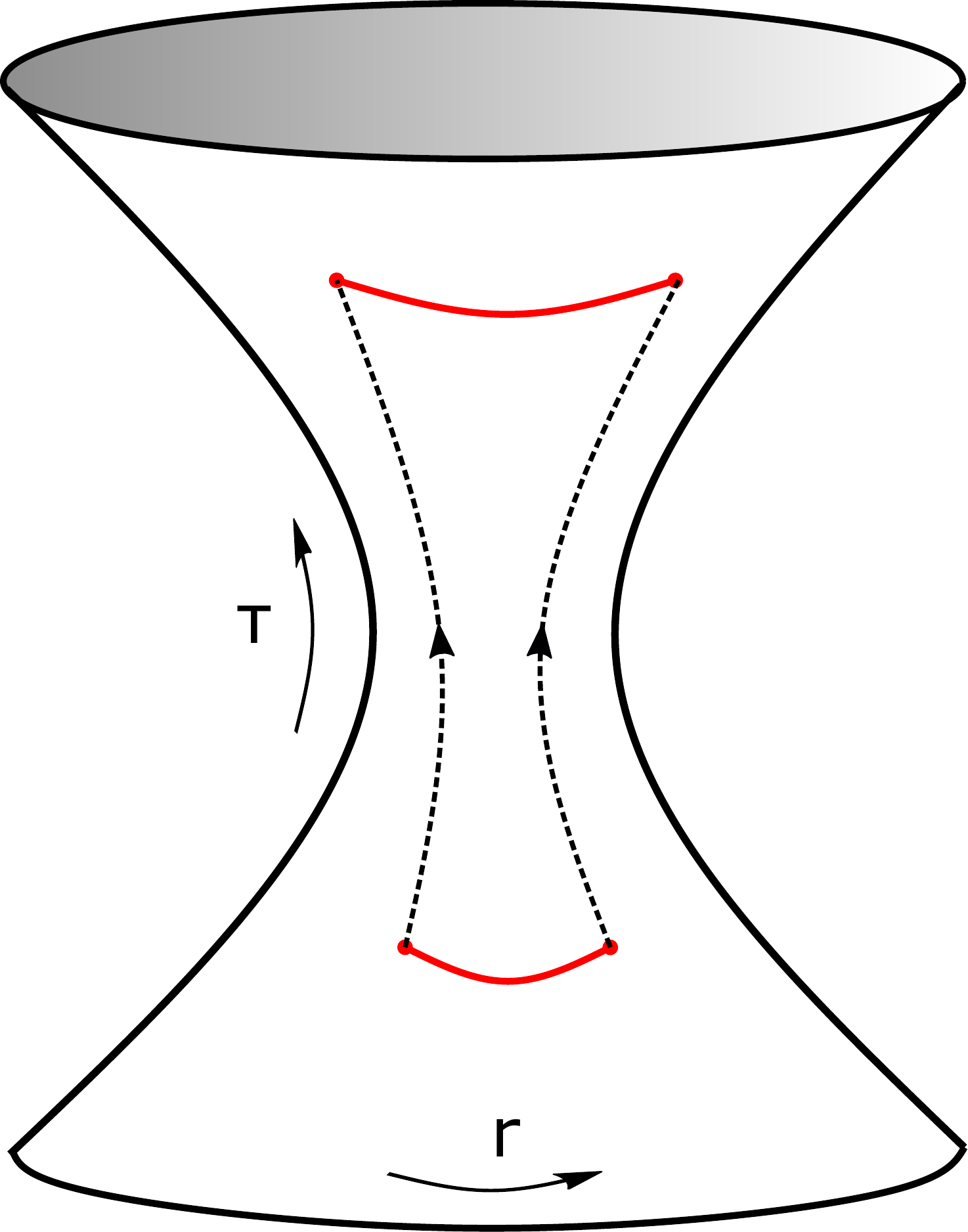} 
 \end{center}
 \caption{
 The hyperboloid represents the de Sitter-like spacetime 
 that arises from the saddle-point solution.
Small fluctuations of the bi-local collective variables propagate in the spacetime,
obeying the dynamics controlled by a local effective theory.
The dashed lines represent the world lines of the end points of the bi-local fields.
 }
 \label{fig:bilocal}
 \end{figure}

We can write the effective theory in the gradient expansion. 
To organize the gradient expansion, 
it is useful to note that 
$\bs_\bk$ and $\td t_\bk$ are both determined 
from  $\bp_\bk$ and $\bq_d$ through 
Eqs. (\ref{eq:sk}) and (\ref{eq:tk}).
Since the spatial metric is directly related to $\bU_\bk$ 
through   \eq{eq:Sgmn}, 
it is  convenient to  take $\bU_\bk$and $\bq_d$
as independent variables, and write
$\bs_\bk$, $\td t_\bk$, $\bp_\bk$ 
as functions of $\bU_\bk$ and $\bq_d$ as
\bqa
\bs_\bk = S[ \bU_\bk, \bq_d], ~~
\td t_\bk =  T[ \bU_\bk, \bq_d], ~~
\bp_\bk =  P[ \bU_\bk, \bq_d].
\eqa
$\bU_\bk$ is an analytic function of $\bk$.
In real space, $\bU^{\br_1, \br_2}$ decays exponentially  
over the coordinate scale that corresponds to 
a unit proper distance\footnote{
This is how the metric is defined in  \eq{eq:Sgmn}.}.
Consequently,  $\bs, \td t, \bp$ all decay exponentially 
in real space in the same manner.
Expanding 
$\bU_\bk$,
$\bs_\bk$,
$\td t_\bk$,
$\bp_\bk$,
$y_\bk$ 
in $\bk$, we write
\bqa
\bU_\bk &=& \bU^{[0]} + \bU^{[2]} g^{\mu \nu} k^\mu k^\nu + O(k^4), \nn
\bs_\bk &=& \bs^{[0]} + \bs^{[2]} g^{\mu \nu} k^\mu k^\nu + O(k^4), \nn
\td t_\bk &=& \td t^{[0]} + \td t^{[2]} g^{\mu \nu} k^\mu k^\nu + O(k^4), \nn
\bp_\bk &=& \bp^{[0]} + \bp^{[2]} g^{\mu \nu} k^\mu k^\nu + O(k^4), \nn
\by_\bk &=&  \by^{[0]} + \by^{[2]} g^{\mu \nu} k^\mu k^\nu + O(k^4),
\eqa 
where
$\bU^{[2]} = \left. 
 \frac{1} {8 \ta \bU_\bk} \right|_{\bk=0}$,
$\bs^{[0]} = S[\bU^{[0]}, \bq_d]$,
$\bs^{[2]} = \left. 
 \frac{1} {8 \ta \bU_\bk} \frac{ \partial  S[\bU_\bk, \bq_d] }{ \partial \bU_\bk} \right|_{\bk=0}$,
$\td t^{[0]} = T[\bU^{[0]}, \bq_d]$,
$\td t^{[2]} =
 \left. 
 \frac{1} {8 \ta \bU_\bk} \frac{ \partial  T[\bU_\bk, \bq_d] }{ \partial \bU_\bk} \right|_{\bk=0}$,
$\bp^{[0]} = P[\bU^{[0]}, \bq_d]$,
$\bp^{[2]} = \left. 
 \frac{1} {8 \ta \bU_\bk} \frac{ \partial  P[\bU_\bk, \bq_d] }{ \partial \bU_\bk} \right|_{\bk=0}$,
 $\by^{[0]} = -\frac{2}{\bq_d} \left( \bar s^{[0]} - < \bar s > \right)$
 and
 $\by^{[2]} = 
 -\frac{1}{\bq_d} 
 \left. 
 \frac{1} {4 \ta \bU_\bk} \frac{ \partial  S[\bU_\bk, \bq_d] }{ \partial \bU_\bk} \right|_{\bk=0}$.
$\by_\bk$ is the shift tensor in \eq{eq:gauge1}
written in the momentum space.
$\bs^{[0]}, \bs^{[2]}, \td t^{[0]}, \td t^{[2]}, \bp^{[0]}, \bp^{[2]}, \by^{[0]}, \by^{[2]}$
are all functions of $\bq_d(\tau)$ and $\bU^{[0]}(\tau)$.
In the coordinate system in  \eq{eq:coord},
the effective theory for the propagating bi-local modes becomes
\bqa
&& S_{2} 
  =  
N \int  d \tau d \br_A d \br_B ~
 \Biggl\{
 s'(\rot) \partial_\tau  q'(\rto)
+ \sum_{\sigma=\pm}  t'_\sigma(\rot) \partial_\tau  p'_\sigma(\rto) \nn
&&~~~~~ -   s'(\rot)   s'(\rot)
- \ta q'(\rot)  \left( \bU^{[0] 2} - 2 \bU^{[0]}  \bU^{[2]} \nabla^2_B \right) q'(\rot) \nn
&&~~~~~
- \sum_{\sigma=\pm} 
\Bigl[
    4 p'_\sigma(\rot)  \left( \td t^{[0]} - \td t^{[2]} \nabla^2 \right) t'_\sigma(\rot)  \nn
    &&  ~~~~~~~~~~~~~~   +  4 t'_\sigma(\rot)  \left( \td t^{[0]} - \td t^{[2]} \nabla^2 \right) p'_\sigma(\rot) \nn
&&~~~~~~~~ ~~~~~ +  2 t'_\sigma(\rot)  \left( \bp^{[0]} - \bp^{[2]} \nabla^2 \right) t'_\sigma(\rot)  \nn
&& ~~~~~~~~~~~~~~
 + \frac{1}{2}  
  \left(  \frac{1}{\bp^{[0] 2}} + 2 \frac{\bp^{[2]}}{  \bp^{[0] 3} }  \nabla^2_A \right) p'_\sigma(\rot)
  \left( \frac{1}{\bp^{[0] }} + \frac{ \bp^{[2]} }{ \bp^{[0] 2} } \nabla^2_B \right) p'_\sigma(\rot)
   \Bigr]  \nn
&&
 - 2 \ta Q'(\rot)  \left( \bU^{[0]} -    \bU^{[2]} \nabla^2 \right) U'(\rot)
 - U'(\rot)  \left( \frac{1}{ \bU^{[0] } } +  \frac{ \bU^{[2]} }{  \bU^{[0] 2}  } \nabla^2 \right) U'(\rot) \nn
 &&
 - q'(\rto)  \left( \by^{[0]} -    \by^{[2]} \nabla^2_A \right) s'(\rot)
 - 2 \sum_{\sigma=\pm} 
  p_\sigma'(\rto)  \left( \by^{[0]} -    \by^{[2]} \nabla^2 \right) t_\sigma'(\rot)
 \Biggr\} \nn
\label{Sfinal3} 
\eqa
up to two derivative order,
where 
$\nabla_A^2 = g^{\mu \nu} \frac{\partial^2}{\partial r_A^\mu \partial r_A^\nu}$,
$\nabla_B^2 = g^{\mu \nu} \frac{\partial^2}{\partial r_B^\mu \partial r_B^\nu}$
and $\nabla^2 = \frac{\nabla_A^2 + \nabla_B^2}{2}$.
Here $\bar Q = \frac{1}{\td \alpha \bU}$ is used.
In \eq{Sfinal3}, all bi-local fields depend on $\tau$.
The propagating degrees of freedom are kinematically non-local,
namely, bi-local in this case\cite{Das:2003vw,Koch:2010cy}.
Nonetheless, the theory is dynamically local\cite{PhysRevLett.114.031104} 
in that the theory does not allow bi-local objects to suddenly jump from one location to another location non-locally.  
The locality of the effective theory is guaranteed by the fact
 that the saddle-point configurations of collective variables decay
exponentially in the relative coordinate 
due to the local structure.
As a result, the gradient expansion is well defined.
At length scale larger than the scale set by the local structure, 
the higher order derivative terms are negligible. 

The effective action includes terms that are cubic and higher order in the bi-local fields.
For example, the cubic action takes the form of 
\bqa
S_3 \sim N \int d\tau dr_A dr_B dr_C ~  f_1(r_A, r_B) f_2(r_B,r_C) f_3(r_C,r_A),  
\label{cubicterm}
\eqa
where $f_i(r_A,r_B)$ represents one of the bi-local fields and their derivatives.
Again the higher derivative terms are suppressed by the scale associated with the local structure,
and the interaction terms are also local.
In the large $N$ limit, 
the interactions are suppressed,
and the bi-local fields are weakly interacting.
To the leading order in $1/N$, 
the bi-local objects freely propagate in the spacetime
that is determined from the saddle-point configuration
of the collective variables.
This is illustrated in \fig{fig:bilocal}.

While the effective theory in \eq{Sfinal3}  is local,
it is not a local field theory of point-like particles.
If one starts with a state with one bi-local excitation, 
its size can grow in time without a bound  
because there is no potential 
that confines the end points in the large $N$ limit.
This is because two end points of the bi-local fields
propagate freely to the leading order in the $1/N$ expansion.
The sub-leading interactions 
such as the one in \eq{cubicterm}
can create bound states 
which behave as point particles.
Only at length scales larger than the length scale of the bound state,
one may use a local field theory description of point particles.
However, the legnth scale of the bound state scales with $N$ in the large $N$ limit.
For this reason, the present theory is far from the theory of pure gravity.

As is discussed in \ref{subsec:path},
only $L(L+1)+2$ phase space variables represent physical
degrees of freedom.
One possible gauge choice is to determine
$t_+'$ and the traceless part of $s'$
to solve the Hamiltonian constraint and the momentum constraint, respectively.
Furthermore, one can fix 
$p_+'$ and the traceless part of $q'$ 
to fix the gauge associated with the Hamiltonian
and the momentum constraints, respectively.
This leaves $t'_-$, $p'_-$  and the trace parts of $q'$ and $s'$ 
 as physical degrees of freedom. 


\section{Summary and Discussion}
\label{sec:conclusion}

In this paper, we present a model for a 
background independent quantum gravity
in which dimension, topology and geometry are all dynamical.
The fundamental degree of freedom is a rectangular matrix.
Matrix elements within each column define a local Hilbert space.
The pattern of entanglement across local Hilbert spaces
determines a spatial manifold if it has a local structure.
A state has a local structure if it is short-ranged
entangled when viewed as a state defined on a Riemannian manifold
in which the column indices of the matrix is embedded.  
The theory does not have a manifold with a fixed dimension,
and the spatial diffeomorphism of the general relativity is generalized to
a larger group that includes diffeomorphism in arbitrary dimensions in the large $L$ limit.
The Hamiltonian that generates a background independent dynamics is relatively local
in that the effective interaction between sites is determined by the
pre-existing entanglement between the sites.
The generalized momentum and Hamiltonian constraints obey 
a first-class constraint algebra 
that is reduced to that of the hypersurface deformation algebra of 
the general relativity in a special case.
Using the constraints,
we express the projection of a state with a finite norm
to a gauge invariant state 
as a path integration of collective variables
that describe fluctuating spacetime.
In the limit that the size of the matrix is large,
the path integration can be replaced with a saddle-point 
that satisfies the classical equation of motion.
The equation of motion is solved both numerically and analytically 
for a state that has a three-dimensional local structure
with a translational symmetry.
We obtain a solution that describes a series of $(3+1)$-dimensional 
de Sitter-like spacetimes with the Lorentzian signature
which are bridged by $4$-dimensional Euclidean spaces in between.   
We find that the dynamical phase transitions at which the signature of spacetime changes
are triggered by Lifshitz transitions of the collective variables 
that control the pattern of entanglement for the underlying matrix.
The gravitational degrees of freedom can be identified 
as a composite of the collective variables.
It is shown that the geometry encodes only a partial information 
on the entanglement of the underlying quantum matter.
There exists non-geometric entanglement 
that is encoded  in higher spin fields of the theory\footnote{
This might be phrased as 
$EPR > ER$.
}.
The effective theory describes bi-local excitations 
that propagate in the spacetime formed by the condensate 
of the collective variables.
We conclude with discussions on the connection to the dS/CFT and AdS/CFT correspondences
and open problems.

\subsection{dS/CFT}

The present construction can be viewed as a non-perturbative formulation 
of the de Sitter space and conformal field theory (dS/CFT) correspondence\cite{
Hull_1998,
Bousso_1999,
Balasubramanian_2001,
2001JHEP...10..034S,
NOJIRI2001145,
ABDALLA2002435,
Anninos:2017aa}.
To illustrate the connection through a concrete example, 
let us consider the projection of
a normalizable state to the gauge invariant state,
$\cb 0 \rb = \int  D \Phi ~  \cb \Phi \rb$.
For the semi-classical normalizable state
given by \eq{eq:st} with \eq{eq:wp},
the projection 
is written as
\bqa
\lzc \chi \rb
& =& 
\int D q D \phi D \varphi D s Dt ~ ~ e^{  i  \tr{ N s q 
+ t_1 ( \phi^T \phi) 
+ t_2 ( \varphi^T \varphi) 
}   
-
i N 
\tr{
 \qstar s
+  \pcstar  t_c
}
-\frac{
\sum_{i,\alpha} \left[   s^i_{~\alpha}  -  \sstar^{i}_{~\alpha} \right]^2
+ \sum_{c,ij} \left[ t_c^{ij}-   \tcstar^{ij} \right]^2
}{
\Delta^2
}
}. \nn
\eqa
Upon integrating over $s, t_c$, 
the projection is expressed as a matrix integration,
\bqa
\lzc \chi \rb
& =& 
\int D q D \phi D \varphi   ~ e^{ 
iN 
\tr{
  \sstar ( q-  \qstar )
+   \mathcal{t}_1 \left( \frac{ \phi^T \phi}{N} - \mathcal{p}_1  \right)
+   \mathcal{t}_2 \left( \frac{ \varphi^T \varphi}{N} - \mathcal{p}_2  \right)
}} \times \nn
&& 
e^{-  \frac{N^2 \Delta^2}{4} 
\tr{
 (q -  \qstar)^T (q -  \qstar)
+ 
\left( \frac{ \phi^T \phi}{N} -  \mathcal{p}_1  \right)^2
+
\left( \frac{ \varphi^T \varphi}{N} -  \mathcal{p}_2  \right)^2
}
}.
\label{eq:0chids}
\eqa  
Suppose that $\tcstar$ and $\pcstar$ has a local structure.
Namely, there exists a mapping from sites to a Riemannian manifold 
such that $\mathcal{t}_{c,ij}$ and $\mathcal{p}^{ij}_c$ decays exponentially
in the proper distance between $r_i$ and $r_j$, 
where $r_i$ is the mapping from site $i$ to the manifold.
In this case,  \eq{eq:0chids}
can be viewed as the generating function
of a local (non-unitary) field theory
defined on the Riemannian manifold.
The equivalence between  \eq{eq:0chids}
and \eq{eq:pathint} shows that the generating function
of the non-unitary field theory is given by
the path integration of the quantum gravity
in which the de Sitter-like spacetime
emerges at the saddle-point.

\subsection{AdS/CFT}

In order to make a connection with 
the AdS/CFT correspondence\cite{
Maldacena:1997re,Witten:1998qj,Gubser:1998bc},
one should consider a unitary field theory 
defined on a  manifold with the Lorentzian signature
instead of the Riemannian manifold.
If $\tcstar$ and $\pcstar$ in \eq{eq:0chids} exhibit a local structure, 
the local structure determines
the signature of the manifold 
on which the field theory is defined.
If there is a translational invariance,
the signature of each direction in the manifold
is determined by
the sign of the second derivative of the bi-local fields
in the momentum space.
If the second derivative of the collective variable 
has the opposite sign in one direction compared with other directions,
this gives rise to a Lorentzian metric. 
In this case, \eq{eq:0chids} corresponds to the generating function of 
a Lorentzian field theory in the continuum limit.
It is useful to continue to think of the generating function
as an overlap of two wavefunctions, one given by $\Psi_1=1$ and the other given by $\Psi_2=e^{iS}$,
where $S$ is the action of the Lorentzian quantum field theory\cite{Lee2016}.
The wavefunctions are defined on spacetime (not just in space).
The Hamiltonian is then replaced
with a generator of coarse graining.
The fact that $\Psi_1$ is invariant under the transformation generated by the coarse graining 
implies that $\Psi_1$ is a fixed point action
that is scale invariant.
In particular, $\Psi_1=1$ represents the trivial insulating fixed point.

Just as the overlap between a gauge invariant state and a state with a finite norm
is invariant under the gauge transformation generated by the Hamiltonian and momentum constraints
in \eq{eq:overlap2}, the generating function is invariant under
an insertion of $e^{-i \hat H \tau}$, where $\hat H$ is the generator of coarse graining\cite{Lee2016}.
An evolution generated by successive applications of coarse graining
gives rise to the exact Wilsonian renormalization group (RG) flow\cite{Polchinski:1983gv}.
The exponent of $\Psi_2(\tau) =e^{-i \hat H \tau} \Psi_2$ 
represents the effective action defined at a length scale $\tau$. 
In the effective action, 
all couplings that are allowed by symmetry are generally turned on.
While the flow of $\Psi_2(\tau)$ can be tracked in terms of the classical couplings
that parameterize the renormalized action,
it is more natural to view the RG flow as an evolution of the wavefunction
in the vector space.
In particular, one can choose a set of basis wavefunctions that 
spans the full space of wavefunctions. 
Because the set of single-trace operators generate all multi-trace operators, 
the basis states can be chosen so that their wavefunctions  
only include the single-trace operators.
As a result, $\Psi_2(\tau)$ can be represented as a linear superposition
of states whose actions  include only single-trace operators.
In this way, the classical Wilsonian RG flow
defined in the space of all couplings 
is replaced with a quantum evolution 
of wavefunction defined in the space
of the single-trace couplings\cite{Lee:2013dln}.
The full Wilsonian renormalization group flow 
is projected to the space of single-trace operators 
at the expense of promoting 
the sources for the single-trace operators to dynamical variables.
Since the source for the single-trace energy-momentum tensor
is also promoted to dynamical variable, 
the path integration of the dynamical source represents a dynamical gravity in the bulk.
This may eventually provide a non-perturbative formulation of 
quantum gravity in the anti-de Sitter space.
One missing piece in the puzzle is to construct 
a UV finite coarse graining operator such as the one in \eq{Hx} 
that gives rise to a background independent gravity in the bulk.

 \subsection{Open questions}

Here we list some open questions.

\begin{itemize}
\item Physical spectrum

In general, saddle-point configurations 
do not have temporal Killing vector. 
It would be interesting to find a saddle-point configuration
that support a temporal Killing vector
and extract the physical energy spectrum of the propagating modes
in this theory\cite{PhysRevD.65.124013,Dittrich2007}

\item Local field theory of point particles

Background independent theories of quantum gravity
include kinematically non-local objects. 
In the present theory, it is the bi-local field. 
The ultimate goal is to construct a background independent theory
that reduces to a field theory of a small number of point-like particles at low energies.
However, the present theory does not achieve this goal 
because point-like particles emerge only through
the interactions between bi-local fields that are suppressed by $1/N$.
This is because the present theory is similar to vector models
\cite{2002PhLB..550..213K,
Das:2003vw,
Koch:2010cy,
Douglas:2010rc,
2013arXiv1303.6641P,
Leigh:2014tza,
2015PhRvD..91b6002L,
2014arXiv1411.3151M,
Vasiliev:1995dn,
Vasiliev:1999ba,
Giombi:2009wh,
Vasiliev:2003ev,
Maldacena:2011jn,
Maldacena:2012sf,
2014PhRvD..90h5003S,
Lee2016}
although the microscopic degree of freedom is a matrix.
One index of the matrix is used to generate an emergent space,
and the internal symmetry acts only on the remaining one index.
In order to construct a quantum theory of gravity that reduces
to the Einstein's general relativity with a small number of additional fields, 
one needs a mechanism that keeps dynamical objects finite    
to the leading order in $1/N$.
For this, tensor models may be a natural direction\cite{doi:10.1142/S0217732391001184,doi:10.1142/S0217732391003055,GROSS1992144,Gurau:2011aa}.
In tensor models with rank greater than $2$,
where one index labels sites
and the remaining indices label internal flavour (or color),
one expects to have multi-local fields as emergent degree of freedom. 
In this case, a non-zero tension may naturally arise 
to keep kinematically non-local objects finite.  
In relation to a non-perturbative formulation of the AdS/CFT correspondence for gauge theories, 
we expect that the background independent coarse grainer  
takes a form of a relatively local tensor model with rank $3$.

\end{itemize}

\section*{Acknowledgments}

I thank Bianca Dittrich for discussions.
The research was supported by
the Natural Sciences and Engineering Research Council of
Canada.
Research at the Perimeter Institute is supported
in part by the Government of Canada
through Industry Canada,
and by the Province of Ontario through the
Ministry of Research and Information.

 \bibliography{references}

\newpage


\begin{appendix}

\section{Derivation of \eq{eq:SVD} }
\label{app:SVD}

Using the singular value decomposition,
a symmetric matrix can be written as 
$v = O_v^T D_v O_v$,
where $O_v$ is an orthogonal matrix
and $D_v = \left( 
\begin{array}{ccc}
d_1 &         &     \\
       & d_2  &     \\
       &         & \ddots 
\end{array} \right)$ is a diagonal matrix.
If $v$ is non-singular, $d_i \neq 0$. 
One can then write the diagonal matrix
as
$D_v = n_v X_v S_v X_v$, 
where
$n_v = |\prod_i d_i  |^{\frac{1}{L}}$,
$X_v = \frac{1}{| \prod_i d_i |^{\frac{1}{2L}}}  \left( 
\begin{array}{ccc}
\sqrt{|d_1|} &         &     \\
       & \sqrt{|d_2|}  &     \\
       &         & \ddots 
\end{array} \right)  
$
and 
$S_v = \left( 
\begin{array}{ccc}
\sgn{d_1} &         &     \\
       & \sgn{d_2}  &     \\
       &         & \ddots 
\end{array} \right)  
$.
Therefore, $v$ can be written as
$v = n_v g_v^T S_v g_v$,
where $g_v = X_v O_v \in  SL(L, \mathbb{R})$.

\section{Non-normalizability of gauge invariant state}
\label{app:nonnormalizability}

In this appendix, we prove that all gauge invariant states have infinite norm
with respect to the norm defined in \eq{eq:norm}.
Consider a gauge invariant wavefunction $\Psi(\Phi^A_{~~i})$ defined in the space of $\{ \Phi^A_{~~i} \}$.
Suppose that the wavefunction has a support away from the origin in the $ML$-dimensional space.
Without loss of generality, let us choose the frame such that a point in the support is given by
$\Phi^A_{~~i} = \Phi_0 \delta^A_1 \delta^1_i$, where $\Phi_0$ is a positive constant.
Under the \SLL transformation generated by $\hat G^1_{~~2} + \hat G^2_{~~1}$,
the point can be mapped into any point on the hyperbola defined by
$( \Phi^1_{~~1})^2 - (\Phi^1_{~~2})^2 = \Phi_0^2$ with $\Phi^1_{~~1} > 0$. 
In order for the wavefunction to be invariant under the transformation,
the wavefunction must have amplitude $\Phi_0$ along the entire hyperbola.
Because the hyperbola is unbounded, the norm of the wavefunction is infinite.
The only state which does not have support away from the origin is 
$\Psi(\Phi^A_{~~i}) \propto \Pi_{A,i} \delta(\Phi^A_{~~i})$.
However, this has an infinite norm as well because 
the wavefunction is constant in the conjugate space.

\section{Computation of the commutator of the Hamiltonians}
\label{app:HH}

The commutator between $\hat H_u$ and $\hat H_v$ can be written as a sum of two contributions,
\bqa
\left[ \hH_u, \hH_v \right]  & = & \hat C_1 +  \hat C_2,
\eqa
where
\bqa
 \hat C_1 & = & - \frac{\ta}{M^2} 
\Biggl[ \tr{ 
  \hPi \hPi^T u},
\tr{  \hPi \hPi^T  \hPhi^T  \hPhi \hPi \hPi^T v } 
\Biggr]
+ \frac{\ta}{M^2} 
\Biggl[ \tr{ 
  \hPi \hPi^T v},
\tr{  \hPi \hPi^T  \hPhi^T  \hPhi \hPi \hPi^T u } \Biggr], \nn
 \hat C_2 & = &  \frac{\ta^2}{M^4}
\Biggl[
\tr{  \hPi \hPi^T  \hPhi^T  \hPhi \hPi \hPi^T u }, 
\tr{  \hPi \hPi^T  \hPhi^T  \hPhi \hPi \hPi^T v } 
\Biggr]. 
\eqa
From \eq{eq:PhiPi} 
we obtain
\bqa
 \hat C_1 & = & -  \frac{4 i \ta}{M^2}
\tr{ \left(
 \hPi \hPi^T u  \hPi \hPi^T v -  \hPi \hPi^T v  \hPi \hPi^T u 
\right) \hat {\bf G}
}.
\eqa
Generators of  $GL(L, \mathbb{R})$ can be decomposed into generators of $SL(L, \mathbb{R})$
and a global dilatation generator in \eq{eq:GG}.
The contribution of the latter vanishes
due to the cyclic property of the trace.
This gives
\bqa
 \hat C_1 & = & - \frac{4 i \ta}{M^2}
\tr{ \left(
 \hPi \hPi^T u  \hPi \hPi^T v -  \hPi \hPi^T v  \hPi \hPi^T u 
\right) \hat G
}.
\label{eq:C1}
\eqa
Similarly, $ \hat C_2$ is given by
\bqa
 \hat C_2 & = & 
4 i \frac{ \ta^2}{M^4} u_{nk} v_{n'k'} 
\Bigl[
- (\hPi \hPi^T)^{kl} 
 (\hPhi^T  \hPhi)_{lm} 
 (\hPi \hPi^T)^{k'l'}   
  \hat {\bf G}^{[m}_{[l'} \delta^{n]}_{m']} 
  (\hPi \hPi^T)^{m'n'}  \nn
&& \hspace{2.5cm} + (\hPi \hPi^T)^{kl} 
  \hat {\bf G}^{[k'}_{[l} \delta^{l']}_{m]} 
 (\hPhi^T  \hPhi)_{l'm'} 
 (\hPi \hPi^T)^{m'n'}   
  (\hPi \hPi^T)^{mn}  \nn
&& \hspace{2.5cm}   + (\hPi \hPi^T)^{kl} 
      (\hPi \hPi^T)^{k'l'}   
 (\hPhi^T  \hPhi)_{l'm'} 
  \hat {\bf G}^{[m'}_{[l} \delta^{n']}_{m]} 
  (\hPi \hPi^T)^{mn}  \nn
&&\hspace{2.5cm} - (\hPi \hPi^T)^{k'l'} 
  \hat {\bf G}^{[k}_{[l'} \delta^{l]}_{m']} 
 (\hPi \hPi^T)^{m'n'}   
 (\hPhi^T  \hPhi)_{lm} 
  (\hPi \hPi^T)^{mn}
\Bigr].
\eqa
In the above expression, the  $GL(L, \mathbb{R})$ generators can be pushed to the far right
using the following commutators,
\bqa
\left[  \hat {\bf G}^{[i}_{[k} \delta^{j]}_{l]},    (\hPi \hPi^T)^{mn} \right]
= 2 i \delta^{[i}_{[k} \delta^{<m}_{l]}   (\hPi \hPi^T)^{j] n>}, \nn
\left[  \hat {\bf G}^{[i}_{[k} \delta^{j]}_{l]},    (\hPhi \hPhi^T)_{mn} \right]
= - 2 i \delta^{[i}_{[k} \delta^{j]}_{<m}   (\hPhi \hPhi^T)_{l] n>}, 
\eqa  
where pairs of indices enclosed by $[ ~~]$ and $<~ >$ are symmetrized,
e.g., 
$
\delta^{[i}_{[k} \delta^{<m}_{l]}   A^{j] n>}
= \frac{1}{8} \left(
  \delta^{i}_{k} \delta^{m}_{l}   A^{jn}
+\delta^{i}_{k} \delta^{n}_{l}   A^{jm}
+ \delta^{i}_{l} \delta^{m}_{k}   A^{jn}
+\delta^{i}_{l} \delta^{n}_{k}   A^{jm}
+ \delta^{j}_{k} \delta^{m}_{l}   A^{in}
+\delta^{j}_{k} \delta^{n}_{l}   A^{im}
+ \delta^{j}_{l} \delta^{m}_{k}   A^{in}
+\delta^{j}_{l} \delta^{n}_{k}   A^{im}
\right)$.
This allows us to write $\hat C_2$ as 
\bqa
 \hat C_2 =  \hat C_{21} +  \hat C_{22},
\eqa
where
\bqa
 \hat C_{21} &=& 
4 i \frac{ \ta^2}{M^4} u_{nk} v_{n'k'} 
\Biggl[ - ( \hPi \hPi^T)^{kl} ( \hPhi^T \hPhi)_{li}  ( \hPi \hPi^T)^{k' i'}  ( \hPi \hPi^T)^{j' n'} \delta^n_j   \nn &&
\hspace{2.5cm} + ( \hPi \hPi^T)^{ki'} ( \hPhi^T \hPhi)_{jl}  ( \hPi \hPi^T)^{l n'}  ( \hPi \hPi^T)^{j' n}  \delta^{k'}_i  \nn &&
\hspace{2.5cm} + ( \hPi \hPi^T)^{ki'}  ( \hPi \hPi^T)^{k' l} ( \hPhi^T \hPhi)_{l i}   ( \hPi \hPi^T)^{j' n}   \delta^{n'}_j   \nn&&
\hspace{2.5cm} - ( \hPi \hPi^T)^{k' i'}  ( \hPi \hPi^T)^{j' n'} ( \hPhi^T \hPhi)_{jl}   ( \hPi \hPi^T)^{l n}  \delta^k_i   
\Biggr] \hat {\bf G}^{[i}_{[i'} \delta^{j]}_{j']}, \nn
 \hat C_{22} & = &  
 \frac{\ta}{M^2} 
 \Biggl[ 
 (M-2) \tr{  (v  \hPi \hPi^T u - u  \hPi \hPi^T v ) \hat H } \nn
 && ~~~~~~~~
 +  4 \left(  \tr{  \hPi \hPi^T v} \tr{ \hat H u} -   \tr{  \hPi \hPi^T u} \tr{ \hat H v} \right) 
\Biggr] \nn
&& + 4 i \frac{\ta^2}{M^4} u_{nk} v_{n'k'} 
\Biggl[
  M ( \hPi \hPi^T)^{ki'}  ( \hPi \hPi^T)^{j' n'} \delta^{k'}_{i} \delta^{n}_{j}  
  + ( M+2) ( \hPi \hPi^T)^{k i'}  ( \hPi \hPi^T)^{k' n} \delta^{j'}_{i} \delta^{n'}_{j}  \nn &&
\hspace{2.5cm} +  2 ( \hPi \hPi^T)^{k i'}  ( \hPi \hPi^T)^{j' n} \delta^{k'}_{i} \delta^{n'}_{j}  
  -2  ( \hPi \hPi^T)^{k n'}  ( \hPi \hPi^T)^{k' i'} \delta^{j'}_{i} \delta^{n}_{j}  \nn &&
\hspace{2.5cm}   -2 ( \hPi \hPi^T)^{k n'}  ( \hPi \hPi^T)^{j' i'} \delta^{k'}_{i} \delta^{n}_{j}  
  -2 ( \hPi \hPi^T)^{nk}  ( \hPi \hPi^T)^{n' i'} \delta^{j'}_{i} \delta^{k'}_{j}  
\Biggr]  \hat {\bf G}^{[i}_{[i'} \delta^{j]}_{j']}. 
\eqa
By writing 
$ \hat {\bf G}^{[i}_{[i'} \delta^{j]}_{j']} =  \hat  G^{[i}_{[i'} \delta^{j]}_{j']}
+ \hat G_0 \delta^{[i}_{[i'} \delta^{j]}_{j']}$ following  \eq{eq:GG},
the contributions from the $SL(L, \mathbb{R})$ generators
and the global dilatation can be separated.
The terms that are proportional to $\hat G_0$ are canceled between $C_{21}$ and $C_{22}$,
and we obtain
\bqa
 \hat C_{2} &=& 
4 i \frac{ \ta^2}{M^4} u_{nk} v_{n'k'} 
\Biggl[
 - ( \hPi \hPi^T)^{kl} ( \hPhi^T \hPhi)_{li}  ( \hPi \hPi^T)^{k' i'}  ( \hPi \hPi^T)^{j' n'} \delta^n_j   \nn &&
\hspace{2.5cm} + ( \hPi \hPi^T)^{ki'} ( \hPhi^T \hPhi)_{jl}  ( \hPi \hPi^T)^{l n'}  ( \hPi \hPi^T)^{j' n}  \delta^{k'}_i  \nn &&
\hspace{2.5cm} + ( \hPi \hPi^T)^{ki'}  ( \hPi \hPi^T)^{k' l} ( \hPhi^T \hPhi)_{l i}   ( \hPi \hPi^T)^{j' n}   \delta^{n'}_j   \nn&&
\hspace{2.5cm} - ( \hPi \hPi^T)^{k' i'}  ( \hPi \hPi^T)^{j' n'} ( \hPhi^T \hPhi)_{jl}   ( \hPi \hPi^T)^{l n}  \delta^k_i    \nn &&
\hspace{2.5cm} + M ( \hPi \hPi^T)^{ki'}  ( \hPi \hPi^T)^{j' n'} \delta^{k'}_{i} \delta^{n}_{j}  
  + ( M+2) ( \hPi \hPi^T)^{k i'}  ( \hPi \hPi^T)^{k' n} \delta^{j'}_{i} \delta^{n'}_{j}  \nn &&
\hspace{2.5cm} +  2 ( \hPi \hPi^T)^{k i'}  ( \hPi \hPi^T)^{j' n} \delta^{k'}_{i} \delta^{n'}_{j}  
  -2  ( \hPi \hPi^T)^{k n'}  ( \hPi \hPi^T)^{k' i'} \delta^{j'}_{i} \delta^{n}_{j}  \nn &&
\hspace{2.5cm}   -2 ( \hPi \hPi^T)^{k n'}  ( \hPi \hPi^T)^{j' i'} \delta^{k'}_{i} \delta^{n}_{j}  
  -2 ( \hPi \hPi^T)^{nk}  ( \hPi \hPi^T)^{n' i'} \delta^{j'}_{i} \delta^{k'}_{j}  
\Biggr] \hat  G^{[i}_{[i'} \delta^{j]}_{j']} \nn
&& 
%
+ \frac{\ta}{M^2} 
 \Biggl[ 
 (M-2) \tr{  (v  \hPi \hPi^T u - u  \hPi \hPi^T v ) \hat H } \nn
 && ~~~~~~~~
 +  4 \left(  \tr{  \hPi \hPi^T v} \tr{ \hat H u} -   \tr{  \hPi \hPi^T u} \tr{ \hat H v} \right) 
\Biggr].
\label{eq:C2f}
\eqa
Due to the cyclic property of the trace,
$ \tr{   ( v  \hPi \hPi^T u  - u  \hPi \hPi^T v )  \hat H } $  vanishes.
Combining \eq{eq:C2f} with \eq{eq:C1},
we obtain \eq{HHc}.

\section{Poisson brackets of the momentum and Hamiltonian}
\label{app:Poisson}

From \eq{Poisson}, the Poisson bracket between momentum constraints is given by
\bqa
 \{  \cG_x, \cG_y \}_{PB} & = & 
 \Big\{ 
 \tr{ ( sq + 2 t_c p_c)x},
  \tr{ ( sq + 2 t_{c'} p_{c'} )y}
 \Bigr\}_{PB} \nn
&=& 
 \tr{
 x s q y
- q x y s 
+ 2 (  x t_c p_c y + x t_c y^T p_c ) 
- 2 ( p_c x y t_c + p_c x t_c y^T ) 
} \nn
&=& \cG_{yx -xy}.
 \eqa
By choosing 
$x^{j'}_{i'} = \delta^{j'}_j \delta^i_{i'}$,
$y^{l'}_{k'} = \delta^{l'}_l \delta^k_{k'}$,
we obtain \eq{eq:GGH1}.

To prove \eq{eq:GGH2}, we first note
\bqa
\Big\{  \cG_x,  
\tr{ U v }
\Bigr\}_{PB} & = & 
U_{vx+x^Tv}, \nn
\Big\{  \cG_x,  
\tr{ Q {\td v} }
\Bigr\}_{PB} & = & 
-Q_{x \td v + \td v x^T}.
\label{GUQ }
\eqa
Here
$U = \left( s s^T + \sum_c \left[ 4 t_c p_c t_c - i t_c \right]  \right)$ and
$Q=\left( q^T q +\sum_c  p_c  \right)$.
$v$ and $\td v$ are covariant and contravariant symmetric tensors
that are independent of $s,q,t_c,p_c$, respectively.
Using the distribution rule,
we write the Poisson bracket between the momentum and the Hamiltonian constraints  as
\bqa
 \Bigl\{  \cG_x, \cH_v \Bigr\}_{PB} & = &
  \Bigl\{  \cG_x, ( -U + \ta U QU )_v \Bigr\}_{PB} \nn
  &=& 
   -  \Bigl\{  \cG_x,  U_v  \Bigr\}_{PB} 
 + \ta   \Bigl\{  \cG_x,  U_{QU v}  \Bigr\}_{PB} 
 + \ta   \Bigl\{  \cG_x,  Q_{UvU}  \Bigr\}_{PB} 
 + \ta   \Bigl\{  \cG_x,  U_{vUQ }  \Bigr\}_{PB}.
 \label{eq:D3}
\eqa
Here $ \Bigl\{  \cG_x,  U_{A}  \Bigr\}_{PB} \equiv $
$  \Bigl\{  \cG^i_{~~j},  U^{lm}   \Bigr\}_{PB}
x^j_{~~i}
A_{ml}$
and
$ \Bigl\{  \cG_x, Q_{B}   \Bigr\}_{PB} \equiv$
$  \Bigl\{  \cG^i_{~~j},  Q_{lm}   \Bigr\}_{PB}
x^j_{~~i}
B^{ml}$.
$A$ and $B$ are covariant and contravariant tensors that 
are made of $v,x,s,q,t,p$.
From \eq{GUQ }, we obtain
\bqa
 \Bigl\{  \cG_x, \cH_v \Bigr\}_{PB} & = &
 \left( 
- U + \ta U QU 
\right)_{vx + x^T v}.
\eqa

In order to compute the Poisson bracket between Hamiltonians,
we use
\bqa
\Big\{  \tr {U v_1}, \tr{ U v_2} \Bigr\}_{PB} & = & 0, \nn
\Big\{   \tr{ Qv_1 }, \tr{ Qv_2 } \Bigr\}_{PB} & = & 0, \nn 
\Big\{  \tr {U v_1},   \tr{ Qv_2 }  \Bigr\}_{PB} & = & -4 \cG_{  v_2 v_1 }.
\label{UUQQ}
\eqa
Similar to  \eq{eq:D3}, we write the Poisson bracket between Hamiltonians as
\bqa
 \Bigl\{  \cH_{v_1}, \cH_{v_2} \Bigr\}_{PB} & = & 
\Bigl\{
\tr{ ( -U + \ta U QU ) v_1},
\tr{ ( -U + \ta U QU ) v_2}
\Bigr\}_{PB} \nn
&=&
\ta \left[ 
- \Bigl\{ U_{v_1}, Q_{U v_2 U}           \Bigr\}_{PB} 
+  \Bigl\{ U_{v_2}, Q_{U v_1 U}           \Bigr\}_{PB}  \right] \nn
 && + \ta^2 \Biggl[
  \Bigl\{  U_{QU v_1}, Q_{Uv_2U}     \Bigr\}_{PB}
 +   \Bigl\{  Q_{Uv_1U}, U_{QU v_2}     \Bigr\}_{PB} \nn
&& ~~~~~ + \Bigl\{  Q_{Uv_1U}, U_{v_2UQ }     \Bigr\}_{PB}
 + \Bigl\{  U_{v_1UQ }, Q_{Uv_2U}     \Bigr\}_{PB}
\Biggr].
 \label{eq:GGH3a}
\eqa 
From \eq{UUQQ}, we obtain
\bqa
 \Bigl\{  \cH_{v_1}, \cH_{v_2} \Bigr\}_{PB} & = & 
 -4 ~\mbox{tr} \Bigl\{ \cG
 \Bigl[
- \ta ( U v_2 U v_1 - U v_1 U v_2 ) \nn
 &&
+ \ta^2 ( U v_2 U QU v_1 + U v_2 U v_1 UQ 
- U v_1 U QU v_2 - U v_1 U v_2 UQ ) 
 \Bigr]
 \Bigr\}.
 \label{eq:GGH3b}
\eqa 
By choosing
$(v_1)_{i'j'} = \delta^{ij}_{i'j'}$,
 $(v_2)_{l'k'} = \delta^{lk}_{l'k'}$,
 we obtain 
 \eq{eq:GGH3}.

\section{Derivations of the transformed shift and lapse functions}
\label{app:fields}

In this appendix, we derive Eqs. 
(\ref{eq:GGcc1}),
(\ref{eq:GGcc2}),
(\ref{eq:zetavx}) and
(\ref{Gmunu}).

\subsection{\eq{eq:GGcc1}}
\label{app:fields1}

From \eq{eq:zeta}, the scalar field associated with the shift tensor $yx-xy$ is given by
\bqa
\zeta_{yx-xy}(r_i) & = & \sum_{j,k} ( y^j_{~k} x^k_{~i} -   x^j_{~k} y^k_{~i} ).
\eqa
The sum over $j$ gives 
\bqa
 \zeta_{yx-xy}(r_i) & = & \sum_{k} ( \zeta_y(r_k) x^k_{~i} - \zeta_x(r_k)  y^k_{~i} ).
 \eqa
 Now, we expand $ \zeta_y(r_k)$ and  $ \zeta_x(r_k)$ around $r_i$ to write
 \bqa
 \zeta_{yx-xy}(r_i) & = &  \sum_{k} \Biggl(  \Bigl[ \zeta_y(r_i) + \partial_\mu \zeta_y(r_i) ( r^\mu_k - r^\mu_i  ) + O(\partial^2)  \Bigr]  x^k_{~i} \nn
&&  ~~~~~ -  \Bigl[ \zeta_x(r_i) + \partial_\mu \zeta_x(r_i) ( r^\mu_k - r^\mu_i  )  + O(\partial^2) \Bigr] y^k_{~i} \Biggr) \nn
&=& \xi_x^\mu(r_i)   \partial_\mu \zeta_y(r_i) -  \xi_y^\mu(r_i)   \partial_\mu \zeta_x(r_i)  + O(\partial^2) .
\label{ap:GGcc1}
\eqa 

\subsection{\eq{eq:GGcc2}}
\label{app:fields2}

From \eq{eq:xi}, the vector field associated with the shift tensor $yx-xy$ is given by
\bqa
&& \xi^\mu_{yx-xy}(r_i) =  \sum_{j,k} ( y^j_{~k} x^k_{~i} -   x^j_{~k} y^k_{~i} )( r^\mu_j - r^\mu_i ).
\eqa
Expanding $y^j_{~k}$ and $x^j_{~k}$ around $r_k=r_i$ gives
\bqa
&& \xi^\mu_{yx-xy}(r_i) =
 \sum_{j,k} \left(  \Bigl[ y^j_{~i}  +   \frac{ \partial y^j_{~i}}{\partial r^\nu_i} ( r^\nu_k - r^\nu_i  ) + O(\partial^2) \Bigr]  x^k_{~i} - 
  \Bigl[ x^j_{~i}  +   \frac{ \partial x^j_{~i}}{\partial r^\nu_i} ( r^\nu_k - r^\nu_i  ) + O(\partial^2) \Bigr]     y^k_{~i} \right)( r^\mu_j - r^\mu_i ) \nn
& = &
 \xi^\mu_y(r_i) \zeta_x(r_i) + \xi^\nu_x(r_i) \left( \frac{ \partial \xi^\mu_y(r_i)}{\partial r^\nu_i} + \delta^\mu_\nu \zeta_y(r_i) \right)
 - \xi^\mu_x(r_i) \zeta_y(r_i) - \xi^\nu_y(r_i) \left( \frac{ \partial \xi^\mu_x(r_i)}{\partial r^\nu_i} + \delta^\mu_\nu \zeta_x(r_i) \right)
  + O(\partial^2)
  \nn
&=& \xi^\nu_x(r_i)  \partial_\nu \xi^\mu_y(r_i) 
- \xi^\nu_y(r_i)  \partial_\nu \xi^\mu_x(r_i)
 + O(\partial^2).
\label{ap:GGcc2}
\eqa

\subsection{\eq{eq:zetavx}}
\label{app:fields3}

From \eq{eq:theta}, the lapse function associated with the shift tensor $vx+x^T v$ is given by
\bqa
\theta_{vx + x^T v}(r_i) & = & 2 \sum_{k}  v_{i k} x^{k}_{~i}, \nn
 \lambda_{vx + x^T v}(r_i,r_j) 
&=&  \sum_{k} \left( v_{i k} x^{k}_{~j} + v_{j k} x^k_{~i} \right).
\eqa
Since $\theta(r_i)$ can be obtained from $\lambda(r_i,r_j)$ by setting
$r_j=r_i$, we first consider the latter.
Expanding $v_{ik}$  around $r_k=r_j$, and $v_{jk}$ around  $r_k=r_i$ gives
\bqa
\lambda_{vx + x^T v}(r_i,r_j) 
& = & \sum_{k} \left(
\Bigl[
v_{ij} + \frac{ \partial v_{ij}}{\partial r^\mu_j} ( r^\mu_k - r^\mu_j ) + O(\partial^2)
\Bigr] x^k_{~j}
+
\Bigl[
v_{ji} + \frac{ \partial v_{ji}}{\partial r^\mu_i} ( r^\mu_k - r^\mu_i ) + O(\partial^2)
\Bigr] x^k_{~i}
\right) \nn
&=& 
v_{ij} \left[ \zeta_x(r_j) +  \zeta_x(r_i) \right] 
+  \frac{ \partial v_{ij}}{\partial r^\mu_j} \xi^\mu_x(r_j)
+  \frac{ \partial v_{ij}}{\partial r^\mu_i} \xi^\mu_x(r_i)
+O(\partial^2) \nn
& = & \left[  \zeta_x(r_i) + \zeta_x(r_j) \right] \lambda_v(r_i,r_j) 
+ \xi^\mu_x(r_j)
  \frac{ \partial \lambda_v(r_i,r_j) }{\partial r^\mu_j}
  + \xi^\mu_x(r_i)
 \frac{ \partial \lambda_v(r_i,r_j)}{\partial r^\mu_i} 
 + O(\partial^2). \nn
 \label{eq:lambdat}
\eqa
For $\theta_{vx + x^T v}$, we can simply set $r_j=r_i$  in \eq{eq:lambdat},
\bqa
\theta_{vx + x^T v}(r_i) 
&=& 2 
 \zeta_x(r_i) 
\theta_v(r_i)
+ \xi^\mu_x(r_i)
  \frac{ \partial \theta_v(r_i) }{\partial r^\mu_i}.
\eqa

\subsection{\eq{Gmunu}}
\label{app:fields4}

From \eq{eq:GGH3}, 
we have
\bqa
\{ H_u, H_v \}_{PB} & = & 
\cG^m_{~n} \cC^{ijkln}_m u_{ji} v_{lk},
\eqa
where $ \cC^{ijkln}_m$ is given by \eq{eq:ABC}.
For diagonal $u$ and $v$,
we write
\bqa
\{ H_u, H_v \}_{PB} & = & 
\cG^m_{~n} \cC^{r_1 r_1 r_2 r_2 n}_m \theta_u(r_1) \theta_v(r_2) +  O( \partial^2),
\eqa
where  \eq{eq:theta} is used.
Now we expand $\theta_u(r_1)$ and $\theta_v(r_2)$ near $r_1=r_m$ and $r_2=r_m$ to write
\bqa
\{ H_u, H_v \}_{PB} & = & 
\cG^m_{~n} \cC^{r_1 r_1 r_2 r_2 n}_m 
\Bigl[ \theta_u(r_m) + \nabla_\mu \theta_u(r_m) ( r^\mu_1 - r^\mu_m) \Bigr]
\Bigl[ \theta_v(r_m) + \nabla_\nu \theta_v(r_m) ( r^\nu_2 - r^\nu_m) \Bigr] \nn
&&
+ O( \partial^2).
\label{eq:HuvPB}
\eqa
Because $\cC^{r_1 r_1 r_2 r_2 n}_m  = - \cC^{r_2 r_2 r_1 r_1 n}_m $,
\eq{eq:HuvPB} can be written as
\bqa
\{ H_u, H_v \}_{PB} & = & 
\frac{1}{2} \cG^m_{~n} \cC^{r_1 r_1 r_2 r_2 n}_m 
\Biggl\{
\Bigl[ \theta_u(r_m) + \nabla_\mu \theta_u(r_m) ( r^\mu_1 - r^\mu_m) \Bigr]
\Bigl[ \theta_v(r_m) + \nabla_\nu \theta_v(r_m) ( r^\nu_2 - r^\nu_m) \Bigr]  \nn
&& 
-\Bigl[ \theta_u(r_m) + \nabla_\mu \theta_u(r_m) ( r^\mu_2 - r^\mu_m) \Bigr]
\Bigl[ \theta_v(r_m) + \nabla_\nu \theta_v(r_m) ( r^\nu_1 - r^\nu_m) \Bigr]  
\Biggr\}
+ O( \partial^2). \nn
\eqa
To the leading order in the derivative expansion, 
we obtain
\bqa
\{ H_u, H_v \}_{PB} & = & 
\frac{1}{2} 
\cG^m_{~n} 
\cC^{r_1 r_1 r_2 r_2 n}_m 
 ( r^\nu_2 - r^\nu_1)
\Biggl\{
\theta_u(r_m)  \nabla_\nu \theta_v(r_m) - 
\theta_v(r_m)  \nabla_\nu \theta_u(r_m) 
\Biggr\}
+ O( \partial^2). \nn
\eqa
By using \eq{eq:Gy} and \eq{eq:Gyc},
we obtain \eq{Gmunu}.

\section{Metric in translationally invariant states}
\label{app:Sgmn}

We first note that $C^{ijkln}_m$ in \eq{eq:ABC} reduces to
\bqa
C^{ijkln}_m & = &
-4 \ta \left(
U^{n [l} U^{k][j} \delta^{i]}_m 
-U^{n[j} U^{i][l} \delta^{k]}_m 
\right)
\label{eq:Cr}
\eqa
on the constraint hypersurface that satisfy $ \ta QU = \ta UQ = I$.
Inserting \eq{eq:Cr} to  \eq{Gmunu2} and \eq{eq:metric}, we write
\bqa
-{\cal S} g^{\mu \nu} & = & \frac{1}{4} \sum_{i,l,n} C^{iilln}_m
 \left( r^\mu_{nm} r^\nu_{li} + r^\nu_{nm} r^\mu_{li} \right) \nn
&=&
-2 \ta \sum_{l,n} U^{nl} U^{lm} 
 \left( r^\mu_{nm} r^\nu_{lm} + r^\nu_{nm} r^\mu_{lm} \right), 
 \label{eq:gUU}
\eqa
where $r^{\mu}_{nm} = r^\mu_{n} - r^\mu_m$.
In the presence of the translational symmetry,
we write
\bqa
U^{nl} = \frac{1}{L} \sum_{\bk} e^{ i \bk  ( \br_{n} - \br_l ) } U_\bk
 \eqa
to rewrite the metric as
\bqa
-{\cal S} g^{\mu \nu} & = &
-2 \ta \sum_{l,n} 
\int \frac{ d \bk d \bk' }{(2 \pi)^{2D}}
U_{\bk} U_{\bk'}
\left[
-\left( \frac{\partial}{\partial k_\mu} + \frac{ \partial}{\partial k^{'}_\mu} \right) \frac{\partial}{\partial k^{'}_\nu}
-\left( \frac{\partial}{\partial k_\nu} + \frac{ \partial}{\partial k^{'}_\nu} \right) \frac{\partial}{\partial k^{'}_\mu}
\right]
e^{i \bk \br_{nl} + i \bk' \br_{lm} }. \nn
\eqa
Integrating $\bk$, $\bk'$ by part followed by the sum over $r_l, r_n$ results in
\bqa
-{\cal S} g^{\mu \nu} & = &
4 \ta 
\left( \frac{\partial U_\bk }{\partial k_\mu}  \frac{\partial U_\bk }{\partial k_\nu} + 
U_\bk \frac{ \partial^2 U_\bk }{\partial k_\mu \partial k_\nu}
\right)_{\bk=0}.
\eqa

\section{Transformation of the collective variables under boost}
\label{app:Lorentz}

In this appendix, we derive
\eq{eq:deltaxtpq}.

\subsection{The shift vector}

From \eq{eq:spacetimegauge},
the shift tensor is transformed as
\bqa
 \delta y^n_{~m} & = & \partial_\tau \eta^n_{~m} + 
 \eta^j_{~i} y^{l}_{~k}  \cA^{i k n}_{j  l m} 
 +   \rho_{ji} v_{lk}  \cC^{ijkl n}_{ m} \nn
 & = &
 \left( \partial_\tau \eta + y \eta - \eta y \right)^n_{~m}
 + \rho_{ii} v_{kk}  \cC^{iikk n}_{ m}.
\eqa
The change in the shift vector is given by
\bqa
\delta \xi^\mu(r_m) & = &
\sum_n \delta y^n_{~m} ( r^\mu_n - r^\mu_m). 
\eqa
Since $y$ 
in \eq{eq:gauge1}
is symmetric and $\eta$ is anti-symmetric,
$(y \eta - \eta y)$ is a symmetric matrix.
Furthermore, $(y \eta - \eta y)^n_{~m}$ depends 
on $n$ and $m$ only through 
$r_n-r_m$ due to the spatial translational invariance
of $y$ and $\eta$
in 
\eq{eq:gauge1}
and 
\eq{eq:etaji}.
As a result, 
\bqa
(*) & \equiv &  \sum_n ( y \eta - \eta y )^n_{~m} ( r^\mu_n - r^\mu_m) 
= \sum_n ( y \eta - \eta y )^n_{~0}  r^\mu_n \nn
& = &   \sum_n ( y \eta - \eta y )^{-n}_{~0}  r^\mu_{-n} 
=   \sum_n ( y \eta - \eta y )^{0}_{~n}  r^\mu_{-n} \nn
& = &  -\sum_n ( y \eta - \eta y )^{n}_{~0}  r^\mu_{n} = -(*).
\eqa
Here we choose the label of sites such that $r_{-n} = - r_n$,
and use the fact that 
$(y \eta - \eta y)^n_{~m} = (y \eta - \eta y)^m_{~n}
= (y \eta - \eta y)^{n+k}_{~m+k}$ for any $k$. 
This shows $(*)=0$, and
\bqa
\delta \xi^\mu(r_m) & = &
 \partial_\tau \sum_n \eta^n_{~m} ( r^\mu_n - r^\mu_m)
- 4  \td \alpha \sum_{n,i} 
\left[ U^{ni} U^{im} \rho_{mm} 
- U^{n i} \rho_{ii} U^{im} \right] ( r^\mu_n - r^\mu_m)  \nn
& = & \left[
\varepsilon \  \zeta^\mu  
 + 4 \td \alpha \sum_{n,i} U^{n i} \rho_{ii} U^{im} ( r^\mu_n - r^\mu_m)
 \right] \nn
 & = & \varepsilon \left[ 1 +  \cS  \right]  \zeta^\mu .
 \label{eq:deltaximu}
\eqa
In the first and the second lines, we use
\eq{eq:Cr}
and
$\sum_{n,i} 
 U^{ni} U^{im} \rho_{mm}  ( r^\mu_n - r^\mu_m) = 0$
which follows from the facts that
 $\sum_i U^{ni} U^{im} $ is symmetric under the exchange of $n$ and $m$,
 and depends only on $r_n-r_m$ due to the translational invariance
of the saddle point solution.
 From the second line to the third line, 
 we use 
 $
  \sum_{n,i} U^{n i} \rho_{ii} U^{im} ( r^\mu_n - r^\mu_m)
  =\varepsilon \zeta_\nu \sum_{n,i} U^{n i}  U^{im} r_i^\nu  ( r^\mu_n - r^\mu_m)
 =  \varepsilon \zeta_\nu  \sum_{n,i} U^{n i}  U^{im} (r_i^\nu - r_m^\nu + r_m^\nu) ( r^\mu_n - r^\mu_m)
  = \varepsilon \zeta_\nu \sum_{n,i} U^{n i}  U^{im} (r_i^\nu - r_m^\nu )  ( r^\mu_n - r^\mu_m)
= \varepsilon \frac{\cS}{ 4 \td \alpha}   g^{\nu \mu} \zeta_\nu$
which follows from  \eq{eq:gUU}.
 \eq{eq:deltaximu} vanishes when the spacetime has the Lorentzian signature,
   $\cS =-1$.

\subsection{The lapse function}

Under the boost, 
the diagonal component of the lapse tensor is transformed as
\bqa
\delta v_{nn} & = & 
\partial_\tau \rho_{nn} 
+  \eta^j_{~i} v_{lk}  \cB^{i k l}_{j  n n} 
- \rho_{kl} y^{j}_{~i}  \cB^{ ikl}_{  j nn} \nn
&= & 2  \left( \eta^n_{~n} v_{nn} - y^n_{~n} \rho_{nn} \right) = 0.
\eqa
Here $n$ is not summed over.
This vanishes  
because $\eta$ 
in \eq{eq:etaji} 
is anti-symmetric,
and $y^n_{~n} = 0$ 
for the translationally invariant saddle-point solution
in \eq{eq:gauge1}.
Therefore, $\delta \theta(r) = 0$.

\subsection{ $p_c$}

Under the boost, $p_c$ transforms as
\bqa
\delta p_c & = &  \{ p_c, \cG_\eta \}_{PB} +   \{ p_c, \cH_\rho \}_{PB} \nn
& = &   \left( p_c \eta + \eta^T p_c \right) +   \left( 4 p_c t_c \rho + 4 \rho t_c p_c - i \rho \right)  \nn
& = &  4  \left( p_c \td t_c \rho +  \rho \td t_c p_c \right). 
\eqa 
The first term in the second line vanishes because
$ \left( p_c \eta + \eta^T p_c \right)_{ij}  
=p_{c,il} \eta^l_{~j} + p_{c,jl} \eta^l_{~i}
=p_{c,il} \eta^l_{~j} + p_{c,j (j+i-l)} \eta^{j+i-l}_{~i}
=p_{c,il} \eta^l_{~j} + p_{c,li} \eta^j_{~l}
=p_{c,il} \eta^l_{~j} - p_{c,il} \eta^l_{~j}
=0
$,
where we use the fact that $p_{c,il}=p_{c,li}=p_{c,(i+k)(l+k)}$
and $\eta^i_{~j}= - \eta^j_{~i} =\eta^{i+k}_{~j+k} $  for any $k$. 
From the second line to the third line, we
use \eq{eq:tildet}.

\subsection{ $q$}

Under the boost generated by \eq{eq:Kz}, 
$q$ transforms as
\bqa
\delta q & = &   \{ q, \cG_\eta \}_{PB} +
 \{ q, \cH_\rho \}_{PB} + 
 \{ q, \cT_{\td o_L} \}_{PB} 
  \nn
& = & 
   \left( 
 \bar q \eta + 2  s^T \rho + \td o_L \bar q\right), \label{eq:dqe}
\eqa
Since $\eta$ is anti-symmetric and $\bar q = q_d I$, 
the transformation generated by $\cG_\eta$ can be canceled
by  a $O(L)$ flavour rotation.
By choosing $\td o_L = -\eta$,
we obtain
\bqa 
\delta q = 2  s^T \rho.
\label{eq:netdeltaq}
\eqa

\end{appendix}

\end{document}